\documentclass[a4paper,11pt,twoside,openright, titlepage]{report}

\usepackage[dvips]{graphicx} 
\usepackage{calc,fancyhdr,array,epsfig}
\usepackage{amsmath,amssymb,amsthm}

\usepackage{floatflt}
\usepackage{caption, subfigure, fancybox}

\newcommand{\chapitre}{\newpage\thispagestyle{empty}\chapter}

\newcommand{\marges}[2]{\setlength{\textwidth}{597pt-#2-#1}
				\setlength{\evensidemargin}{-72pt+#2}
				\setlength{\oddsidemargin}{-72pt+#1}}
\marges{90pt}{50pt}

\fancypagestyle{plain}{\fancyhf{}\fancyfoot[C]
					{\bfseries -\ \thepage\ -}}
\marges{90pt}{112pt}

\newcommand{\fl}{\rightarrow}
\newcommand{\ctab}[1]{\multicolumn{1}{c|}{#1}}
\newcommand{\ctabi}[1]{\multicolumn{1}{|c|}{#1}}

\newcolumntype{A}[1]{>{\raggedleft}m{#1in}}
\newcolumntype{B}[1]{>{\centering}m{#1in}}

\begin{document}
 
	\begin{titlepage}
\topmargin = 0pt

\begin{figure}[h]
\mbox{
\subfigure{\includegraphics*[width=5cm]{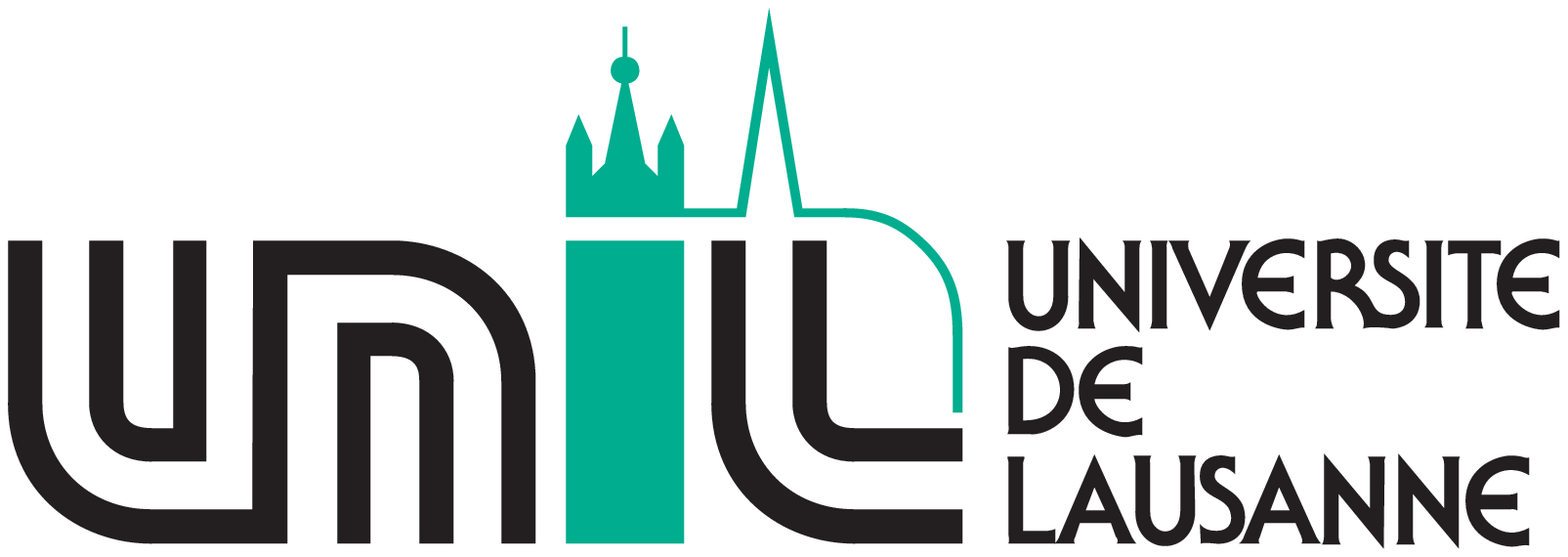}
}
\hspace*{5cm}
\subfigure{\includegraphics*[width=3cm, bb= 0 27 291 225]{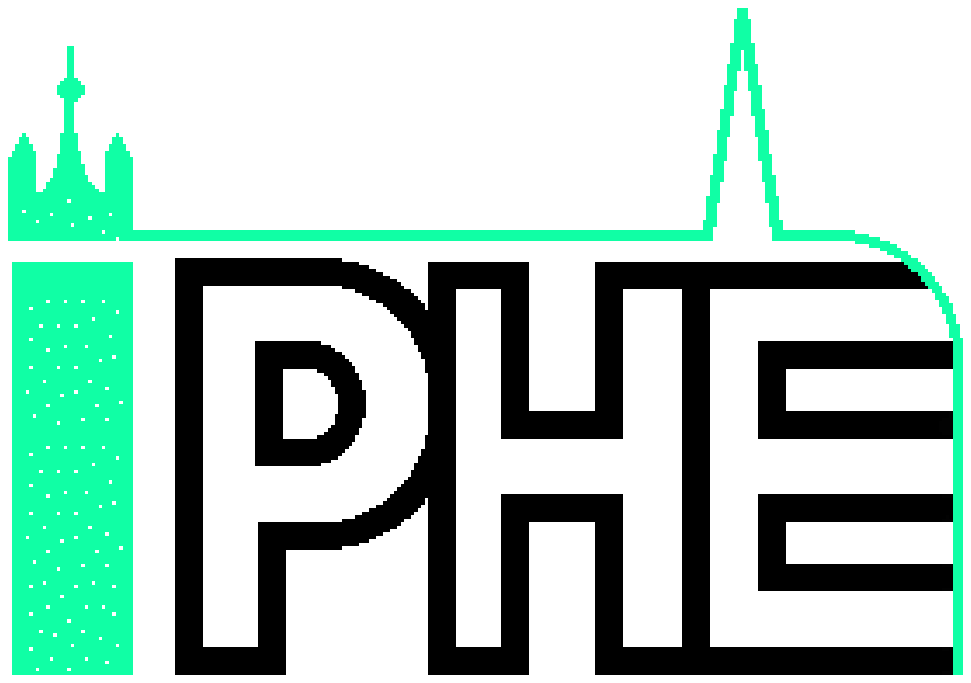}
}
}
\end{figure}

\begin{center}
\vspace*{0.4cm}
\huge Prospects of Measuring \\ $qq \rightarrow qqH$ and  $gg \rightarrow H$
Cross Sections \\ at the LHC with CMS\\ 
\end{center}
\begin{center}
\LARGE{or}\\
\vspace{0.5cm}
\huge{the \emph{Higgs files}}\\
\vspace{1.5cm}
\Large{Diploma thesis} \\
\vspace{1.75cm}
\Large{Anne-Sylvie Nicollerat} \\
\vspace{1.5cm}
\normalsize{performed at CERN} \\
\vspace{0.15cm}
\normalsize{under supervision of} \\
\vspace{0.75cm}
\Large{Dr. Michael Dittmar,} \normalsize{ETHZ} \\
\vspace{0.15cm}
\Large{Prof. Felicitas Pauss,} \normalsize{ETHZ} \\
\vspace{0.15cm}
\Large{Prof. Tatsuya Nakada,} \normalsize{Universit\'e de Lausanne} \\
\vspace{1.75cm}
\large{October 2000 to March 2001} \\
\end{center}

\end{titlepage}

\begin{abstract}

The possibility to observe a Higgs boson
having a mass between 300 and 600~GeV
and to measure its couplings to vector bosons and top quark 
with CMS at the LHC is studied. The different signatures:
\begin{equation}
\nonumber
\begin{split}
H &\fl ZZ \fl \ell^+ \ell^- \ell^+ \ell^-\\
H &\fl ZZ \fl \ell^+ \ell^- \nu \bar{\nu}\\ 
H &\fl ZZ \fl \ell^+ \ell^- q \bar{q}\\
H &\fl ZZ \fl \nu \bar{\nu} q\bar{q}\\
H &\fl WW \fl \ell \nu \ell \nu\\
H &\fl WW \fl \ell \nu q q
\end{split}
\end{equation}
are analyzed.
The possibility to separate the Higgs events produced through weak boson fusion 
($qq \fl qqH$) from
the Higgs produced through gluon fusion ($gg \fl H$)
using the forward going jets emitted in the weak boson fusion
process is discussed for each of these different channels.
The results are then used to determine the possible statistical errors
on the ratio between the two Higgs decay branching 
fractions $H\fl ZZ$ and $H\fl WW$ and on
the weak boson fusion and gluon
fusion cross sections after one year of LHC running.

\end{abstract}

\topmargin = 0.25in

	\tableofcontents\markboth{Table of contents}{Table of contents}       
        \chapitre*{Introduction \\
\small{or how to explain to my friends what I did during four months...}}
\addcontentsline{toc}{chapter}{Introduction}
\markboth{Introduction}{Introduction}

During the late sixties, Salam, Glashow and Weinberg proposed a Model, the
\emph{Standard Model of electroweak interactions}\footnote{
The idea of Salam, Glashow and Weinberg \cite{hist} was to 
propose a gauge theory which unify
weak and electromagnetic interactions using the symmetry groups $SU(2)$
 of 'weak isospin' and $U(1)$ of 'weak hypercharge'.}, which was giving
a new description of the interactions between the basic components of matter. 
Based on the same principles (the gauge symmetries), 
a Model which describes interactions of quarks and gluons was developed
later, the 
so-called Quantum Chromodynamic (QCD)\footnote{
In that theory a new symmetry group was added, describing the strong interactions,
$SU(3)$, completing the Standard Model gauge group: $SU(3)_c \times 
SU(2)_I \times U(1)_Y$. }. 
These models are now gathered under the name of
\emph{Standard Model of Particle Physics} and provide a description of the basic 
components of matter and its interactions.
Since that time, the Standard Model gave predictions which were
well confirmed by observations.
Within that Model and given the experimental results,
all the known matter can be described with 12 fundamental
entities, named fermions (spin 1/2 particles), 
together with some interaction carriers, named vector bosons (spin 1 particles).

These particles can interact via three types of forces\footnote{We 
exclude here the gravity, which 
is not described in that model.}, see Table \ref{13}: The
electromagnetic interaction, the strong interaction (which binds the  
quarks together, e.g. in the protons) and the weak interaction 
(used for the description of radioactivity, $\beta$ decay).

\begin{table}[htb]
\begin{center}
\begin{tabular}{|c|c|c|c|}
\hline
Interaction & Vector boson & Mass (GeV) & Fermions involved\\
\hline
Strong & 8 gluons & 0 & quarks \\
\hline
Electromagnetic  & $\gamma$ & 0 & all charged particles \\   
\hline
Weak  & $W^{\pm},Z^0$ & 80, 91 & all left-handed particles\\
\hline
\end{tabular}
\caption{Interactions of the Standard Model.}
\label{13}
\end{center}
\end{table}

The interactions are carried by particles named vector bosons. 
To 
understand the idea of an interaction carried by a particle, let's 
take the image of two children playing with a ball. 
As they are throwing the ball, they cannot move too far away from each 
other, as the ball will not reach a too long distance because of its weight.
They are 
somehow bound to each other, by the presence of the ball. 
In today used
representation of interactions, the particles can be considered like the children
and vector bosons like the ball.

The twelve fundamental fermions can be sorted out using their different
behavior under the interactions.
For example, quarks are defined as particles which feel the strong interaction
and leptons as particles which does not.
These twelve particles can also be sorted out in three doubleted families,  
each family having the same general behavior under the weak interaction.
These families are shown in Table \ref{12}.
One achievement of the late CERN accelerator, LEP, was to show that
there is no fourth family.

Everyday matter is made out of the components from the first 
family ($electrons$ and $electron$ $neutrino$, $up$ and $down$ quarks).
For example, a proton is made out of one $d$ and two $u$ quarks.
The two other generations ($muon$ and $muon$ 
$neutrino$, $c$ and $s$ quarks, 
$tau$ and $tau$ $neutrino$, $t$ and $b$ quarks) are used to describe high 
energy states like the ones which were present at the beginning of the Universe
and can now be created in the particle accelerators.\\

\begin{table}[htb]
\begin{center} 
\begin{tabular}{|c r|c r|c r|}
\hline
\multicolumn{2}{|c|}{1st family} &
\multicolumn{2}{|c|}{2nd family} &
\multicolumn{2}{|c|}{3rd family} \\
\hline
\multicolumn{4}{|c}{Leptons} & &\\ 
\hline
 & & & & & \\
 &  Mass \hspace*{0.1cm} & &  Mass \hspace*{0.1cm}& & 
  Mass \hspace*{0.1cm}\\
$\left( \begin{array}{c} \nu_{e} \\ e \\ \end{array}\right)$ & 
$\begin{array}{r} \scriptstyle{<0.003\,\mathrm{MeV}} 
\\ \scriptstyle{0.511\, \mathrm{MeV}} \\ 
    \end{array} $ &
$\left( \begin{array}{c}  \nu_{\mu} \\ \mu \\ \end{array}\right)$ & 
$\begin{array}{r} \scriptstyle{<0.17\,\mathrm{MeV}} 
\\ \scriptstyle{106\,\mathrm{MeV}} \\ 
    \end{array} $ &
$\left( \begin{array}{c} \nu_{\tau} \\ \tau \\ \end{array}\right)$  &
$\begin{array}{r} \scriptstyle{<18.2\,\mathrm{MeV}} 
\\ \scriptstyle{1777\,\mathrm{MeV}} \\ 
    \end{array} $ \\
 & & & & & \\ 
\hline
\multicolumn{4}{|c}{Quarks} &&\\ 
\hline
 & & & & & \\
 &  Mass \hspace*{0.1cm} & &  Mass \hspace*{0.1cm}& & 
  Mass \hspace*{0.1cm}\\
$\left( \begin{array}{c} u \\ d \\ \end{array}\right)$ & 
$\begin{array}{r} \scriptstyle{\sim 6\,\mathrm{MeV}} 
\\ \scriptstyle{\sim 3\, \mathrm{MeV}} \\ 
    \end{array} $ &
$\left( \begin{array}{c} c \\ s \\ \end{array}\right)$ & 
$\begin{array}{r} \scriptstyle{\sim 1'250 \,\mathrm{MeV}} 
\\ \scriptstyle{\sim 115\, \mathrm{MeV}} \\ 
    \end{array} $ &
$\left( \begin{array}{c} t \\ b \\ \end{array}\right)$  &
$\begin{array}{r} \scriptstyle{175'000\,\mathrm{MeV}} 
\\ \scriptstyle{\sim 4'250\, \mathrm{MeV}} \\ 
    \end{array} $ \\
 & & &&&\\
\hline
\end{tabular}
\caption{The building blocks of matter given by the Standard Model.}
\label{12}
\end{center}
\end{table}

All these elementary particles have 
very different masses: For instance the top quark is almost about one million times 
heavier than the electron.
The Standard Model as it was initially developed 
cannot explain such a variety in the masses of the 
particles. Even worse, it
predicted all particles to be massless\ldots To solve that problem, a 
British physicist, Peter Higgs, introduced a new field in the Standard Model,
named the Higgs field, which through the mechanism of 
\emph{spontaneous symmetry 
breaking}, would give their mass to the particles \cite{phiggs}. 
To understand better how this works, let's quote David Miller from
the Department of Physics and Astronomy of the University College of 
London: \\
\textit{Imagine a cocktail party of political party workers who 
are uniformly distributed across the floor, 
all talking to their nearest neighbors. The ex-Prime Minister enters and
crosses the room. All of the workers in her neighborhood are strongly attracted 
to her and cluster round her. As she moves she attracts the people she comes 
close to, while the ones she has left return to their even spacing. 
Because of the knot of people
always clustered around her she acquires a 
greater mass than normal, that is she has more momentum for 
the same speed of movement across the room. Once moving she is hard
to stop, and once stopped she is harder to get moving again because the
clustering process has to be restarted.}
\\
In this story, the Higgs field would be represented by the people
taking part to the party and the particles 
would be the ex-Prime Minister.
But now, how can we see that such a field exists ? In Quantum 
Mechanics,
fields can be seen as particles and vice versa. The idea will be
to detect the particle associated to that field, namely the \emph{Higgs 
boson} (a spin 0 particle).
To illustrate this, lets
go on with the story:\\
\emph{(...) Now consider a rumor passing through our room full of uniformly spread 
political workers. 
Those near the door hear of it first and cluster together to get the details,
then they turn and move closer to their next neighbors who want to 
know about it too. A wave of clustering passes through the room. It may spread 
to all the corners or it may form a compact bunch which carries the 
news along a line of workers from the door 
to some dignitary at the other side of the room. Since the information is
carried by clusters of people, and since it was the clustering that gave extra 
mass to the ex-Prime Minister, then the rumor-carrying clusters also have mass.} 
\\
The Higgs boson is indeed predicted to be just such a clustering in the Higgs field. 
We will find it much easier to believe that this field 
describes a physical reality, and that 
the mechanism for giving other particles masses is true, if we actually see the 
Higgs particle itself.

\begin{figure}[htb]
\begin{center}
\includegraphics*[scale=0.6,angle=90]
{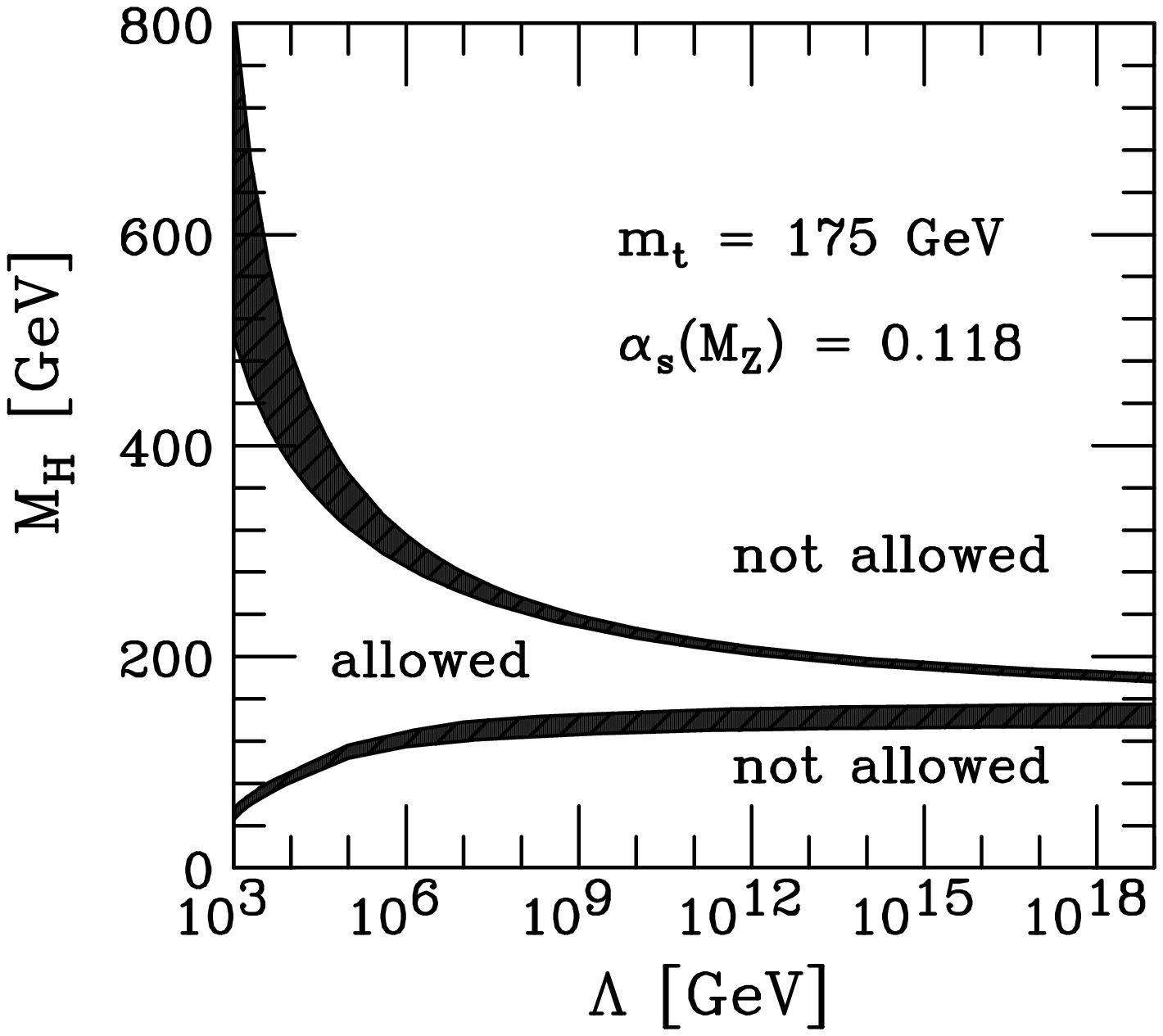}
\end{center}
\caption{The area between black curves shows the allowed Higgs
mass range assuming the validity of the Standard Model up to
an energy scale of $\Lambda$, \cite{smhigg}.}
\label{fig0}
\end{figure}

This particle is the only 
predicted particle of the Standard Model which still remains to be found.
The theory does not make prediction for the mass of that new 
particle. Lower and upper limits to its mass can nevertheless be set
\cite{smhigg}:
The Standard Model is 
assumed to give predictions only up to an energy
$\Lambda$.
Above that limit,
the method used to make calculations cannot be applied any more. 
The mass of the Higgs 
and $\Lambda$ are not independent and a light Higgs 
allows $\Lambda$ to be higher. 
If $\Lambda$ is large,
calculations with the Standard Model will be possible up to higher energies
and more energetic states of the Universe will be described
by that Model.
However arguments on the vacuum stability
suggest a lower Higgs mass limit. Figure \ref{fig0} shows how the Higgs 
mass can be constrained in respect to the parameter $\Lambda$.
For the Standard Model to be valid up to energies as high as possible, 
the Higgs mass should lie between 140 and 180~GeV. A higher or a
lower Higgs mass restricts the region in which the Standard Model
could describe Nature.
Today, physics is tested only up to a mass scale of about 100 to 200~GeV
(notice that the plot begins at energies of 1000~GeV~!) 
and almost all the measurements agree with the Standard Model.
The observation of a Higgs boson is a crucial test of the Standard Model
and at the same time the value of its mass will 
determine up to which energy the Standard Model could make predictions.

Thus today the Higgs boson can be considered like the Holy Grail of Particle 
Physics and if it is eventually found, it will be by far the most expensive 
particle of the Universe, given all the efforts that are and 
will be done to find it. The final answer to the question of its 
existence will probably be given in about five years either at the 
Fermilab accelerator Tevatron or at the CERN's new Collider, 
LHC.\\
Story to be continued\ldots \\[0.3cm]

The work I did started from the hypothesis
that the Higgs boson has a mass between 300 and 600~GeV.
It was interesting to study this mass region as it was not deeply analyzed yet
in opposition to the lighter Higgs
mass sector.
I studied how a heavy Higgs might be discovered and what 
one might learn about its properties.

The first chapter gives a short description of the Large hadron Collider (LHC),
the CERN's new accelerator and of one of its detectors (CMS), with which we hope
to find the Higgs. It also describes the simulation used.

The second chapter reviews the important physical characteristics of the Higgs, 
like its production and decays modes.
The next chapter gives the general ideas used in that study to find the selection
cuts.
The fourth chapter goes through the different Higgs search channels studied
and presents the results obtained.
Finally a discussion is made, analyzing the different properties
of the Higgs which could be measured with the previously obtained results.

	\chapitre{The Large Hadron Collider}
\markboth{The Large Hadron Collider}{The Large Hadron Collider}

CERN started to build a new accelerator, the Large Hadron Collider (LHC),
whose main task will be to find the Higgs boson. Situated in the
27-kilometer of the now former LEP tunnel, 
it will bring protons into head-on collisions at expected center of mass
energies of 14~TeV and should be ready to run in Spring 2006.

\section{LHC characteristics}

An important parameter, used to
characterize an accelerator needs to be defined first, namely, the
luminosity, $\mathcal{L}$. It determines, 
together with the cross section,~$\sigma$, the event rate
produced in a given process:
$$N_{evts}=\sigma \mathcal{L}$$
While the cross section depends on the physical
characteristics of the process, the luminosity depends on the characteristics
of the accelerator. For a circular accelerator, it is given by:
$$\mathcal{L}=\frac{N^2kf\gamma}{4\pi \epsilon_n \beta^*}$$ where
$N$ is the number of particles in each of the $k$ circulating bunches, $f$ the
revolution frequency, $\beta^*$ the value of the betatron function at the 
crossing point and $\epsilon_n$ the emittance corresponding to the one $\sigma$
contour of the beam, normalized by multiplying by the Lorentz factor 
$\gamma = (E/m_0 c^2)$.

In the first years, LHC should reach a luminosity
of $10^{33}\,cm^{-2}s^{-1}$ which represents an integrated
luminosity of $10\,fb^{-1}$ per
year. Then it will be raised till a maximum of $100\,fb^{-1}$ per year.
The designed frequency is about $40\,\mathrm{MHz}$, 
which is equivalent to a bunch crossing
approximatively every 25 nanosecond. 
Each crossing contains about 20 events, as the
total cross-section at hadron colliders is very large, about $100\,\mathrm{mb}$
for the LHC.
\\[0.5cm]
The detection of processes with a signal to total cross section ratio of about
$10^{-12}$, like for a $120\,\mathrm{GeV}$ Higgs decaying into two photons,
will be a difficult experimental challenge.
Note that
the interesting signatures for detecting the Higgs particle
are often characterized
by charged leptons or photons, since its hadronic decays is
overwhelmed by a huge QCD background. Purely leptonic modes lead to very
small branching fractions. In order to observe such signals, a machine with a
high constituents center-of-mass energy and a high luminosity is required.

Four detectors will be built around the LHC : ATLAS (A Toroidal LHC ApparatuS),
CMS (Compact Muons Solenoid), ALICE (A Large Ion Collider Experiment) and
LHCb. The first two are dedicated to Higgs searches and to the exploration of
possible new physics beyond the Standard Model at high masses,
whereas ALICE will analyze the collisions of 
heavy ions.
LHCb will study CP violation and
other rare phenomena in the decay of beauty particles.

\section{The Compact Muon Solenoid (CMS) detector}

Before describing the detector, it is important to say some words about the rapidity.
The rapidity, $y$, of a particle is defined like:
$$y=\frac{1}{2}ln\left(\frac{E+p_z}{E-p_z} \right) $$
The rapidity is defined compared with a direction, $z$, usually given by 
the beam axis.
If $p \gg m$, the rapidity can be approximated in the following way and we 
get the \emph{pseudorapidity}, $\eta$:
$$\eta=-ln(tan(\theta/2))$$

\begin{figure}[htb]
\begin{center}
\subfigure{\includegraphics[width=.45\textwidth]{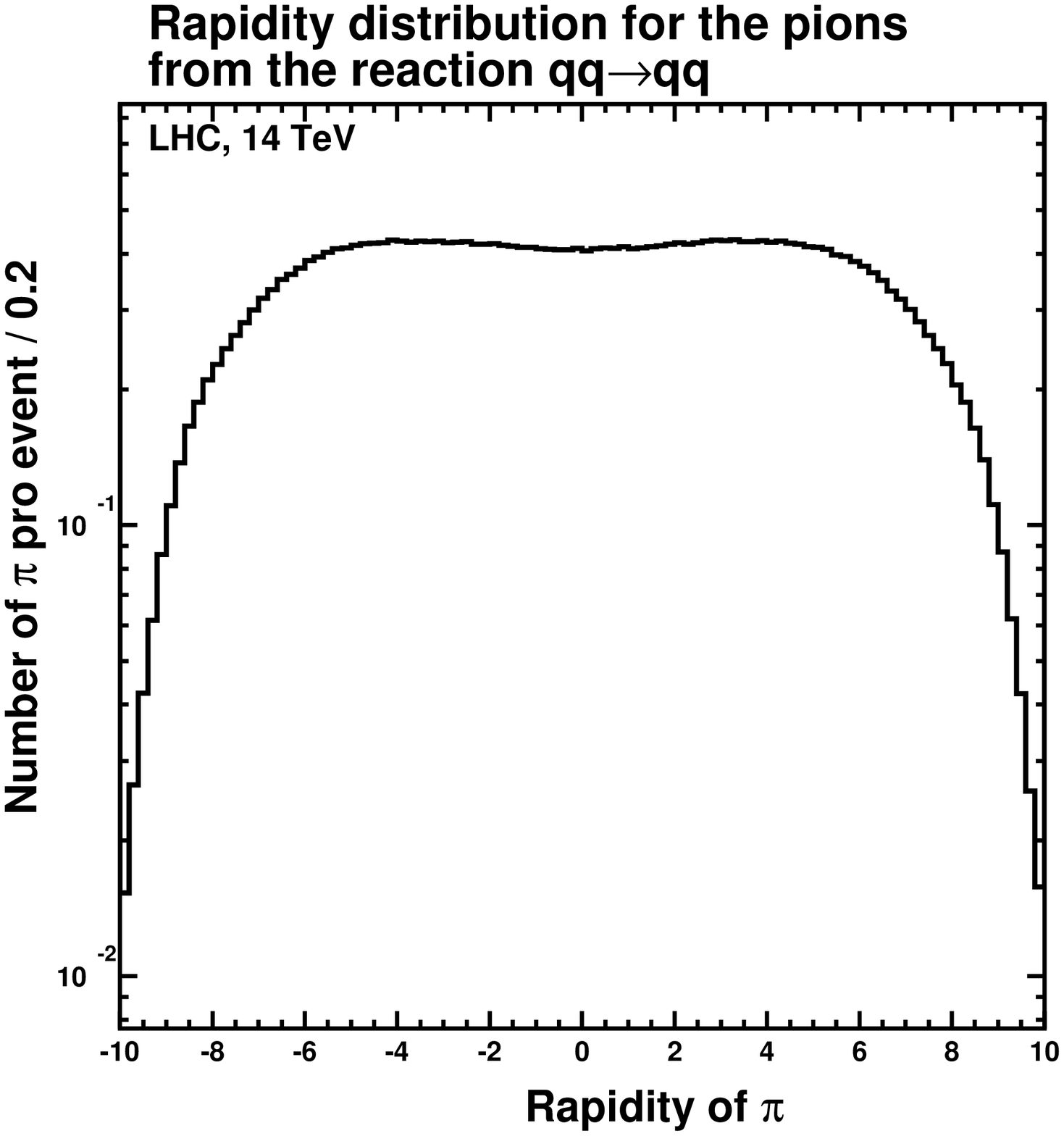}\quad
}
\subfigure{\includegraphics[width=.45\textwidth]{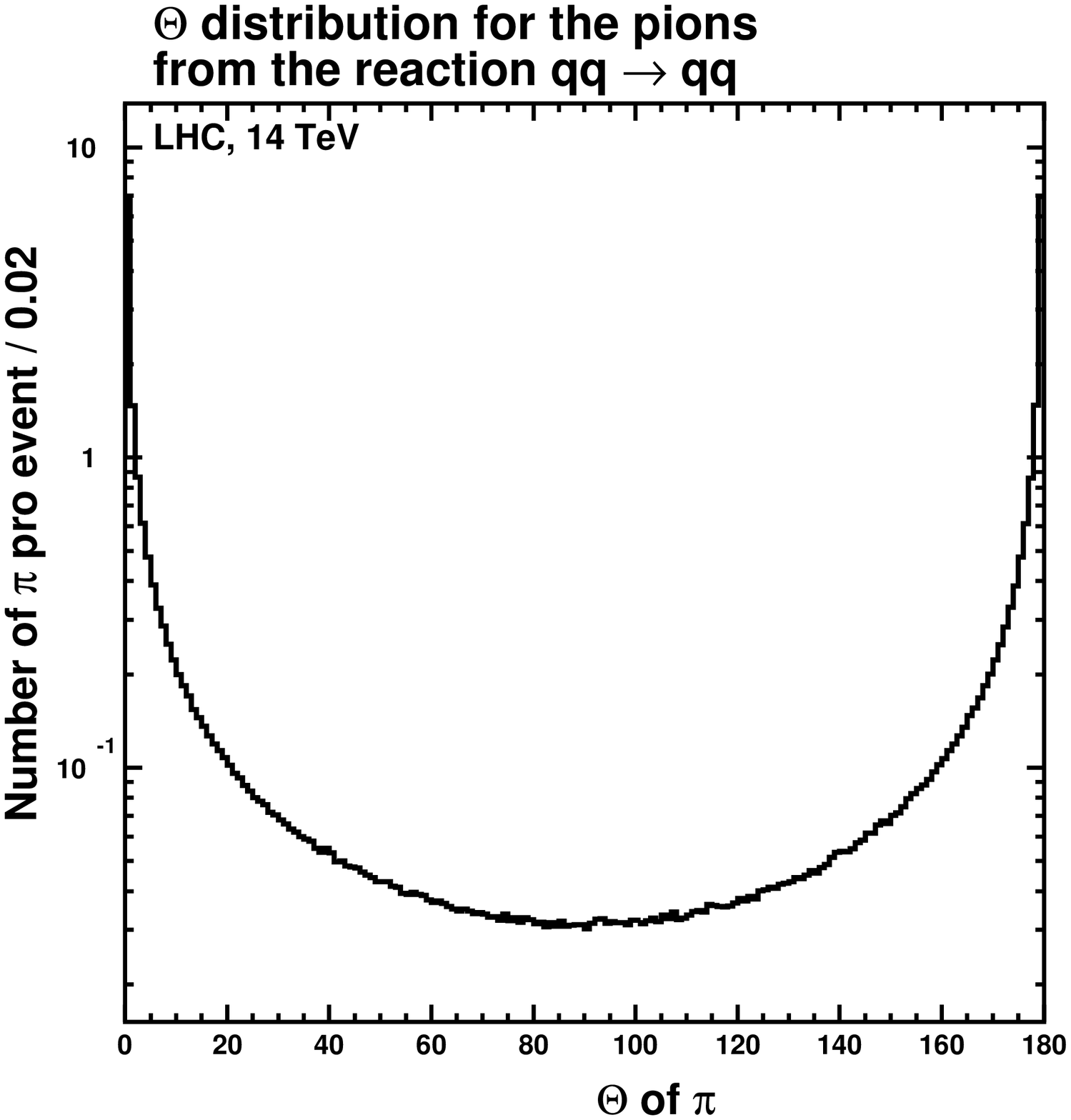}}
\caption{(Left) Rapidity distribution of the pions pro event of the QCD process
$qq \fl qq$ (all $p_t$ and all rapidities allowed) (Right)
$\Theta$ distribution of the pions for the same process.}
\label{raptheta}
\end{center}
\end{figure}

We see that the pseudorapidity is directly linked with the angle  
between the particle momentum and the beam pipe, $\theta$.
A pseudorapidity of 1, represents an angle of 40 degrees, a pseudorapidity of
2.5, an angle of 9 degree and a pseudorapidity of 4.5 an angle of 1 degree.
What is interesting with the rapidity, is that 
in opposition to $\theta$, 
it is only shifted by a constant under Lorentz transformations, leading to
flat distributions.
The density of rapidity is then independent of the chosen referential.
Rapidity is thus preferably used to describe the event geometry than 
$\theta$.
Figure \ref{raptheta} give a comparison 
between the $\eta$ and $\theta$ distribution obtained in the reaction
$qq \fl qq$ for the pions present in event.

The rapidity range in which the particles can be detected is
an important characteristic of
a detector.
\\[0.5cm]
CMS (see Ref. \cite{CMS})
is a detector optimized for the search of 
the Standard Model Higgs over a mass range 
from 100~GeV to 1~TeV. The detector has a length of 21.6~m, a diameter of 
14.6~m and a total weight of 14500~tons. It will be able to identify 
and measure isolated muons, photons and electrons with high precision.
The expected energy 
resolution for the above particles will be less than $1\%$ at 100~GeV.
The proposed CMS detector has a large superconductive solenoid with a radius
of about 2.9~m, generating an uniform magnetic field of 4~T. The choice of a
strong magnetic field leads to a compact design for the muon spectrometer 
without compromising the momentum resolution up to pseudorapidities
of 2.5.

Right around the beam pipe a tracking system will allow the detection of
charged particles. Then comes the electronic calorimeter (ECAL) which will detect
photons and electrons. Around it, 
lies the hadronic calorimeter (HCAL) which 
measures the energy of hadronically interacting
particles ($\pi^{\pm}$, $K^{\pm}$, $K^0_L$, $p$, $\bar{p}$, $n$, $\bar{n}$).
Finally the detector is surrounded by muons chambers which will
detect the muons.
The expected transversal impulse resolution is given, for high $p_t$ particles
by $\Delta p / p \approx 0.1\,p_t$ ($p_t$ in TeV) in the range $|\eta|<2.5$.
As the electromagnetic calorimeter is essentially designed to detect the Higgs
in the two photons channel, it has a small expected energy resolution
($\Delta E/E \approx 2\% / \sqrt{E}$). For the hadronic calorimeter, a resolution
of $\Delta E/E \approx 65\% / \sqrt{E}$ is expected. 
Muons should be detected up to $|\eta|=2.4$ and for
$p_t>4\,\mathrm{GeV}$.
Forward calorimeter will measure the energy of hadrons up to a rapidity
of $|\eta|\approx5$.

\section{The PYTHIA event generator}

PYTHIA is the standard event generator used by the LHC collaborations
to study the physics performance of their detectors.
Like the ancient Greeks who used to consult the Pythia to know their destiny,
modern physicists use PYTHIA to get a feeling for the kind 
of events one may expect/hope
to find, and at which rates.
An event generator simulates data, 
with more or less the same average behavior and fluctuations as the real data.
In PYTHIA,
data are simulated through Monte Carlo techniques.

PYTHIA gives the list of all particles together with their 
physical and kinematical properties 
involved in a given process.
The detector responses are still needed.
For the study
described in the following, a 'fast detector simulation' of CMS was made, where
different requirements were asked for a particle to be detected:
\begin{itemize}
\item Only stable particles can be detected, i.e. $\pi^{\pm}$, $K^{\pm}$, $K^0_L$, 
$p$, $\bar{p}$, $n$, $\bar{n}$ for the hadrons, $e^{\pm}$, 
$\mu^{\pm}$ for the leptons and photons.
\item The detected particles have to have $p_t>0.5\,\mathrm{GeV}$ and $|\eta|<4.5$.
\item Isolated electrons, photons and muons can be detected up to rapidity
of \\ $|\eta|<2.5$.
\item Jets can be detected up to a rapidity of $|\eta|<4.5$.
\end{itemize}
Other important detector characteristics like the 
energy and momentum 
resolutions were not simulated.
However for electrons, photons and muons the resolution effects are small for
the region of interest (about 1\%)
and should not have a large effect on the obtained results.

Jet jet mass reconstruction is more difficult, as it involves the combination
of several detector elements over large rapidity ranges for many particles. 
To take this effect into account,
larger selected $W$ and $Z$ mass ranges ($\pm$~20~GeV)
were taken than the simulated ones ($\pm$~10~GeV).

This fast simulation of the CMS detector should be
sufficient to get a first
idea of the detector capacity and to estimate the possible 
experimental accuracy 
that can be reached at CMS.

        \chapitre{Basic Higgs properties}
\markboth{Basic Higgs properties}{Basic Higgs properties}

Twelve years of direct Higgs searches at LEP show that the Higgs mass
should be higher than 113~GeV\footnote{Last year, LEPII reached that limit 
and observed signs for a Higgs boson of 115~GeV with a $2.9 \; \sigma$
deviation, see \cite{aleph}.}.
Many detailed Higgs studies for the LHC (for example \cite{cmss} and \cite{atlas})
give confidence that the Higgs 
boson with Standard Model - like couplings can be discovered at the LHC.

The experimental observation of one or several Higgs bosons will be fundamental
for a better understanding of the mechanism of electroweak symmetry breaking.
In the Standard Model, one doublet of scalar fields is assumed, 
leading to the existence of one neutral scalar particle, the Higgs,
with an essentially unconstrained mass.
An upper limit of about 1~TeV can be derived from unitarity arguments.
Assuming the Standard Model to be valid up to an energy scale $\Lambda$, 
the allowed Higgs range can be calculated. For $\Lambda$ set at Planck
scale, the Higgs should have a mass lying between 140 and 180~GeV.
However these bounds become weaker if new physics appears at lower mass scale.
If  $\Lambda$
is chosen to be 1~TeV, the Higgs boson mass is constrained 
to be in the range $50\,\mathrm{GeV} < m_H < 800\,\mathrm{GeV}$. 

Assuming the overall
validity of the Standard Model, a global fit to all electroweak data leads to
$m_H=76^{+85}_{-47}\,\mathrm{GeV}$ and its mass is less than 210~GeV
with 95\% confidence level.
However, in a recent paper
\cite{ew}, Peskin and Wells study a variety of new physics models which might
have a heavy Higgs and simultaneously some other new physics.
They review how these Models could
allow a heavy Higgs boson being consistent with the precision electroweak
constraints and come to the conclusion that the electroweak fit constraint
the Higgs to small masses, only if no other new physics enters in.

\section{Higgs production mechanisms}

\begin{figure}[p]
\begin{center}
\includegraphics*[scale=0.6,angle=0,bb=0 180 580 420 ]{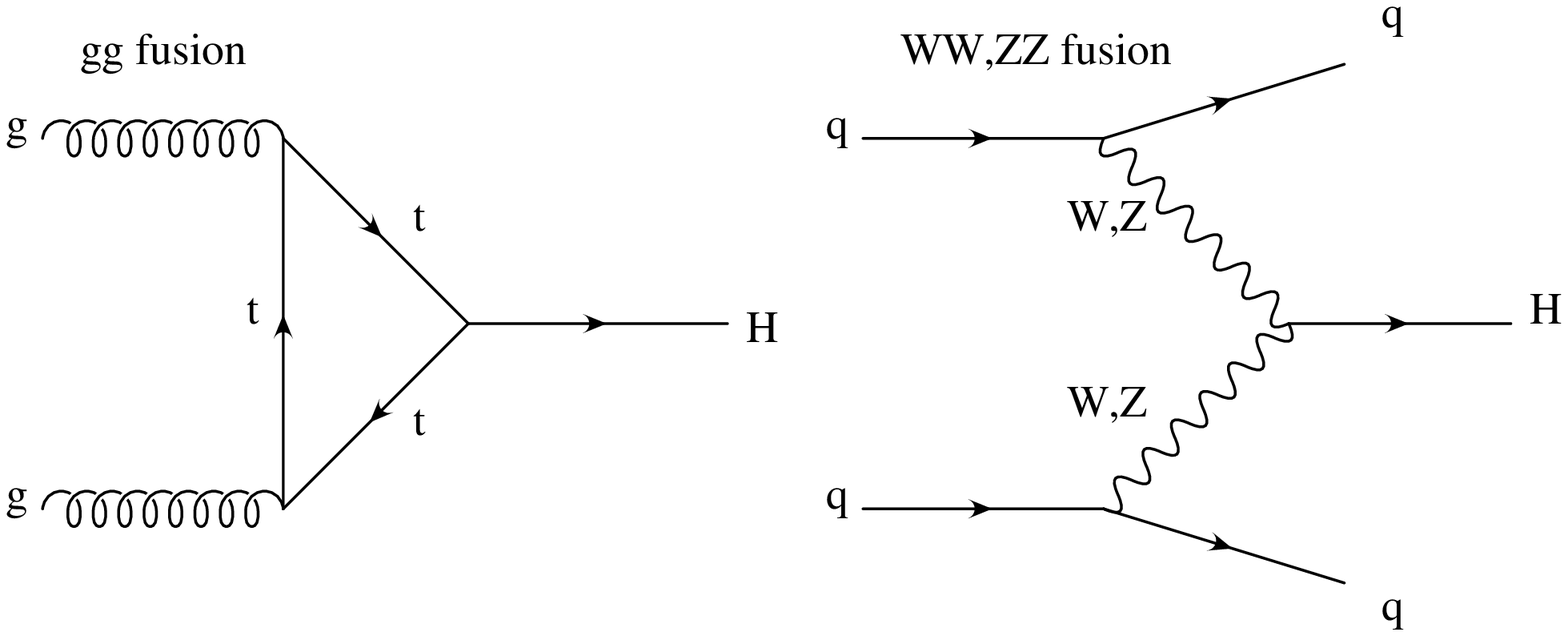}
\caption{Feynman diagrams for two production mechanisms of Higgs.
(Left) gluon fusion (Right) weak boson fusion.}
\label{feynhprod}
\end{center}
\end{figure}

We will review here the different production mechanisms of the Higgs boson
within the Standard Model framework.

\begin{figure}[p]
\begin{center}
\includegraphics*[scale=0.55,angle=-90, bb=28 100 553 771]{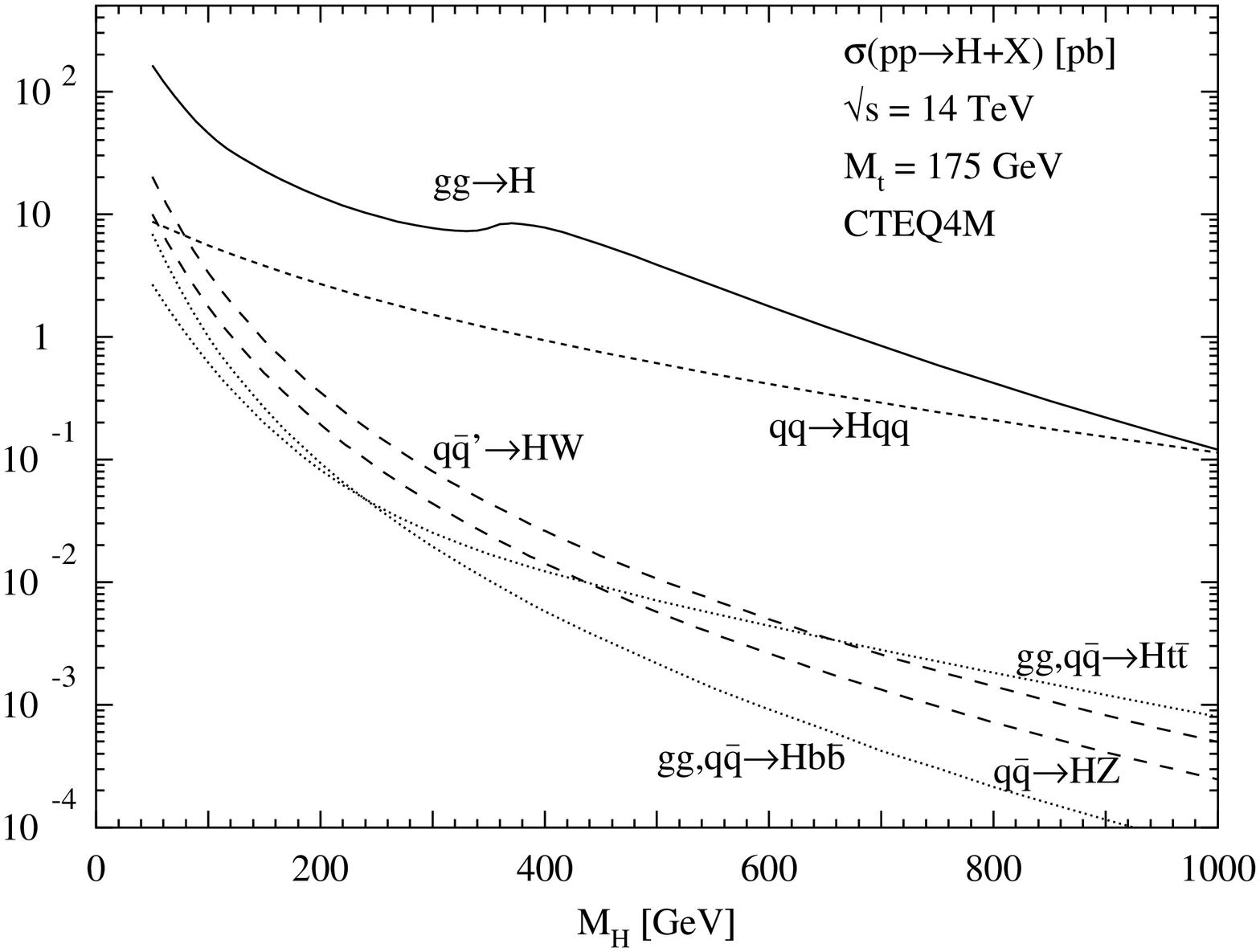}
\caption{Cross sections of different production mechanisms of Higgs as a 
function of the Higgs mass, given at the next to leading order,
Ref. \cite{prodplot}.}
\label{csprod}
\end{center}
\end{figure}

The total production cross section for the Standard Model Higgs at the
LHC is dominated by the \emph{gluon fusion process}
(see Figures \ref{feynhprod} and \ref{csprod}), which largely proceeds via
a top quark loop, directly sensitive to the $t \bar{t}H$ coupling.

The second interesting production process for the heavy Higgs boson
is the \emph{weak boson fusion},
where two quarks emit vector bosons which interact to create a Higgs
(see Figure \ref{feynhprod}). Its cross section is about 5 to 10 times
smaller than the gluon fusion cross section.
This process is then sensitive to the Higgs coupling to
the weak vector bosons.
The ratio of the two cross sections provides the information how the Higgs
couples to fermions and bosons. This is then a crucial test whether the
heavy Higgs is within the Standard Model or beyond.
The possibility to make such a measurement at the LHC
will be one of the main purposes of this study.

Another possible Higgs production mechanisms is
the \emph{associated Higgs production}: $q_i \bar{q_j} \fl WH$
or $q_i \bar{q_j} \fl ZH$,
where an off-shell vector boson is produced and radiates a Higgs. This process has
a sizable cross section for
a Higgs with low masses (see Figure \ref{csprod}). 
Like for the weak boson fusion process, its cross
section depends on the Higgs coupling to vector bosons.

The last production mechanism is the \emph{Higgs Bremsstrahlung}: 
$gg,q_i \bar{q_j} \fl t \bar{t}H$
where top quarks are 
produced and radiate a Higgs. 
Notice that in this case, the cross section
depends on the Higgs coupling at fermions, like for the gluon fusion case.

Unfortunately, in the
mass region considered for this study, these two last
production mechanisms have a cross section about 100
times smaller than the gluon fusion process, leading to essentially undetectable
signals.

\section{Higgs decays}

\begin{figure}[htb]
\begin{center}
\includegraphics[width=8cm]{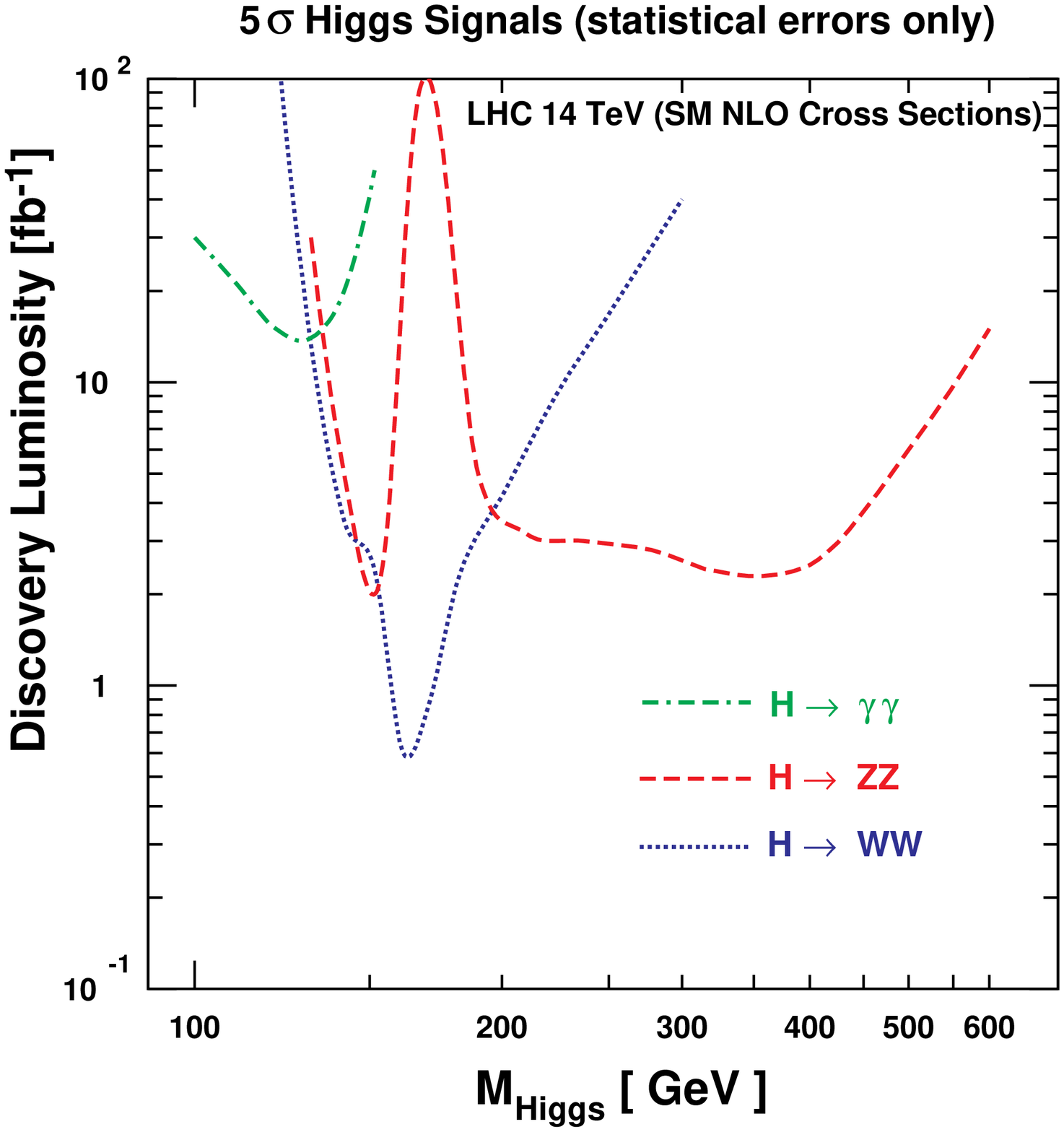}
\caption{Discovery luminosity as a function of the Higgs mass
as given in Ref.~\cite{disclumi}.}
\label{lumidisc}
\end{center}
\end{figure}

The strategy to find the Higgs changes depending on its mass.
Figure \ref{lumidisc} give the necessary luminosity for a five
standard deviations\footnote{Five
standard deviations from the expected background, 
which corresponds to a probability of $5.9\cdot 10^{-7}$,
is the usual requirement to claim for a discovery.},
for different Higgs masses and channels.

For Higgs masses between 110~GeV and 130~GeV, 
the most significant channel is the decay $H \fl \gamma \gamma$.

If the Higgs mass is lying between 130~GeV and 180~GeV,
the expected discovery channel is then
the  $H \fl WW^{(*)} \fl \ell^+ \nu \ell^- \bar{\nu}$ decay as the
$H \fl ZZ^* \fl 4\ell^{\pm}$ has there a very low branching ratio, due to the 
production of two on-shell $W$'s.

For $2 \times m_Z \leq m_H \leq 400\,\mathrm{GeV}$,
the decay $H \fl ZZ \fl 4\ell^{\pm}$
provides the easiest discovery signature as the events should contain 
four isolated leptons with high transversal momentum, $p_t$.

For higher Higgs masses, additional signatures involving hadronic $W$ and $Z$
decays as well as invisible $Z$ decays like 
$H \fl ZZ \fl \ell^+\ell^-\nu \bar{\nu}$
have been investigated and promising signals have been obtained. These channels
will also be studied in detail in that report.

\begin{figure}[htb]
\begin{center}
\includegraphics*[scale=0.7,angle=-90,bb=40 110 545 760]{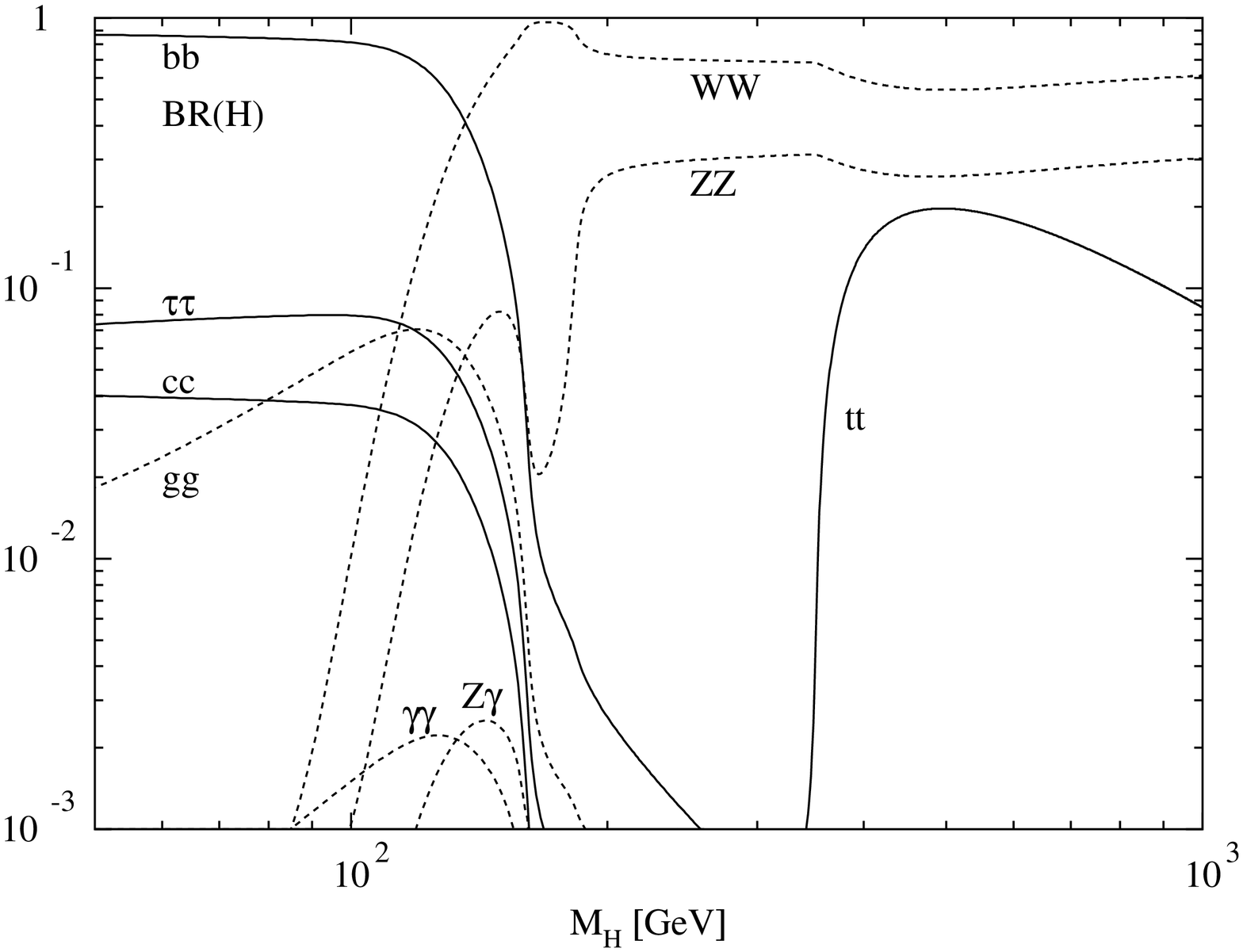}
\caption{Decay branching ratios 
of the Higgs as a function of its mass, Ref.~\cite{hbr}.}
\label{higgsbr}
\end{center}
\end{figure}

For a deeper study of the Higgs discovery channels 
see \cite{atlas} and \cite{searchhiggs}.
\\[0.5cm]
The obvious next question, if we assume that the Higgs has been found,
is then how well the Higgs sector can be 
tested at the LHC. As it was already said before, 
the purpose of this work is to treat that 
question for the case of a heavy Higgs.
As we worked with Higgs masses between 300 and 600~GeV, 
we can see on Figure \ref{higgsbr} that 
the $H \fl WW$ and  $H \fl ZZ$ decay are, within the Standard Model,
by far the dominant branching ratios.

For high Higgs masses, we see on Figure \ref{higgsbr}
that the Higgs can also decay in top quarks, with a branching ratio of
about 10 to 20\%.
However this channel is currently believed to be unmeasurable at the LHC, given
the presence of a too important background.

	\chapitre{General ideas to get a signal}
\markboth{General ideas to get a signal}{General ideas to get a signal}

As discussed in the previous chapter, a SM
Higgs with a mass between 300 and 600~GeV
is produced mainly through weak boson fusion ($q_i q_j \fl q_i q_j H$)
and gluon fusion ($gg \fl H$) and cross sections for other processes
are 100 to 1000 times smaller.
A heavy Higgs will then mainly decay in vector 
bosons through the processes $H \fl WW$ and $H \fl ZZ$. 
The branching ratio for the first decay mode
is about two times higher than that for $H \fl ZZ$.

The branching fractions for the decays of the 
$W$ and $Z$ vector bosons are well known from LEP 
and agree with the Standard Model expectations.
These vector bosons can decay either hadronically 
($W \fl q_i q_j$ or $Z \fl q_i \bar{q_i}$, with a branching ratio of about 70\%)
or leptonically
($W \fl \ell \nu$, $Z \fl \ell^+\ell^-$ or $Z \fl \nu \nu$
with branching ratios of respectively 30\%, 10\% and 20\%, if the taus
are included in the leptons). Hadronic
decays have a higher branching ratio 
but are harder to detect than
leptonic decays, 
characterized by high $p_t$ leptons and/or missing energy coming
from neutrinos.

The decay of a heavy Higgs boson can thus lead to very 
different signatures: only leptons, leptons together with neutrinos and 
only neutrinos,
for the fully leptonic signatures and 
leptons together with jets, neutrinos together with jets and only jets
for the hadronic signatures. These signatures
will have to be studied separately. \\

In the first part of this chapter, 
the different Higgs signatures are described. This is followed by some
words about how vector bosons can be detected and combined to reconstruct the
Higgs. Then,
independently of the channel, we will analyze
how one can use the kinematical properties of an Higgs event
to improve the efficiency of the selection cuts. 
Finally, we will explain the method used to distinguish
the Higgs production process through
weak boson fusion.

\section{Signatures of a heavy Higgs}

Starting from the assumption that the Higgs decays into
either two $W$ or two $Z$ bosons, the different Higgs signatures 
are all
the combinations of the possible 
$W$ and $Z$ decay modes.
However, some channels can be immediately
discarded as they do not give 
measurable signatures, like for instance $H \fl ZZ \fl \nu_i \bar{\nu_i} \nu_j 
\bar{\nu_j}$ 
were nothing about the Higgs
can be reconstructed or $H \fl WW,\,ZZ \fl q_i q_ j q_k q_l$
which is overwhelmed by backgrounds.
The signatures which can possibly be seen in the CMS detector
are summed up in Table \ref{chan}.
Note that in this Table,
$\ell$ refers to electrons and muons, as the taus events are always 
removed by the selection cuts\footnote{
Studies like \cite{Dani} were done about the
possibility to include taus in the leptonic signatures and showed that the 
significance could be raised. However, such treatment will not be
considered in this study.}.

Note that the channel $H \fl ZZ \fl \nu_i \bar{\nu_i} q_i \bar{q_i}$ 
stays at the border of the
observable channels.

\begin{table}[htb]
\begin{center}
\begin{tabular}{|l|c|l|}
\hline
Channel & $\sigma \times BR$ & Main backgrounds studied \\
& \small{(normalized)} & for this channel \\
\hline
$H \fl ZZ \fl \ell^+ \ell^- \ell^+ \ell^-$ & 1 &
$q_i \bar{q}_i \fl ZZ$ \\
\hline
& & $q_i \bar{q}_i \fl ZZ$, $q_i \bar{q}_j \fl WZ$ \\
$H \fl ZZ \fl \ell^+\ell^- \nu_i \bar{\nu_i}$ & 6 & 
$q_i \bar{q}_i \fl g Z$, $q_i g \fl q_i Z$ \\
& & $q_i \bar{q}_i \fl t \bar{t} \fl Wb \, Wb$ \\
\hline
& &
$q_i \bar{q}_i \fl WW$ \\
$H \fl WW \fl \ell \nu \ell \nu$ & 27 & $q_i \bar{q}_i \fl g Z$
, $q_i g \fl q_i Z$ \\
& & $q_i \bar{q_i} \fl Z$ \\
& & $q_i \bar{q}_i \fl t \bar{t} \fl Wb \, Wb$ \\
\hline
& & $q_i \bar{q}_i \fl WW$ \\
$H \fl WW \fl \ell \nu q_iq_j$ & 135 &
$q_i \bar{q}_j \fl g W^+$, $q_i g \fl q_k W^+$ \\
& & $q_i \bar{q}_i \fl t \bar{t} \fl Wb \, Wb$ \\
\hline
& & $q_i \bar{q}_i \fl ZZ$ \\
$H \fl ZZ \fl \ell^+\ell^- q_i\bar{q_i}$ & 21 &
$q_i \bar{q}_i \fl g Z$, $q_i g \fl q_i Z$ \\
& & $q_i \bar{q}_i \fl t \bar{t} \fl Wb \, Wb$ \\
\hline
& & $q_i \bar{q}_i \fl ZZ$ \\
$H \fl ZZ \fl \nu_i \bar{\nu_i} q_j\bar{q_j}$ & 64 &
$q_i \bar{q}_i \fl g Z$, $q_i g \fl q_i Z$ \\
& & $q_i \bar{q}_i \fl t \bar{t} \fl Wb \, Wb$ \\
& & QCD events\\
\hline
\end{tabular}
\caption{Heavy Higgs decay channels together
with their potential sources of background.
Only the channels which could give a detectable signature are given here.
$\ell$ refers to electrons and muons.}
\label{chan}
\end{center}
\end{table}

The following criteria have to be considered in order to judge the feasibility
of detecting a Higgs signal for a given decay channel:
\begin{itemize} 
\item In which way is it possible to reconstruct the Higgs mass peak
and what is the obtained mass resolution~? 
This is especially important
for a discovery where the exact Higgs mass is not known.
\item Is the cross section of the channel big enough to get a signal~?
\item Has the channel a lot of backgrounds, is it easy to get rid of them~?
\end{itemize}
Given these criteria, we can compare the different channels.
For instance, the four leptons channel has an advantage as
the Higgs mass peak can be nicely reconstructed, but 
its branching fractions are small.
On the contrary, all channels involving hadronic decays have higher
branching fractions, but also have 
backgrounds with huge cross sections
that have to be suppressed with tighter cuts.
For instance, in the $H \fl ZZ \fl \ell^+ \ell^- q_i \bar{q_i}$ channel, the Higgs
mass peak can be reconstructed and its branching fraction is twenty
times higher than the one for four leptons channel, but too
many sources of background prevent it to be a 
really significant Higgs discovery channel.
\\[0.5cm]
Background is clearly a crucial issue in this study. 
Table \ref{chan} gives of list of these
different backgrounds for the channels we studied.

The sources of background can be divided in three
categories with different physical properties:
\begin{itemize}
\item The \emph{continuum backgrounds}:
$q_i \bar{q}_i \fl ZZ$, $q_i \bar{q}_i \fl WW$ and
$q_i \bar{q}_j \fl WZ$.
\item The \emph{single boson production with multi jets}:
$q_i \bar{q}_j \fl g W^+$, $q_i g \fl q_k W^+$,
$q_i \bar{q}_i \fl g Z$ and $q_i g \fl q_i Z$.
\item The \emph{Top-antitop production}:
$q_i \bar{q}_i \fl t \bar{t} \fl Wb \, Wb$.
\end{itemize}
Different cuts will have to be found to reduce these three types of backgrounds.

\section{Reconstruction of the event}

The signal is reconstructed in two steps:

First we try to find
the two $Z$'s or the two $W$'s 
present in the event. This first step would be for example
useful against \textit{Single boson production backgrounds} but not against
\textit{Continuum backgrounds} in which anyway two vector
bosons are produced. The Higgs is then reconstructed in combining the
two $W$'s or the two $Z$'s. If possible, a cut on the mass of the reconstructed 
Higgs was done.

Subsequently the kinematics of the Higgs event is exploited and
cuts on the $p_t$'s of
the reconstructed vector bosons and/or the Higgs are added.

\subsection*{Detecting the $Z$'s in the event 
}
The $Z$ has the following decay modes (note that in the following,
$\ell$ stands for electrons, muons \emph{and} taus):
\begin{itemize}
\item $Z \fl \ell^+\ell^-$ with a branching ration of $3\times 0.033$
\item $Z \fl jet \, jet$ with a branching ration of 0.70
\item $Z \fl \nu \bar{\nu}$ with a branching ration of 0.20
\end{itemize}

\begin{figure}[htb]
\begin{center}
\mbox{
\subfigure{\includegraphics*[width=.5\textwidth]{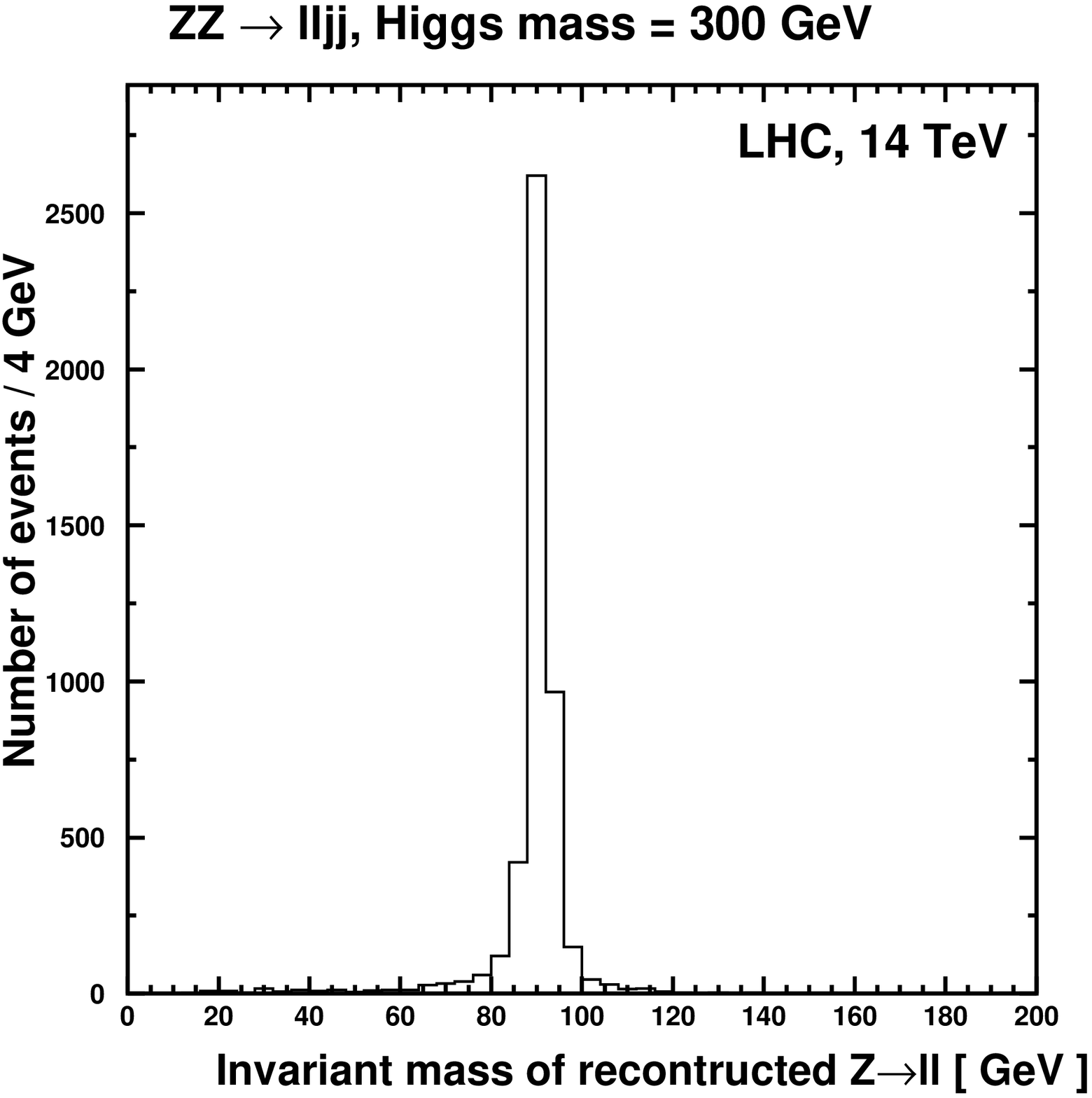}\quad
}
\subfigure{\includegraphics*[width=.5\textwidth]{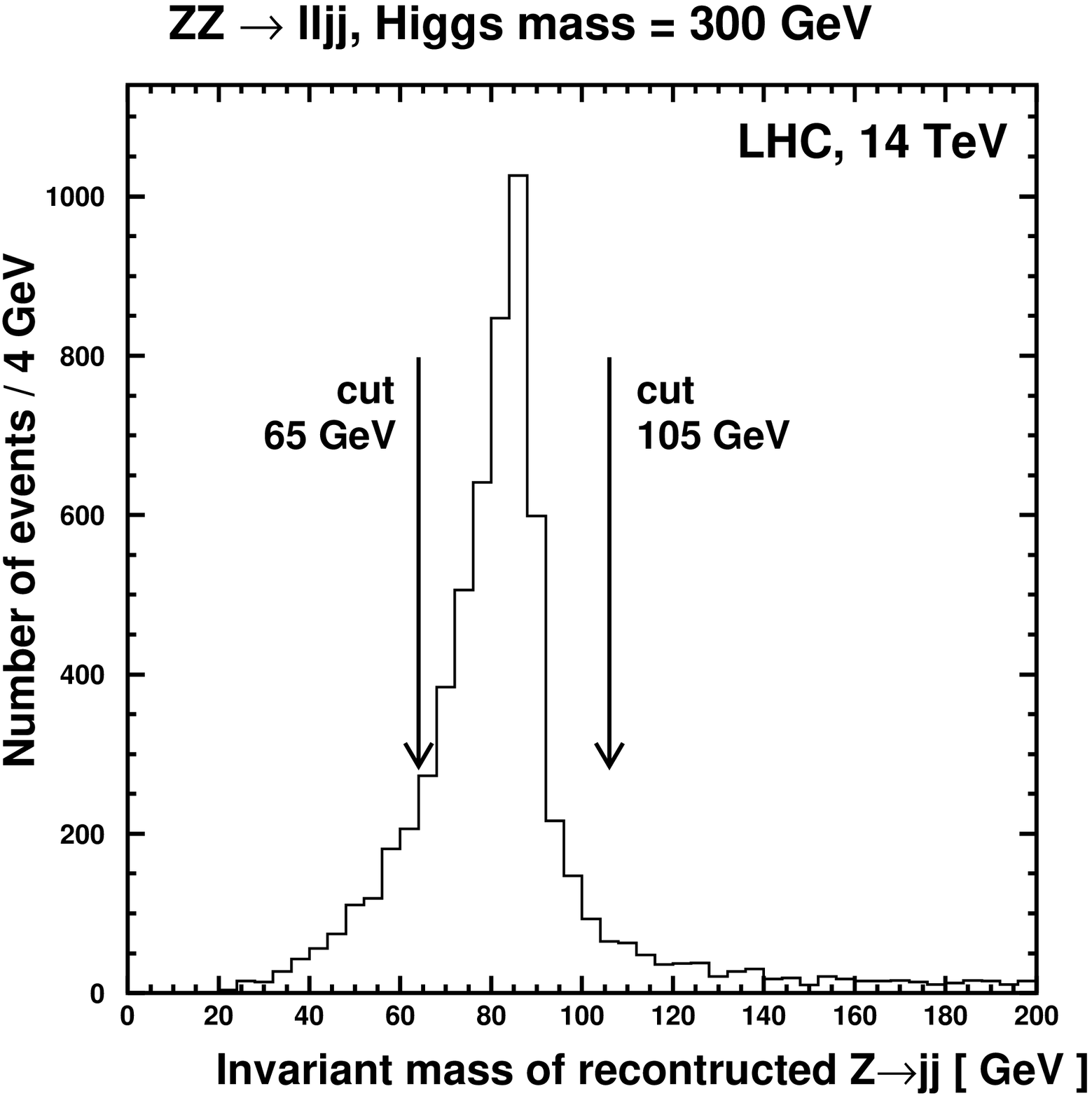}
}
}
\caption{Reconstructed $Z$'s in the channel $gg \fl H \fl ZZ \fl 
\ell\ell jj$ for a
300~GeV Higgs after minimal cuts (e.g. asking two isolated leptons 
and more than two jets). 
(Left) $Z \fl \ell \ell$ (Right) $Z \fl jj$.}
\label{zlljj}
\end{center}
\end{figure}

When it decays into leptons, the $Z$ can simply be identified by asking
two isolated leptons\footnote{
For this study an \emph{isolated lepton} or more generally an 
\emph{isolated particle}
is a particle
with a rapidity smaller than 2.5, a transverse momentum higher than 
10~GeV and whose energy represents 90\% of the energy
within a
cone of $\Delta R=\sqrt{\Delta \eta^2+\Delta \phi^2}=0.5$ around it.
Even more, the total mass of all particles present in this cone, should 
be smaller than 2~GeV and only one other particle with a $p_t$ higher
than 0.1~GeV can be next to the 'isolated' particle.}
in event
which form an invariant mass within an interval of
10 or 20~GeV (depending on the resolution we had in the simulation, see 
for instance Figure
\ref{zlljj}, left) centered in the $Z$ mass (91~GeV).

When the $Z$ decays in hadrons generating jets\footnote
{The jets are reconstructed in the following way: 
We take a stable particle 
with a transversal momentum larger than 5~GeV. Than we take 
all stable particles with a transversal momentum higher
than 500 MeV (this is the smallest energy we assume to detect) 
in a cone of $\Delta R=0.5$ around the first particle,
and add up all these particles. This gives us the jet. Moreover 
the jet
has to have a transversal momentum higher than 20~GeV and a rapidity smaller
than 4.5.}, 
we have basically two problems: 
The invariant mass resolution is not as good as for leptons, as it
can be seen on Figure \ref{zlljj}
and the $Z$ mass peak is shifted from 91~GeV to 85~GeV.
As we cannot detect particles with an energy 
smaller than 500 MeV, this means that there will be missing particles 
in the jet and this latter will have a smaller 
reconstructed invariant mass. 
The jet jet invariant mass is asked to be
within an interval of 40~GeV 
centered in 85~GeV.
As it can be seen on Figure \ref{zlljj}, this mass window could
be chosen to be smaller. However the measured jet resolution
is expected to be lower than the one given by the Monte Carlo. In
order to be closer to the real experiment, a wider
mass range was chosen here.

When more than two jets are reconstructed in one event,
all the possible jet mass combinations were done 
and only the ones
whose mass was in an interval of 40~GeV 
centered in the shifted $Z$ mass, 85~GeV, were kept. 
From the remaining
combinations, the hypothetical $Z$ with the highest $p_t$ was selected as
the 'true' $Z$, as this particle is supposed to have a large $p_t$.

The reconstruction of
$Z$ decaying into neutrinos is more problematic as its
decay products cannot be detected.
An event with large
missing transverse momentum 
is then required
(the value we chose for it depended
on the channel studied, as well as on the Higgs mass generated).
The $Z \fl \nu \bar{\nu}$ four vector is reconstructed in the following way:
$$\left(p_x^{Z_{\nu\nu}}\;;\;
p_y^{Z_{\nu\nu}} \;;\; p_z^{Z_{\nu\nu}}\;;\;E^{Z_{\nu\nu}}\right)=
\left(\boldsymbol{-}\!\sum_{lep,jets}p_x^i\;;\;
\boldsymbol{-}\!\sum_{lep,jets}p_y^i\;;\;0.\;;\;
\sqrt{\left(p_t^{Z_{\nu\nu}}\right)^2+m_Z^2}\right)$$
The $p_x$ and $p_y$ coordinates of the $Z$ 
are the opposed ones of the vector obtained by adding up the four vectors of
all the jets and leptons present in the event. The $p_z$ is set to zero and the
energy is calculated by assuming a $Z$ mass of 91~GeV.

Sources of background
come as well from events which contain neutrino, as from events where the
jets or the leptons were not accurately measured, leading to an artificial missing 
energy in the detector.

\subsection*{Detecting the $W$'s in the event}

The $W$ has the following decay modes :
\begin{itemize}
\item $W \fl jet \, jet$ with a branching ration of 0.69
\item $W \fl \ell \nu$ with a branching ration of $3 \times 0.11$
\end{itemize}
When the $W$ decays in jets, the same criteria were used to reconstruct it
than the ones for the $Z$ with the
invariant mass of the jets to be within an interval of 40~GeV 
centered in 73~GeV,
which is where the shifted $W$ mass peak is sitting. 

In the channel $WW \fl \ell \nu j \, j$, when one $W$ decays to a lepton 
and a neutrino, the neutrino four vector is reconstructed exactly in the same way
than for the $Z \fl \nu \bar{\nu}$.
\label{wlnu}
But in that case, we can besides try to
determine its $p_z$ component: Assuming that the
reconstructed $W$'s mass is 83~GeV
and that the neutrino mass is zero and after some algebra,
we have two solutions for the neutrino's $p_z$:
\begin{eqnarray*}
\alpha & \stackrel{\mathrm{def}}{=} & 
\frac{m_W^2}{2}+p^{lept}_x p^{\nu}_x+p^{lept}_y p^{\nu}_y \\
p^{\nu}_z & = & \frac{\alpha \cdot p^{lept}_z \pm \sqrt{\alpha^2 (p^{lept}_z)^2-
[(E^{lept})^2-(p^{lept}_z)^2][(E^{lept})^2(p^{\nu}_t)^2-\alpha^2]}}
{(E^{lept})^2-(p^{lept}_z)^2}
\end{eqnarray*}
The question is now, which $p_z$ to chose. As we expect the event to be central,
the smallest absolute value for the $p^{neut}_z$ was chosen.
But this criterion is not very good since we only get the right solution in 50\%
of the cases~! Unfortunately no smarter criterion was found.
However we showed
that this did not really affect the final mass peak.

When the Higgs decays in leptons and neutrinos only, as in the 
channel $WW \fl \ell \nu \ell \nu$, we cannot reconstruct a $W$
mass peak. We then had to find special kinematic properties of that channel
to get a signal.

\subsection*{Knowing that the vector bosons come from the Higgs}

Now that we 'know' that there are two vector bosons in the event, we can reconstruct
the Higgs out of them and try to use the expected physical properties 
of a Higgs event to enhance a signal.

\begin{figure}[htb]
\begin{center}
\mbox{
\subfigure{\includegraphics[width=.48\textwidth]{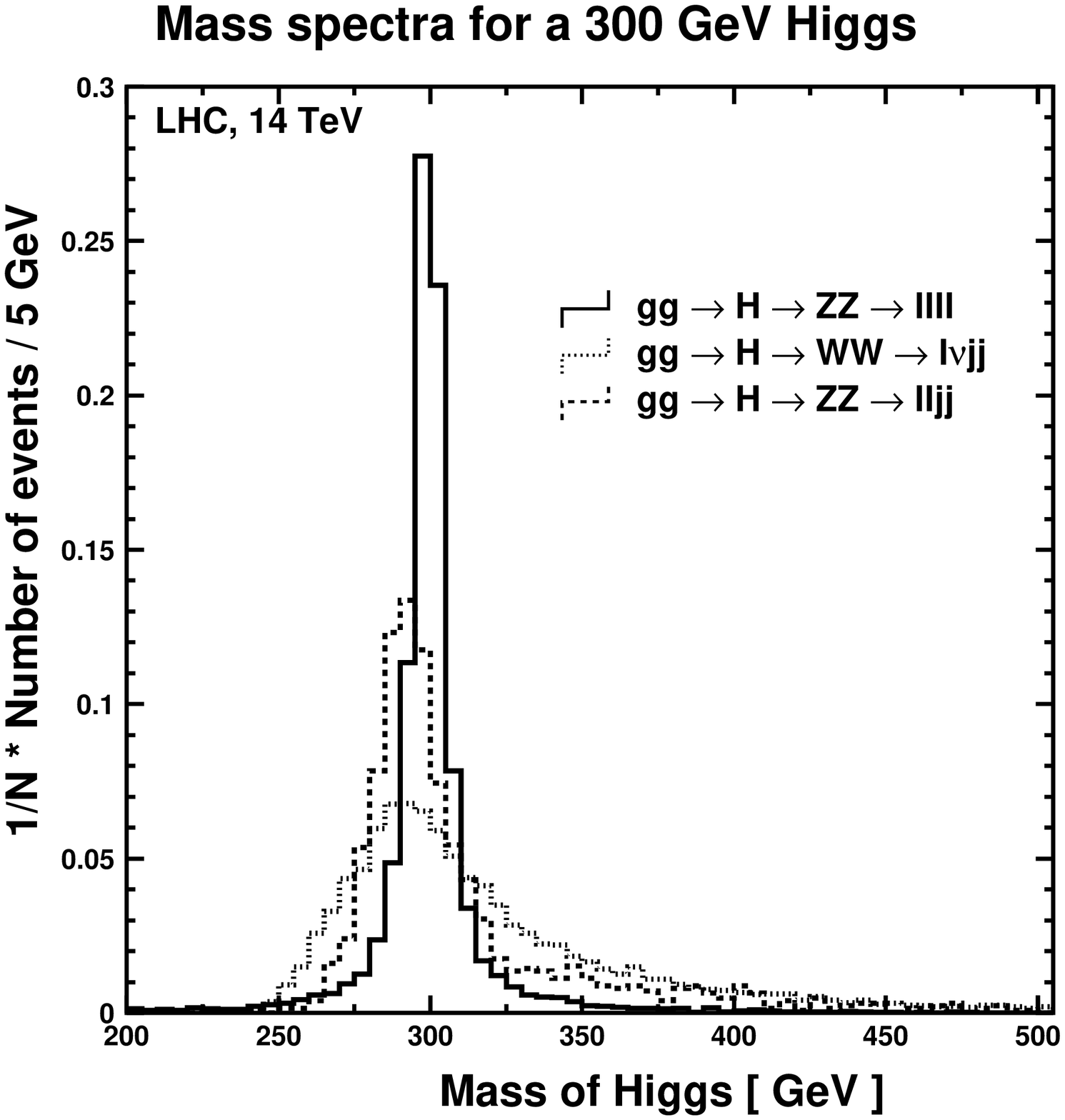}
\quad}
\subfigure{\includegraphics[width=.48\textwidth]{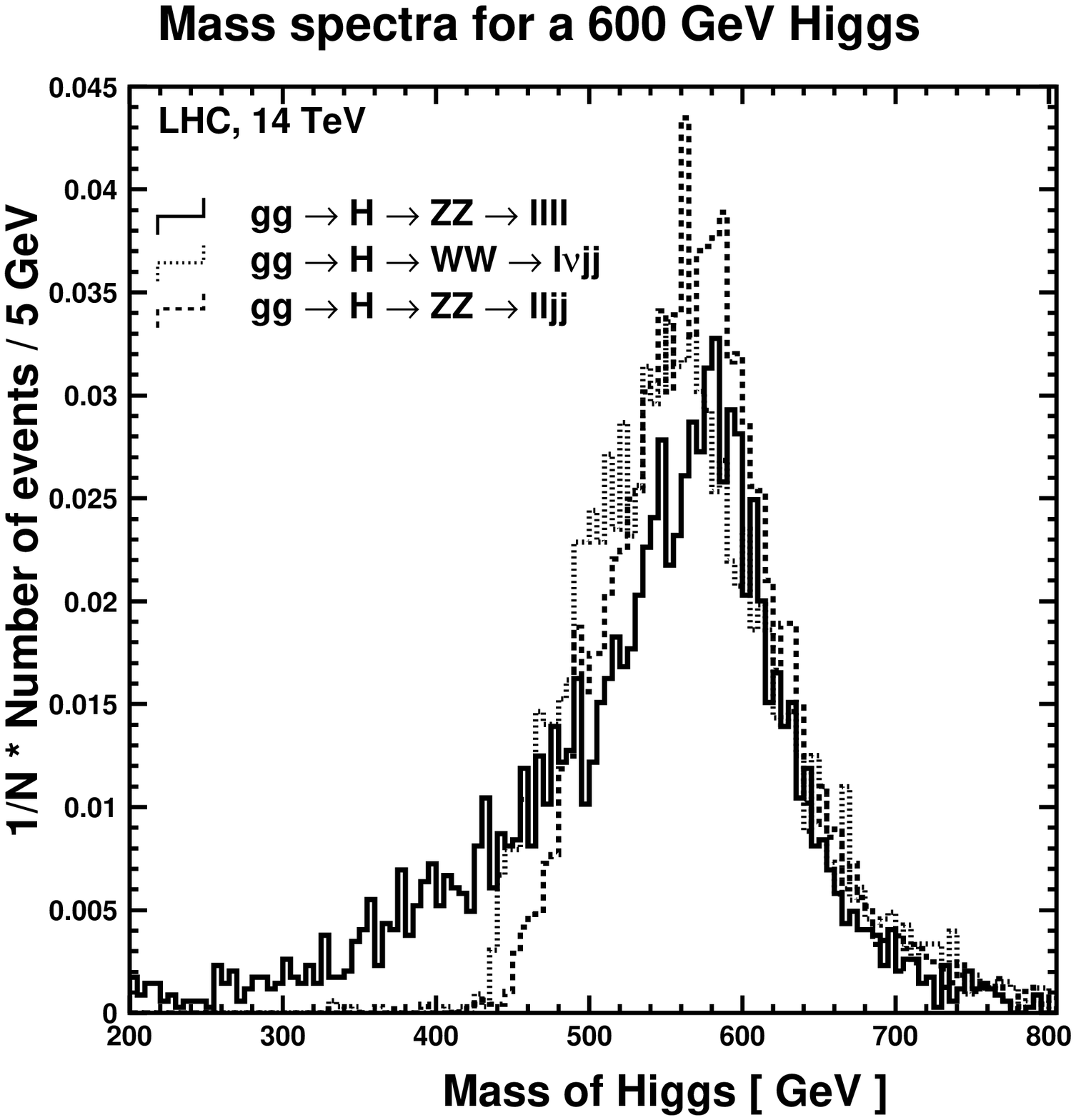}}
}
\caption{Comparison between the mass resolution the reconstructed 
Higgs mass spectrum for different signatures. The plots are made
after the reconstruction cuts.
The Higgs is produced through the gluon fusion.
(Left) for a 300~GeV Higgs
(Right) for a 600~GeV Higgs.}
\label{mres}
\end{center}
\end{figure}

Figure \ref{mres} shows the invariant mass of the two vector bosons
for the signal in different channels.
All the plots were normalized so that the integral of the curves gives 1.
As expected, the best mass resolution is obtained with the four leptons
channels for a 300~GeV Higgs. Broader peaks are seen for the 
other channels.  
The $H\fl WW \fl \ell \nu jj$ channel gives a large tail for the higher mass
region in the case of
a 300~GeV Higgs, 
due to the presence of the neutrino which cannot be well 
reconstructed.
As already pointed out, the jets are not
as well reconstructed as in that simulation and the 
expected measured mass peak should be wider.

The natural width of the Higgs increases with its mass. Then,
the different channels give more or less
the same resolution. The natural width of the Higgs here dominates the 
uncertainty in the reconstruction.
For the high Higgs masses, the four leptons channel
has a longer tail in the low mass region compared to the other channels.
This is due to the large value of the $p_t$'s 
cuts applied in these other channels, whose effects are to cut the low mass region in
the reconstructed Higgs mass spectrum.

The reconstruction of the Higgs mass peak is the best way to discover 
this particle.
However, other criteria can improve the signal significance.
Furthermore, when the backgrounds are too important,
new cuts have to be found.

\begin{figure}[htb]
\begin{center}
\includegraphics[scale=0.6]{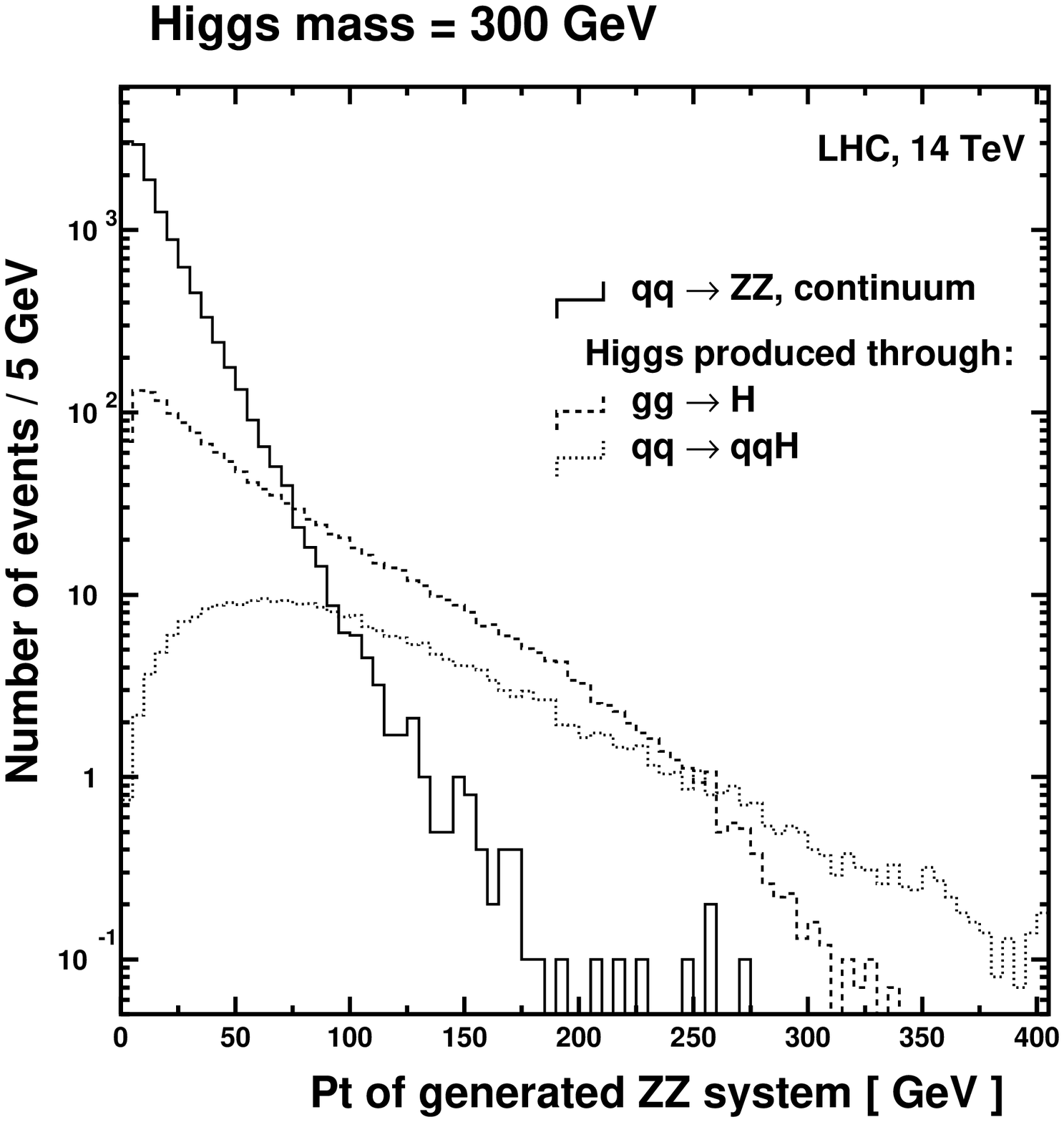}
\caption{Generated transverse momentum for a 300~GeV Higgs for the
two signal processes (with K-factors) and for the $ZZ$ continuum.}
\label{pt}
\end{center}
\end{figure}

One possibility is to cut on the $p_t$ of the vector bosons and/or of the
Higgs.  More specifically, in the weak boson fusion processes, the Higgs is
always produced together with jets, which allows it to have a large $p_t$. As we
can see on Figure \ref{pt}, events produced through weak boson fusion 
have a higher mean transverse momentum than events produced
through gluon fusion.\label{coucou}
As it usually does not contain jets, the $ZZ$
continuum background leads to small mean values for the $p_t$ of the
generated $ZZ$ system. 

Notice that the mean $p_t$ for that background is even smaller than the 
mean $p_t$ for gluon fusion, as gluons can radiate jets with higher
probability than quarks.
A cut on the
reconstructed Higgs $p_t$ will be very efficient against the continuum
backgrounds. For instance in the channel
$ZZ \fl \ell\ell\ell\ell$, where
the only background is  $qq \fl ZZ$, a cut on the Higgs
transverse momentum was found to improve the signal significance considerably.

\begin{figure}[p]
\begin{center}
\subfigure{\includegraphics[scale=0.4]{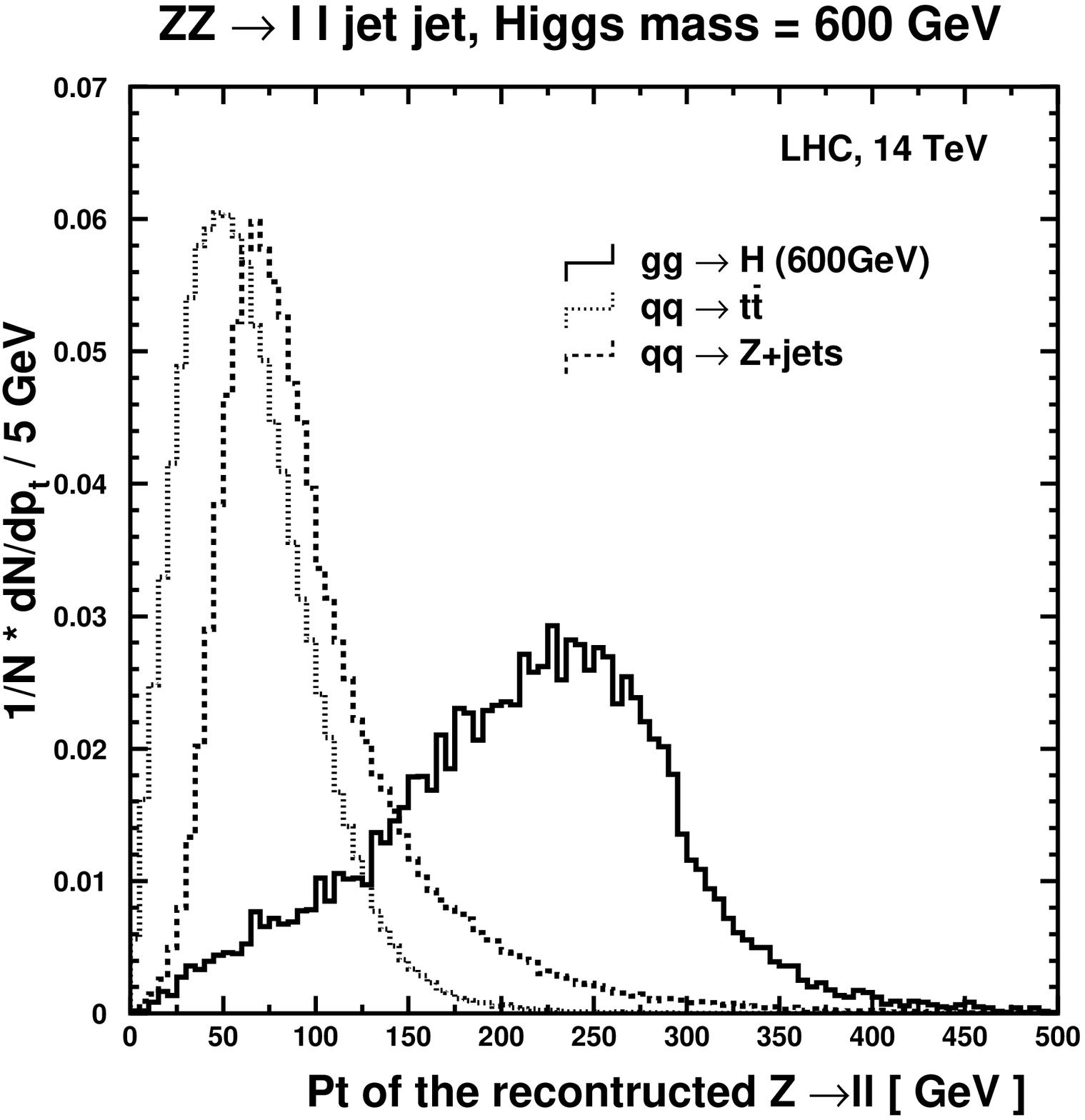}}
\\
\subfigure{\includegraphics[scale=.4]{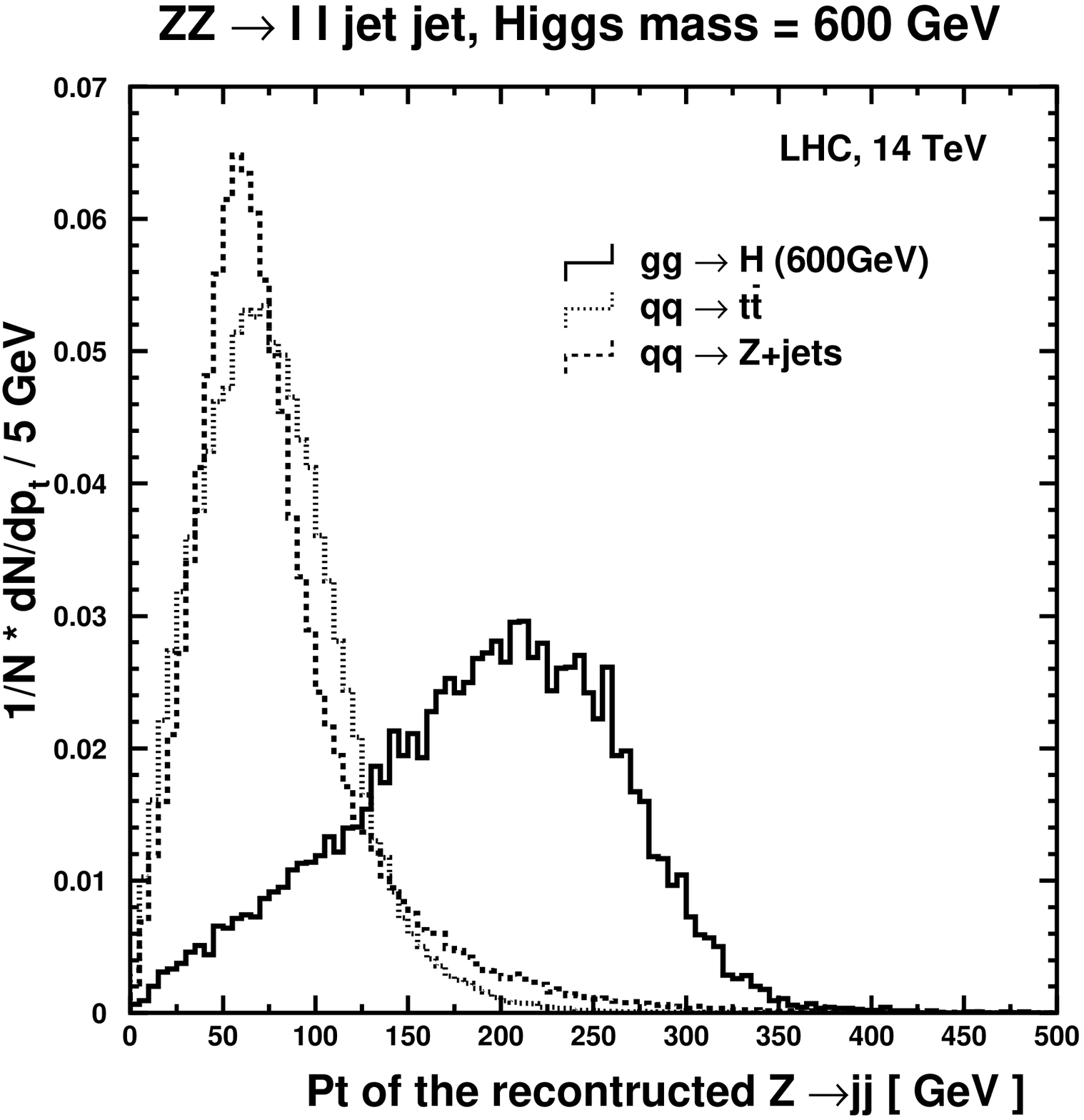}}
\caption{$p_t$ spectra of the reconstructed $Z$ for a 600~GeV Higgs and
for different backgrounds in the $H\fl ZZ \fl \ell \ell jj$ channel, before
the selection cuts and
normalized to the total number of events.
(Up) $p_t$ spectrum for the reconstructed $Z$ decaying leptonically
(Down) $p_t$ spectrum for the reconstructed $Z$ decaying hadronically.}
\label{pt2}
\end{center}
\end{figure}

However, we found more significant to cut on the $p_t$ of the reconstructed
$W$ or $Z$, as soon as backgrounds like $W$, $Z$ production and $t\bar{t}$
were important. 
Figure \ref{pt2} shows the different $p_t$ distributions for the signal
and these backgrounds for a 600~GeV Higgs in
the channel $H\fl \ell \ell jj$.

	\section{Reconstruction of the $qq\fl qqH$ signal}

Up to now, to isolate a signal, only the Higgs decay characteristics
were used in the selection cuts.
It is also interesting to consider the 
way the Higgs was produced, especially when it is produced through
weak boson fusion. This production process leads to a very
specific signature, which can be used
to get a signal in some particular Higgs search channels,
which contains jets and where the reconstruction of the
event only through the Higgs decay products is not sufficient,
and also to isolate that process
from the other Higgs production processes (in our case, gluon fusion).

\begin{figure}[htb]
\begin{center}
\mbox{
\subfigure{\includegraphics*[width=.5\textwidth]{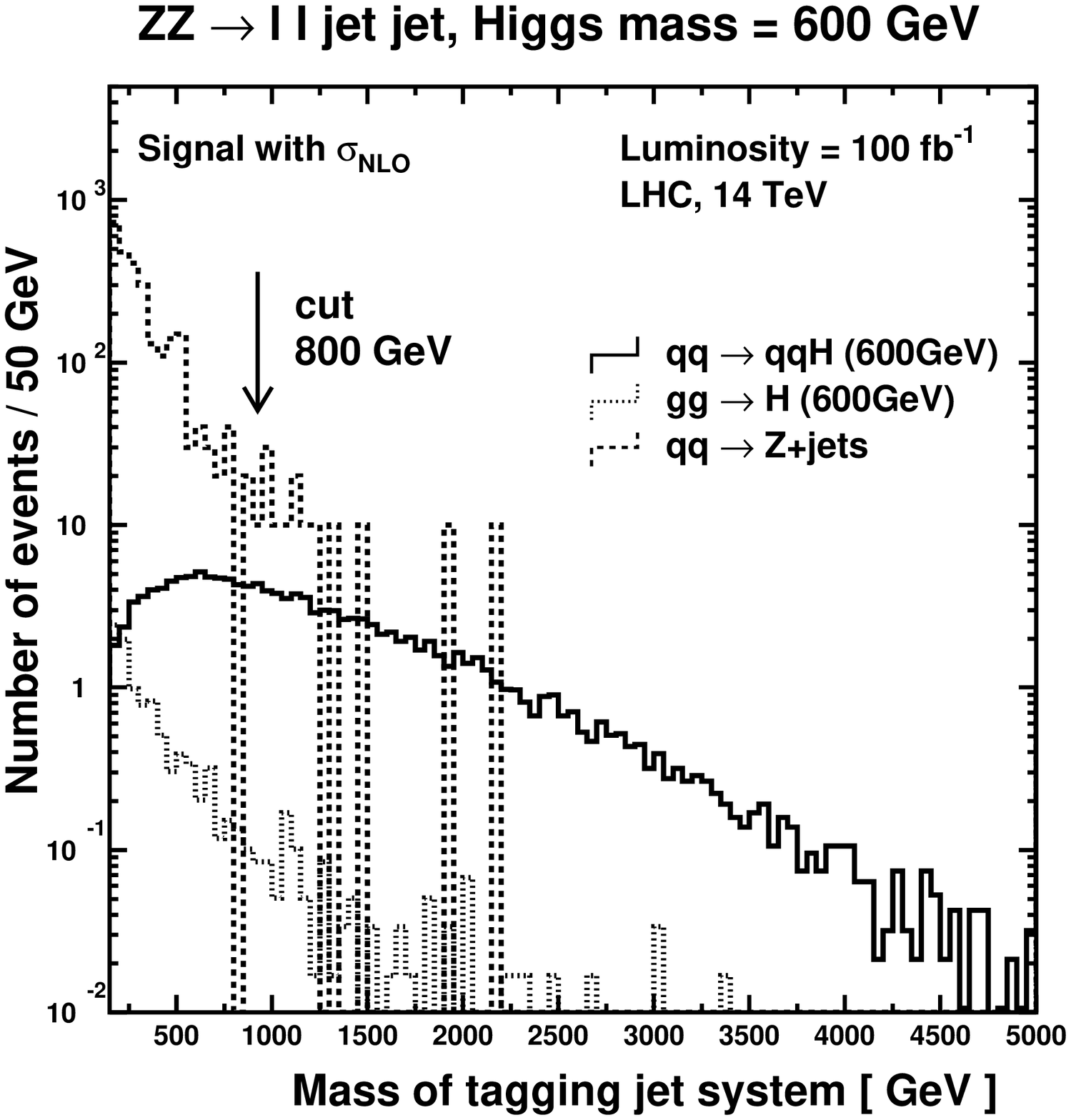}\quad
}
\subfigure{
\includegraphics*[width=.5\textwidth]{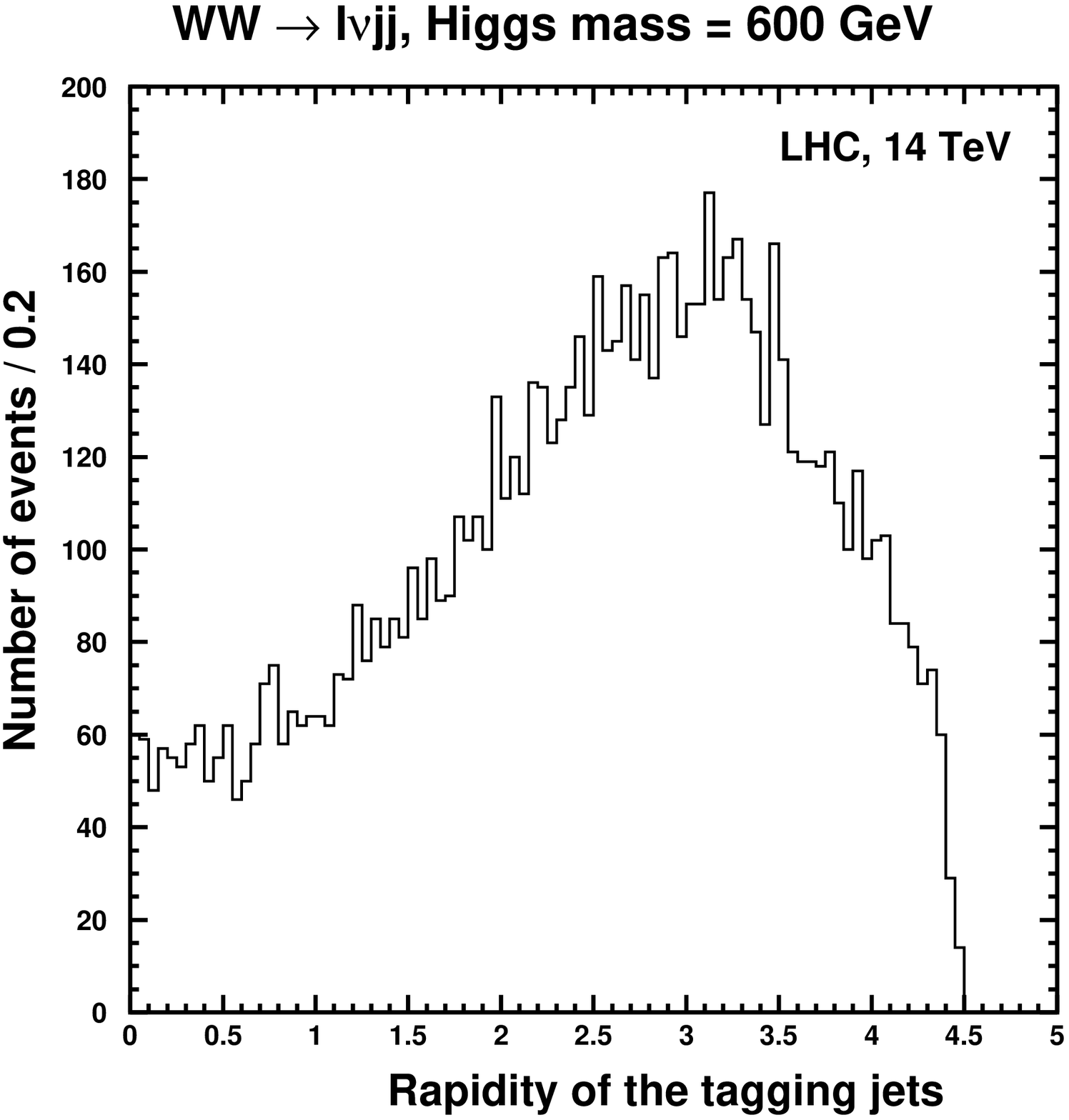}
}
}
\caption{Characteristics of the tagging jets system:
(Left) Mass of the tagging jet system for different processes in the channel
$qq \fl qqH \fl ZZ \fl \ell \ell jj$
(Right) Rapidity of the tagging jets for the channel $qq \fl qqH \fl WW
\fl \ell \nu jj$.}
\label{mjj}
\end{center}
\end{figure}

One of the characteristics of the $qq\fl qqH$ signal is that the 
Higgs tends to be produced centrally and yields central decay products.
In contrary,
the two quarks which emitted the vector bosons enter the detector
at large rapidities and have a transverse momentum in the
20 to 100~GeV range, thus leading to two observable forward tagging jets.
To isolate such type of events, we can cut on the invariant mass of the
two jets, asking it to be higher than 800~GeV. \label{tagging}
This cut will be called in the following,
the \emph{jet tagging technique}.
It is very efficient since the distribution of the invariant mass of the two jets is
drastically different for $qq \fl qqH$ than for all the other backgrounds, as
it is rare for the backgrounds to have two jets emitted with such a high mass.
The tagging jets mass spectrum, as well as their $p_t$ 
distribution is shown on Figure
\ref{mjj} for a 600~GeV Higgs.

\subsection*{Suppressing the top-antitop production background}
\label{cuttop}
As soon as we are dealing with jets and $W$'s, a main source of background comes from 
the top-antitop production : $qq \fl t \bar{t} \fl Wb\, Wb$.
The top quark decays into a $W$ and a $b$ quark with a branching ratio of
almost 100\%. This background will thus be characterized by two $W$'s
and two jets.
It is an important background source for the channels
$qq \fl qqH \fl WW$ and $WW\fl \ell \nu \ell \nu$ or 
$WW\fl \ell \nu jj$.

However, since the $W$ and the jet are decay products of a top
quark, their invariant mass has to give the top quark mass, that is 175~GeV.
A good way to eliminate top-antitop production is to 
cut on the mass of top decay products, which are here
a $W$ and a jet. As we have two reconstructed $W$'s in event and two
supplementary jets, we have several possibilities to reconstruct the top.
Out of the 
masses we get from the different $W$-jet combinations,
the mass closest to the top mass was chosen, which was called $m_{Wjet}$.
In order to remove as much $t\bar{t}$ events as possible, we
demand $m_{Wjet}$ much more larger than the top mass.
For the signal events, always high $m_{Wjet}$ are found as the tagging
jets are emitted with a high rapidity. 

\begin{figure}[htb]
\begin{center}
\includegraphics*[scale=0.7]{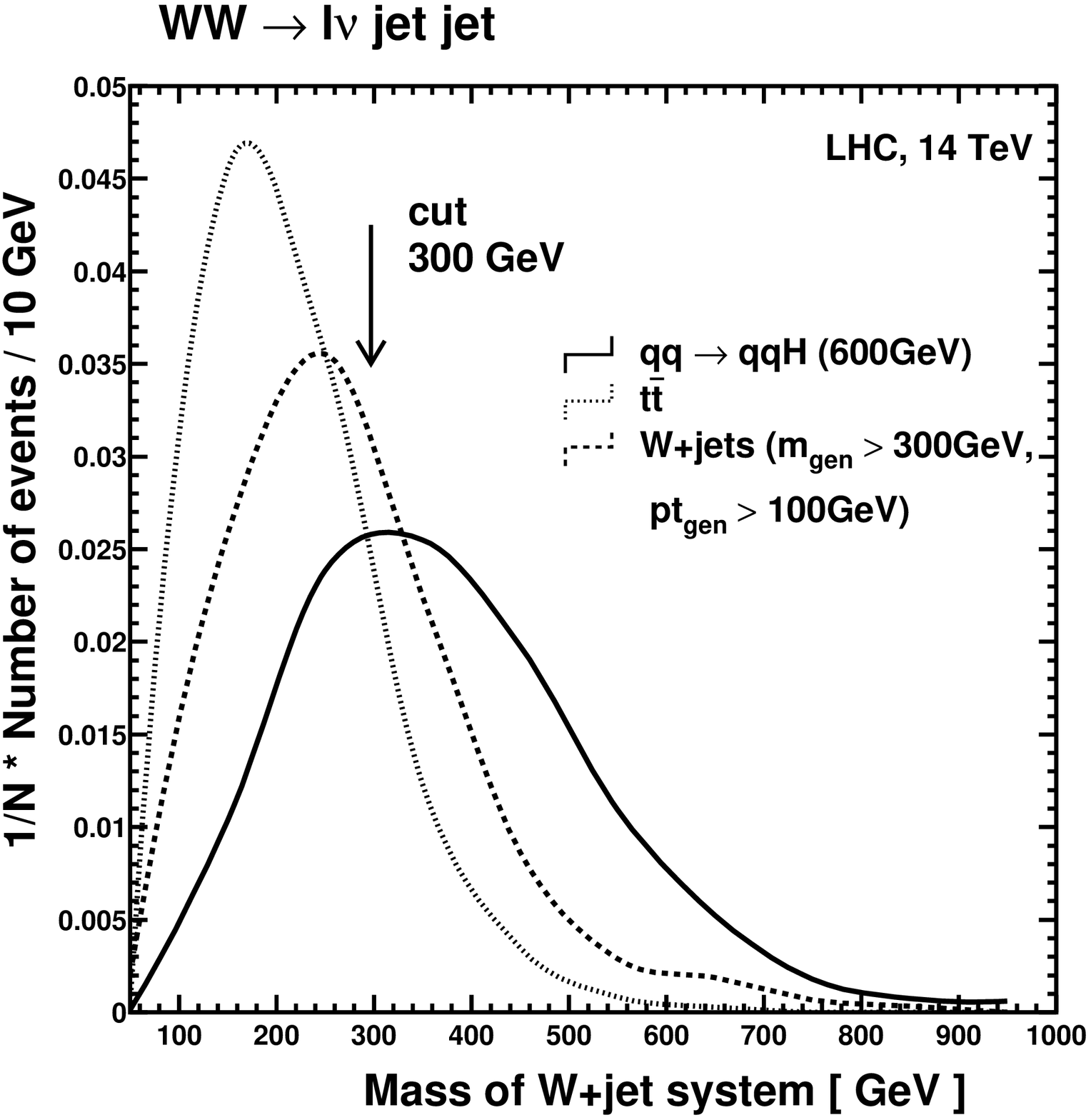}
\caption{Distribution of the $m_{Wjet}$ candidates for the signal, the top-antitop
and $W$ + jets backgrounds.
The curves are normalized to 1.}
\label{top}
\end{center}
\end{figure}

An example of the efficiency of this cut 
in the $H \fl WW \fl \ell \nu jj$ channel
is given in Figure \ref{top}. Note that the curves were normalized to 1.
On this Figure, notice also that the single $W$ production background
is reduced by a factor 3. 
Actually this last background was generated 
with a cut on the mass of the total system
(as we will explain later), namely
asking it to be higher than 300~GeV. This requirement
cuts the low masses of the $m_{Wjet}$ mass spectrum.
The cut on $m_{Wjet}$ is even more efficient against single $W$ production
if this background is simulated without kinematical cuts.

	\chapitre{Analysis}
\markboth{Analysis}{Analysis}

The cuts used to isolate a given signature are always applied in two steps:
First we try to get a signal
without taking into account the way the Higgs was produced. 
Subsequently,
we concentrate on signal events where the Higgs was produced 
through weak boson
fusion and combine the usual selection cuts with a cut using the presence of
the forward emitted jets in that process, like it was described
on page \pageref{tagging}.

In the text,
these two ways of getting a Higgs signal are differentiated
by saying 'with' and 'without tagging'.

This chapter begins with a few words about how
the signal and the different backgrounds
were simulated.
In the next sections,
the results for the different channels studied are presented separately.

\section{Simulating the different processes}

\subsection*{Running PYTHIA}

Version 6.1 \cite{pyt} of PYTHIA set at a center of mass energy of 14 TeV
was used for this study.

GRV (LO) \cite{strucfunc} was used for the proton structure function and
the top quark mass was set to 175~GeV.

Before starting the event selection,
we always ask no events with
high $p_t$ isolated photons or hadrons.

\subsection*{Determination of the cross sections of the processes}

At the end of every simulation, PYTHIA gives an estimate of the cross 
section of the process which was simulated. However,
all PYTHIA cross sections and branching ratio estimates are based
on leading order (LO) calculations. Often these leading order cross sections
differ significantly from the next to leading order (NLO)
QCD calculations. The ratio of $\sigma(NLO)/\sigma(LO)$ defines the so-called
K-factor.

For a Higgs produced in the
gluon fusion process, 
it has been shown \cite{kfactkin} that the next to leading order 
corrections consist mainly in soft gluon radiation. That means that 
the expected event kinematics with NLO estimates should not differ
too much from LO distributions.
A simple scaling with the K-factor should thus be a good approximation
of the NLO corrections.
Ref \cite{crossec} gives the values for the 
K-factors for different production mechanisms and Higgs masses.
We see that for gluon fusion processes, these K-factors are rather 
important (about 1.6)
in contrast to weak boson fusion where K-factors of
about one have been calculated.
To simulate correctly the different signal processes, we took the cross
section given by PYTHIA and multiplied it by the corresponding K-factor.
\label{ici}
Table \ref{kfa} gives the K-factor for the two Higgs production processes
and for different Higgs masses.

\begin{table}[htb]
\begin{center}
\begin{tabular}{|c|c|c|}
\hline
& weak boson fusion & gluon fusion \\
& $qq \fl qqH$ & $gg \fl H$ \\
\hline
Mass of Higgs & K-factor & K-factor \\
\hline
300 GeV & 0.88 & 1.48 \\
450 GeV & 0.94 & 1.66 \\
600 GeV & 1.06 & 1.69 \\
\hline
\end{tabular}
\caption{K-factors for the two Higgs production mechanisms
and for different Higgs masses, as given in \cite{crossec}.}
\label{kfa}
\end{center}
\end{table}

\begin{figure}[htb]
\begin{center}
\includegraphics[scale=0.4]{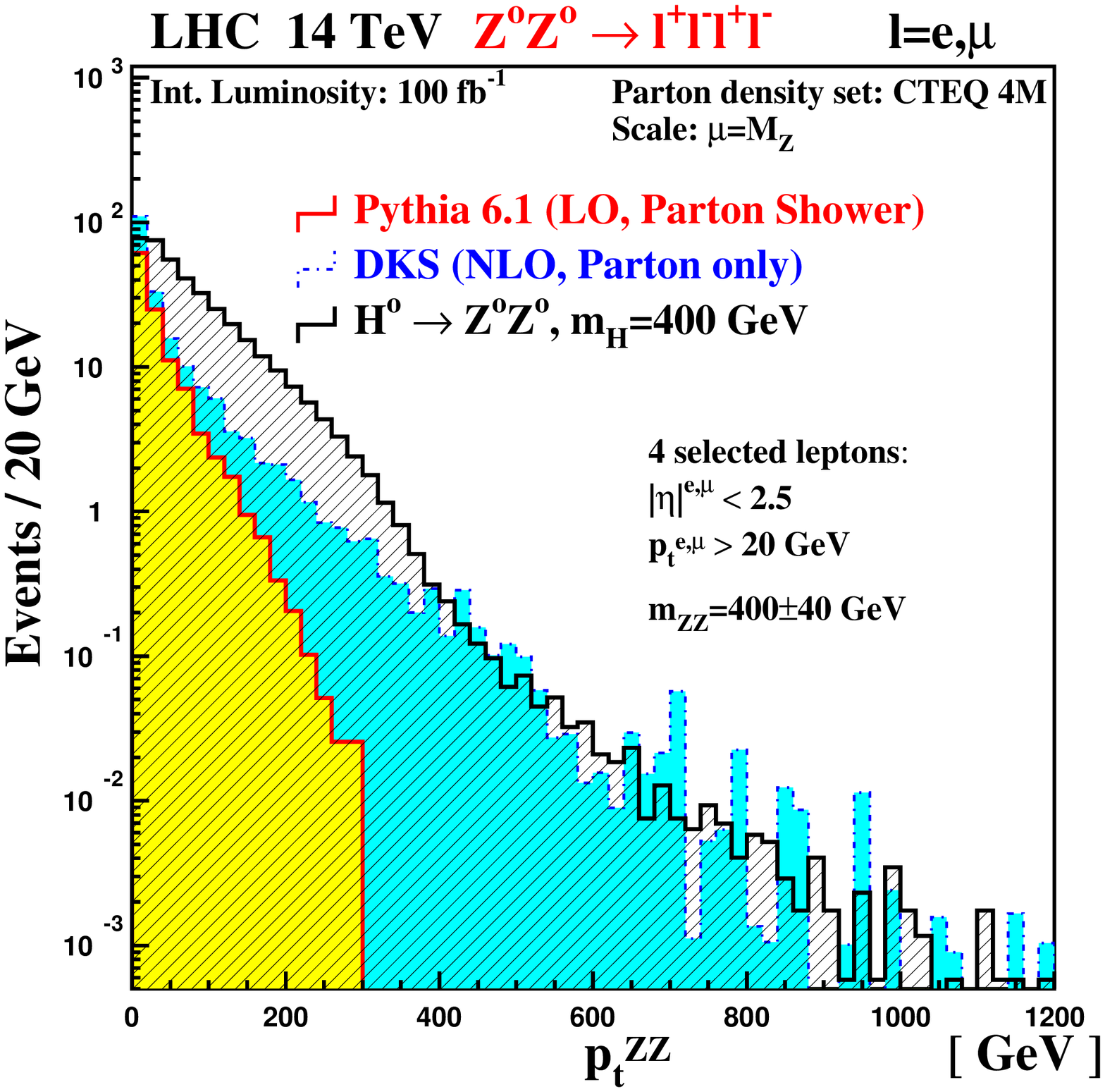}
\caption{Transverse momentum of the $ZZ$ system in the four leptons channel
for signal and background (LO and NLO) \cite{dario}.}
\label{NLO}
\end{center}
\end{figure}

Not all backgrounds have been calculated in NLO,
as some of the calculations are quite complex and until today, 
no consistent simulations 
for all backgrounds are available.
This could lead to some analysis modifications in the
future, as some 
kinematics of the background
events could be changed if NLO corrections are added.
For example, as we will show later, in the four leptons channel, a cut on 
the $p_t$ of the reconstructed Higgs is proposed. Such a selection
in a LO Monte Carlo has been found to improve the experimental sensitivity
by a relatively large factor. However, as has been pointed out (Ref. 
\cite{crossec} and \cite{dario}),
the NLO backgrounds have
higher statistic for the large $p_t$'s 
of the reconstructed $ZZ$ system
than the corresponding 
LO results. Thus, a hard cut on the $p_t$ of the reconstructed $ZZ$ system
would not be as efficient as expected
(if for example $p_t>100\,\mathrm{GeV}$ is asked). However, the efficiency of
a soft $p_t$ cut (for example $p_t>30\,\mathrm{GeV}$), as we proposed in sections
4.2, 4.3 and 4.6,
should not be too much influenced
by the introduction of NLO corrections
in the future (see Figure \ref{NLO}).

Note that as the signal significance is proportional to $1/\sqrt{B}$, 
the addition of a K-factor higher than one
in the background will reduce the significance by a 
factor $\sqrt{K_B}$, whereas for the signal, the introduction of a K-factor
higher than one increases the significance by a factor $K_S$.

For that study, no K-factors were taken for the backgrounds.
However this is
something which will need to be studied in more details in the following years.

\subsection*{Simulating the Higgs with PYTHIA}

The Higgs signals were generated using the PYTHIA processes
123 and 124 for
the weak boson fusion, $qq \fl qqH$, and 102 for
the gluon fusion, $gg \fl H$.

\begin{figure}[p]
\begin{center}
\mbox{
\subfigure{\includegraphics*[width=.45\textwidth]{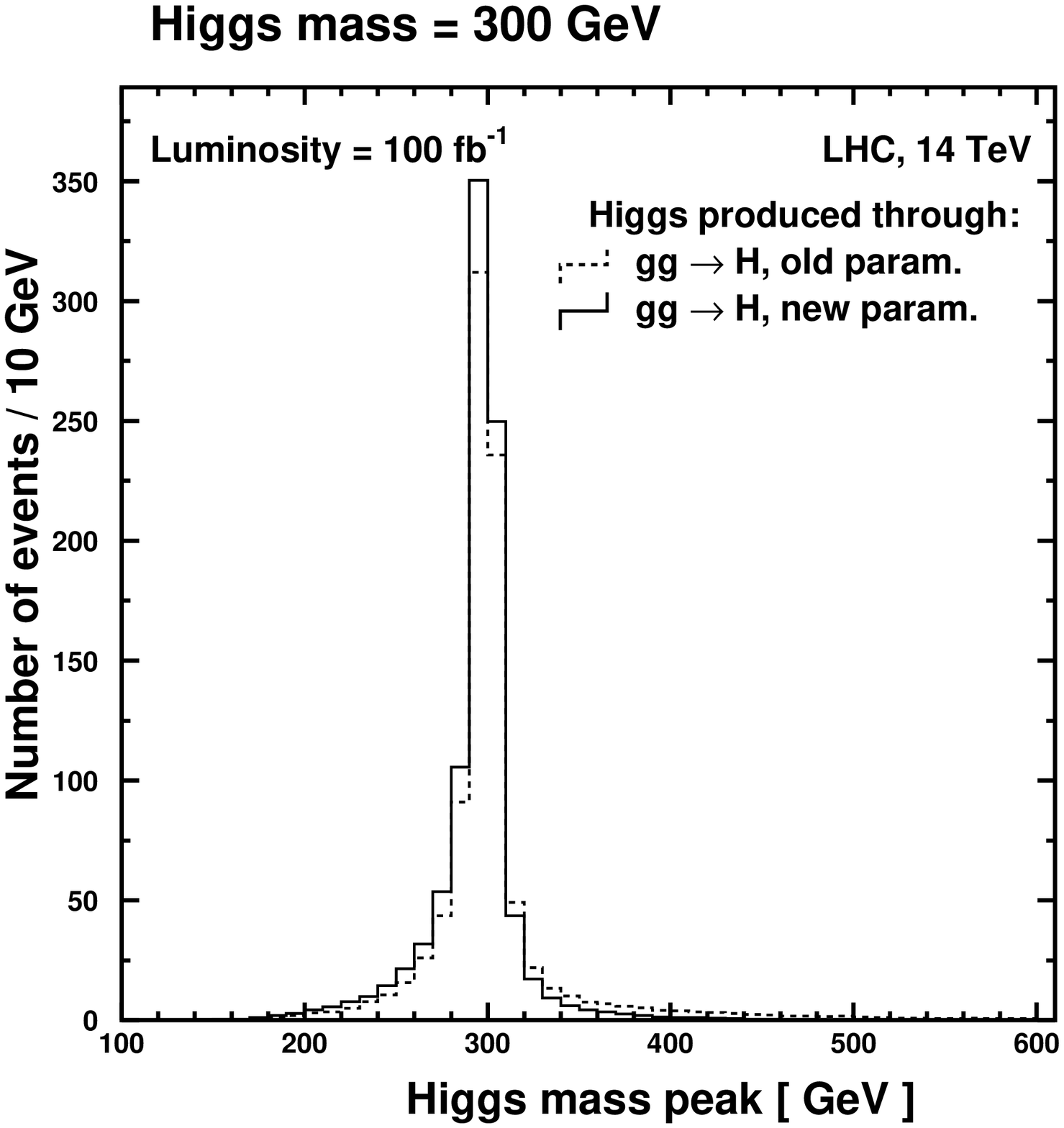}\quad
}
\subfigure{\includegraphics*[width=.45\textwidth]{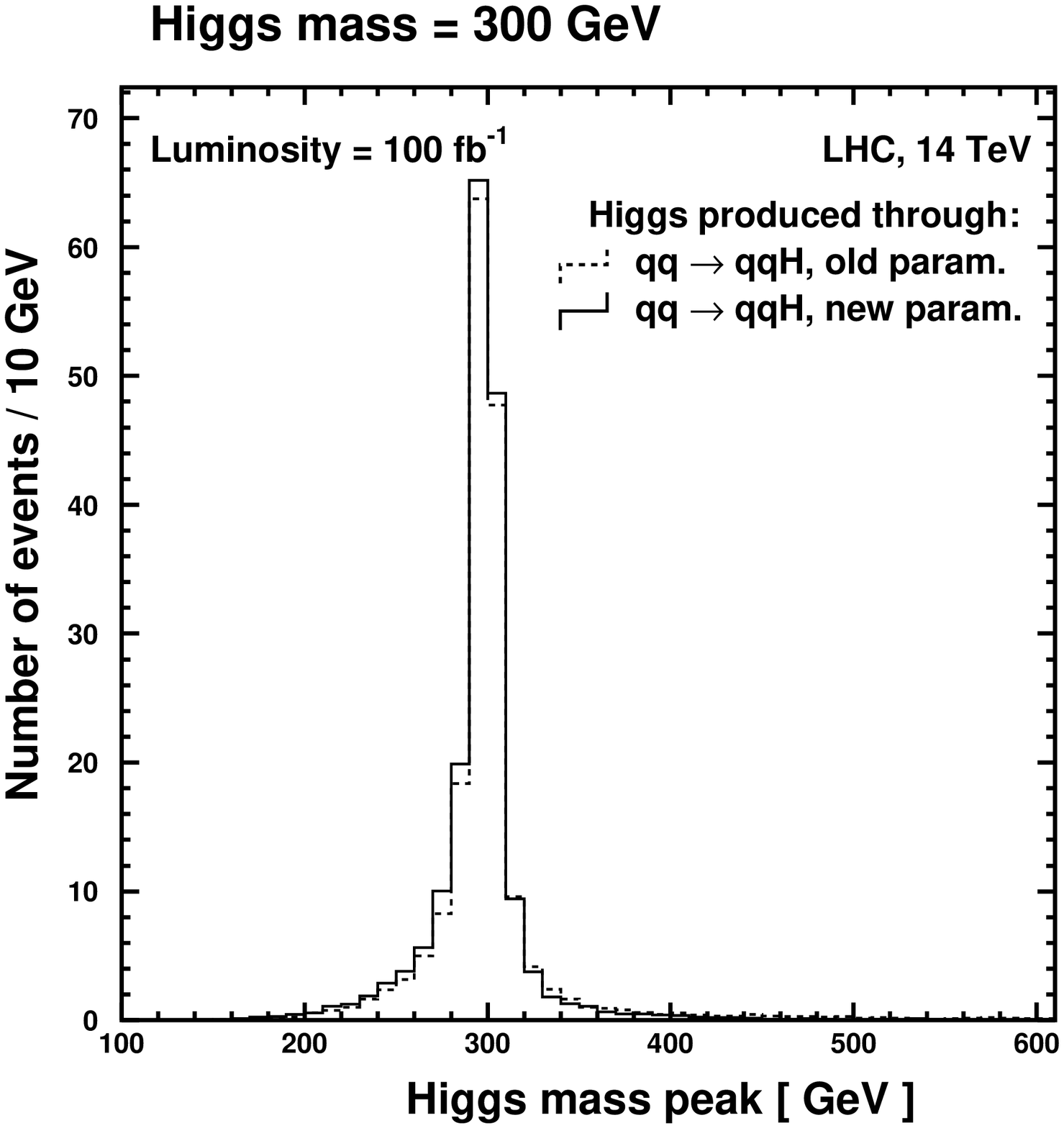}
}
}
\\
\mbox{
\subfigure{\includegraphics*[width=.45\textwidth]{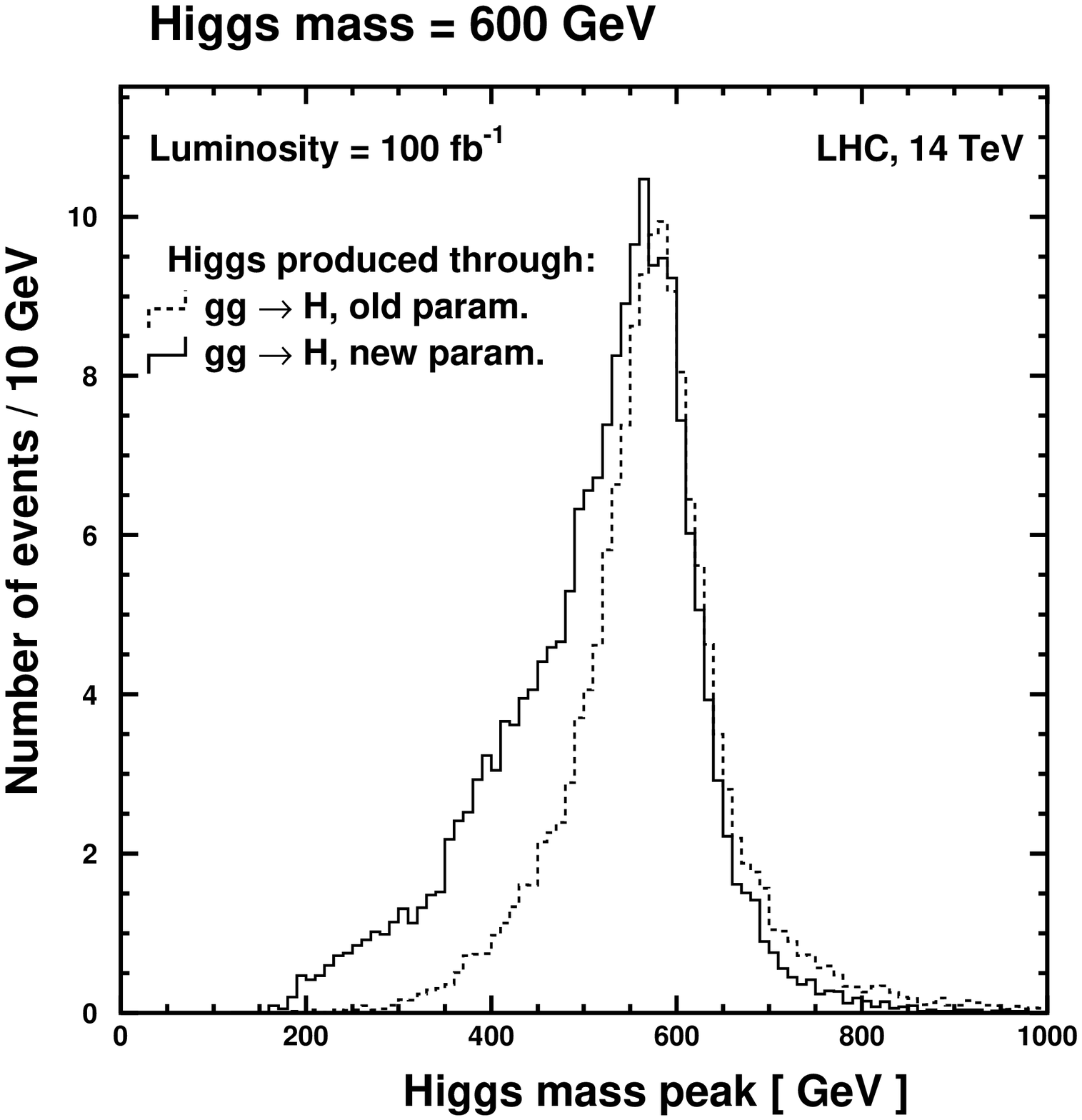}\quad
}
\subfigure{\includegraphics*[width=.45\textwidth]{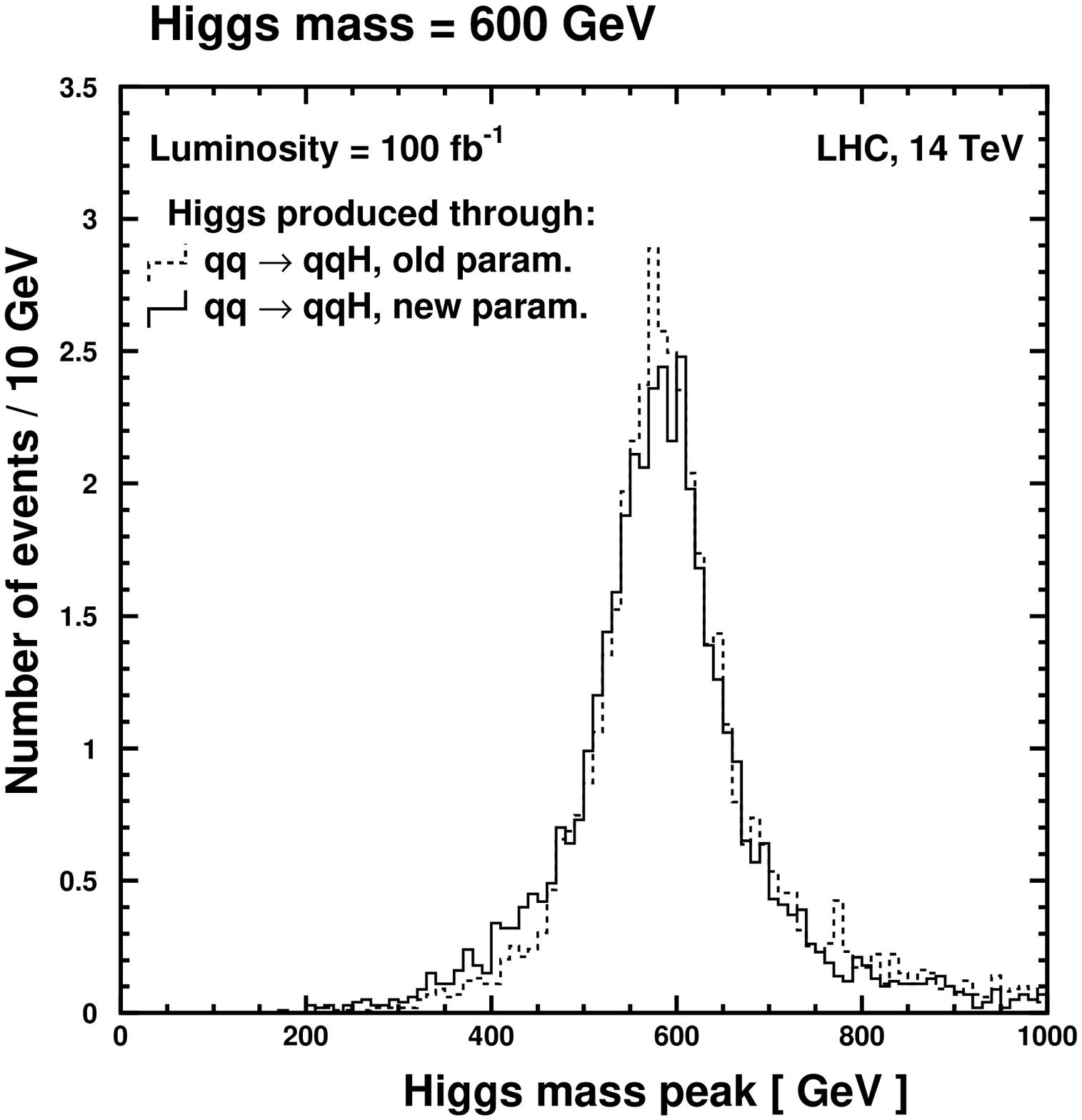}
}
}
\caption{Difference between the generated Higgs mass peak obtained with
the old and the new parameterization. Upper plots
correspond to a 300~GeV Higgs and lower plots to a 600~GeV Higgs.
The Higgs events on the left plots were
produced through
gluon fusion and on the right plots through weak boson fusion.
The plots are made on the generator level, without any selection cuts
and NLO cross sections are used.}
\label{param}
\end{center}
\end{figure}

In contrast to previous studies, like for instance \cite{atlas},
in all the simulations a new parameterization for the
Higgs width (described in \cite{higgswidth}) was used.
Figure \ref{param} gives a comparison of the Higgs mass spectra obtained
with the old and the new parameterizations for different Higgs masses and the two
Higgs production processes.

The use of the new parameterization increases
up to 10\% the signal
cross section. The plots are thus made for a number of events corresponding to
a luminosity of $100 fb^{-1}$ and NLO cross sections. 
There is no real difference between the
two parameterizations for a 300~GeV Higgs. A noticeable
difference between the two parameterizations appears for a 600~GeV Higgs
produced through gluon fusion. Actually this new parameterization
tends to give
larger tails to the Higgs lineshape.
Moreover the simulated mass spectrum is
the convolution between the parton
distribution function and the natural Higgs width. 
As the gluons have a
smaller mean momentum
than the valence quarks, the
tail in the Higgs mass distribution will be more significant for lower
Higgs masses as soon as gluons are involved. 
The plot for a 600~GeV Higgs produced through gluon fusion clearly shows that:
More events with a mass around 400~GeV are expected
with the new parameterization compared to what older simulations give.

What we conclude from this is that this effect
should be taken into account for high Higgs masses 
and when gluons are involved in the production mechanism.

\subsection*{Simulating the continuum vector boson production with PYTHIA}

The continuum boson production was generated using the PYTHIA processes
22 for $qq \fl ZZ$, 23 for $qq \fl WZ$
and 25 for $qq \fl WW$.

\begin{figure}[htb]
\begin{center}
\includegraphics[scale=0.4]{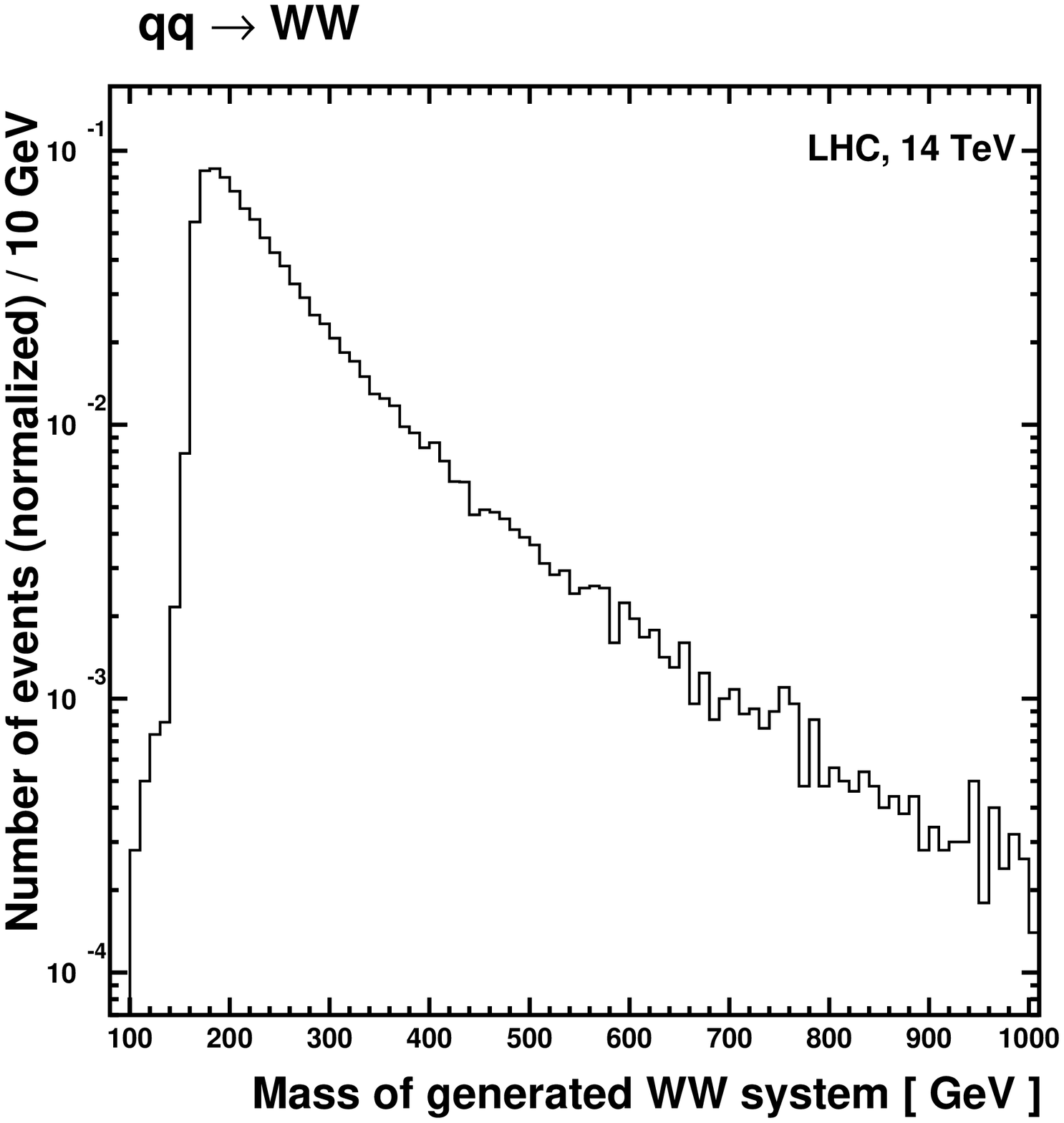}
\caption{Mass spectrum of the 
generated $WW$ system, produced through $qq \fl WW$.}
\label{masscont}
\end{center}
\end{figure}

It is interesting the consider the shape of the generated $ZZ$, $WW$ or $WZ$ mass
spectrum.
Figure \ref{masscont} shows the mass spectrum of the generated
$qq \fl WW$ process.
The mass spectrum shows a maximum around $2 \, m_W$ and tends then to decrease 
exponentially with the $WW$ mass.
Therefore there are much more events around 300~GeV
than around 600~GeV. More 
background events of that kind will also be
present around 300~GeV than around
a 600~GeV. But as the Higgs width is smaller 
when its mass is 300~GeV than
600~GeV, we compensate this effect by taking narrower mass windows
in the selection cuts.

\subsection*{Simulating the single boson production with PYTHIA}

The single $W$,$Z$ production with multi jets was generated using the
PYTHIA processes 15 and 30 for
$qq \fl Z+jets$ and 16 and 31 for
$qq \fl W+jets$.

These processes have matrix elements 
which are divergent for $p_t \fl 0$ and therefore should not be used down
to the low-$p_t$ region. 
We then generated them asking the $p_t$'s of the initial system being larger than
50~GeV to avoid that type of problems.  \label{cutbak}

As the cross section of those kind of processes 
are too large to allow a simulation of all events required,
a way of generating less background events had to be found.
A solution was to generate the
single boson production with a cut on the total generated system mass,
$\widehat{m}$, and on the $p_t$ of the total
system, $\widehat{p_t}$. It lowers the background cross section and allows
less events to be generated for the same luminosity. 
By simulating smaller background samples, we verified than
the events which were not
generated in the analysis
would not have survived the cuts anyway.
For instance,
in the channel $H \fl WW \fl l \nu jj$, the backgrounds 
$qq \fl g W^+$ and $ qg \fl q W^+$ have to be generated.
If it is done with $\widehat{m}>300\,\mathrm{GeV}$ and
$\widehat{p_t}>100\,\mathrm{GeV}$, their cross section falls from $3600\,pb$ with
$\widehat{m}>100\,\mathrm{GeV}$ and 
$\widehat{p_t}>50\,\mathrm{GeV}$ (the minimal requirements)
to $360\,pb$. We need thus to generate ten 
times less events for the same final
number of events !

Even with those generation cuts, these background
cross sections are very high.
Sometimes a number of events corresponding to
a luminosity of only $10\,fb^{-1}$ instead of $100\,fb^{-1}$ 
could be generated. And even with this luminosity,
the number of events to generate was about $10^7$~!
\label{reff}

\subsection*{Simulating the top antitop production with PYTHIA}

The top antitop production, $qq \fl t\bar{t} \fl Wb\,Wb$, 
was generated using the processes
81 and 82 in PYTHIA.

The cross section for this process is also very high ($\approx 600\,pb$)
and the presence of
many jets in the events slowed a lot the simulation. We could not
generate the number of events requested for
a luminosity of $100 fb^{-1}$ and therefore we have the same problem 
for this background than
for the single $W$, $Z$ production backgrounds.

To limit the number of background events to generate, we also forced
the vector bosons to decay only in the dangerous channels
for the signature studied.
For example,
in the channels where only leptonic decays were present, like in the
$H \fl WW \fl \ell \nu \ell \nu$ one, we could reduce the
number of $t\bar{t}$ events to generate
by a factor 10 by forcing the $W$'s to decay into leptons.
The assumption was made that the events were
the $W$'s did not decay leptonically were
excluded by the selection cuts, as the event had
to contain isolated leptons.

	\section{$H \fl ZZ \fl \ell^+ \ell^- \ell^+ \ell^- $}
The first signature studied is the so-called
\emph{gold plated channel}, namely the
$H \fl ZZ \fl \ell^+ \ell^- \ell^+ \ell^- $ signature.
It takes its name from the fact that
a Higgs mass peak can be nicely reconstructed, as leptons 
can be detected with a very good resolution.
Essentially only one source of background 
needs to be studied in detail, the $ZZ$ continuum production.
The different cross sections 
for the signal and the background are given in Table \ref{csllll}.

We give the
cross sections times branching ratio
directly at the next to leading order, as the
corresponding K-factors are given on page \pageref{ici}. 
Note that here and also for the other channels, $Z$ and $W$ and can decay
into $\tau$'s, even if almost all the signal events get lost with the selection cuts.
However, for backgrounds, $Z\fl \tau^+ \tau^-$ and $W\fl \tau \nu$ 
branching ratios
can be a danger and need to be simulated. Therefore, if not explicitely specified,
$\ell$ will refer to electrons, muons \emph{and} taus.

\begin{table}[htb]
\begin{center}
\begin{tabular}{|c|c||c|}
\hline
\multicolumn{3}{|c|}{\emph{Signal}} \\
\hline
Channel: & $qq \fl qqH \fl ZZ \fl \ell \ell \ell \ell$ &
$gg \fl H \fl ZZ \fl \ell \ell \ell \ell$ \\
\hline
Mass of Higgs &$\sigma \times BR$, NLO 
& $\sigma \times BR$, NLO \\
(GeV) & ($fb$) & ($fb$)  \\
\hline
300 & 4.3 \hspace*{0.3cm} & 22.1 \hspace*{0.3cm} \\
450 & 1.7 \hspace*{0.3cm} & 15.3 \hspace*{0.3cm} \\
600 & 1.0 \hspace*{0.3cm} & 4.5 \hspace*{0.3cm}  \\
\hline
\hline
\multicolumn{3}{|c|}{\emph{Background}} \\
\hline
\multicolumn{2}{|c|}{Channel} & 
$\sigma \times BR$ ($fb$) \\ 
\hline
\multicolumn{2}{|c|}{$qq \fl ZZ \fl \ell \ell \ell \ell$} & 163 \\ 
\hline
\end{tabular}
\caption{Cross sections  times branching ratio 
for the signal (NLO) and background in the channel 
$H \fl ZZ \fl \ell^+ \ell^- \ell^+ \ell^- $.}
\label{csllll}
\end{center}
\end{table}

The following cuts were used to isolate the Higgs signal 
when no jet tagging was made:
\begin{itemize}
    \item There must be four isolated leptons in the event
($e$ or $\mu$ with $p_t>10\,\mathrm{GeV}$
and $|\eta|<2.5$).
    \item The two $Z$ are reconstructed by pairing the leptons so that
the chosen combination out of the possible ones 
give masses lying as close as possible to the real $Z$ mass (91~GeV).
The mass of the
dilepton, which is closest to the $Z$ one, has to be in an interval of 10~GeV
centered in 91~GeV. The other reconstructed $Z$ mass has to be
in an interval of 20~GeV centered in 91~GeV.
    \item Only events where the mass of the reconstructed $ZZ$ system is within 
an interval of 16, 60 and 200~GeV centered in respectively 300, 450 and 
600~GeV were kept.
Note that the half of the size of the interval chosen,
8, 30 and 100~GeV, correspond to one 
sigma of variation around the Higgs reconstructed mass peak.
    \item As the only background is the continuum $ZZ$ production, we will 
make a soft cut on the reconstructed Higgs $p_t$,
rather than on the $p_t$'s of the $Z$'s alone.
(see page \pageref{coucou}),
$p_{t}(ZZ)>30\,\mathrm{GeV}$ was chosen.
\end{itemize}

The cuts used to isolate the Higgs produced through weak boson fusion,
with the jet tagging technique are the following:
\begin{itemize}		
    \item The events must survive the same cuts as explained before,
except for the $p_t$ cut on the reconstructed Higgs, since almost all the
continuum background events are removed with the jet tagging requirements.
    \item Exactly 2 jets must be in the event, with $p_t>20\,\mathrm{GeV}$
and $|\eta|<4.5$. Their invariant mass has to be higher than 800~GeV.
\end{itemize}

The expected results for $\mathcal{L}=100\,fb^{-1}$ are given in Table \ref{resllll}.
We give first
the number of events left after the minimal reconstruction cuts, ie.
when no cuts on the Higgs $p_t$ is done. Then the number of events left
after all cuts without and with the jet tagging is given.
We see that very good signal to background ratios
can be obtained.
As pointed out before, the cut on the reconstructed 
Higgs $p_t$ removes 90\% of the background in the considered
Higgs mass window, while
only 50\% of the signal events are removed, improving considerably
the signal to background ratio.

The jet tagging removes almost all the background events, while the
weak boson fusion signal events are remaining, but the statistics 
is small (between 5 and 15 events for a luminosity of $100\,fb^{-1}$).

\begin{table}[htb]
\begin{center}
\begin{tabular}{|l||r|r|r||r|}
\hline
\multicolumn{5}{|c|}{$H \fl ZZ \fl \ell^+ \ell^- \ell^+ \ell^- $}\\
\hline
\hspace*{0.7cm} Channel & \multicolumn{4}{c|}{Number of events} \\
& Generated & Min. rec. cuts    
& $p_t$ cut & Jet tag. \\
\hline
\multicolumn{5}{|c|}{$m_{Higgs}=300\,\mathrm{GeV}$} \\ 
\hline
$qq \fl qqH \fl \ell\ell\ell\ell$ & 430 & 50 & 50 & 15 \\
\hline
$gg \fl H \fl \ell\ell\ell\ell$ & 2'210 & 180 & 90 & 2 \\
\hline
\hline
$qq \fl ZZ  \fl \ell\ell\ell\ell$ & 16'300 & 50 & 4 & 0 \\
\hline
\multicolumn{5}{|c|}{$m_{Higgs}=450 \,\mathrm{GeV}$} \\
\hline
$qq \fl qqH \fl \ell\ell\ell\ell$ & 170 & 20 & 20 & 6 \\
\hline
$gg \fl H \fl \ell\ell\ell\ell$ & 1'530 & 120 & 70 & 1 \\
\hline
\hline
$qq \fl ZZ  \fl \ell\ell\ell\ell$ & 16'300 & 30 & 6 & 0 \\
\hline
\multicolumn{5}{|c|}{$m_{Higgs}=600 \,\mathrm{GeV}$}\\
\hline
$qq \fl qqH \fl \ell\ell\ell\ell$ & 100 & 15 & 15 & 5 \\
\hline
$gg \fl H \fl \ell\ell\ell\ell$ & 450 & 40 & 25 & 0 \\
\hline
\hline
$qq \fl ZZ  \fl \ell\ell\ell\ell$ & 16'300 & 30 & 10 & 0 \\
\hline
\end{tabular}
\caption{Expected results for $\mathcal{L}=100\,fb^{-1}$ and for different Higgs masses
in the $H \fl ZZ \fl \ell^+ \ell^- \ell^+ \ell^- $ channel,
NLO cross sections.
After the number of generated events,
the third column shows
the number of events left after all cuts except the one
on the reconstructed Higgs $p_t$ (Min. rec. cuts). 
The two last columns give the
number of events left after all cuts with and without the use of the jet tagging
technique.}
\label{resllll}
\end{center}
\end{table}

We can now calculate the necessary luminosity to get a 5 standard 
deviation.
We find, for example,
for a 600~GeV Higgs, a luminosity of approximatively
$30\,fb^{-1}$ without jet tagging.
The number of expected events in the signal for this luminosity
is 12 and in the background 3.
The probability to observe 15 events when the expected number of events
is 3 is about $10^{-6}$, assuming that the number of events follows a 
Poisson distribution.
This probability corresponds approximately
to the probability of a $5 \sigma$
deviation in a Gaussian distribution and this is the usual 
requirement for a discovery.

These discovery luminosities are given in Table \ref{ldiscllll}.

\begin{table}[htb]
\begin{center}
\begin{tabular}{|c|r|r|r|r|}
\hline
\multicolumn{5}{|c|}{$H \fl ZZ \fl \ell^+ \ell^- \ell^+ \ell^- $}\\
\hline
Higgs mass (GeV)& Signal & Background & 
S/B & $\mathcal{L}_{disc}$ ($fb^{-1}$)\\
\hline
300 & 140 & 4 & 35 & 4 \\
\hline
450 & 90 & 6 & 15 & 8 \\
\hline
600 & 40 & 10 & 4 & 30 \\
\hline
\end{tabular}
\caption{Expected discovery luminosities for the channel 
$H \fl ZZ \fl \ell^+ \ell^- \ell^+ \ell^-  $, without systematic errors taken into
account. The number of events for signal and background 
corresponds to a luminosity of $100\,fb^{-1}$.}
\label{ldiscllll}
\end{center}
\end{table}

We find
4, 8 and $10\,fb^{-1}$ for a 300, 450 and 600~GeV Higgs respectively.
These numbers are in agreement with the previous studies (see
for instance Ref \cite{atlas}). 
Finally one should keep in mind that the detector inefficiencies
are not taken into account.
For isolated leptons, they are expected to be high and therefore 
these luminosities might be increased.
For example, if we assume that the isolated leptons are detected with an efficiency
of 90\%, then the number of measured signal events will be
$N_s^{measured}=0.65 \cdot N_s^{expected}$, which leads to a loss of 19\%
in the significance of the
signal $\frac{N_s^{measured}}{\sqrt{B}}=\frac{0.65 \cdot N_s^{expected}}
{\sqrt{0.65 \cdot B}}$.

\begin{figure}[htb]
\begin{center}
\includegraphics[scale=0.4]{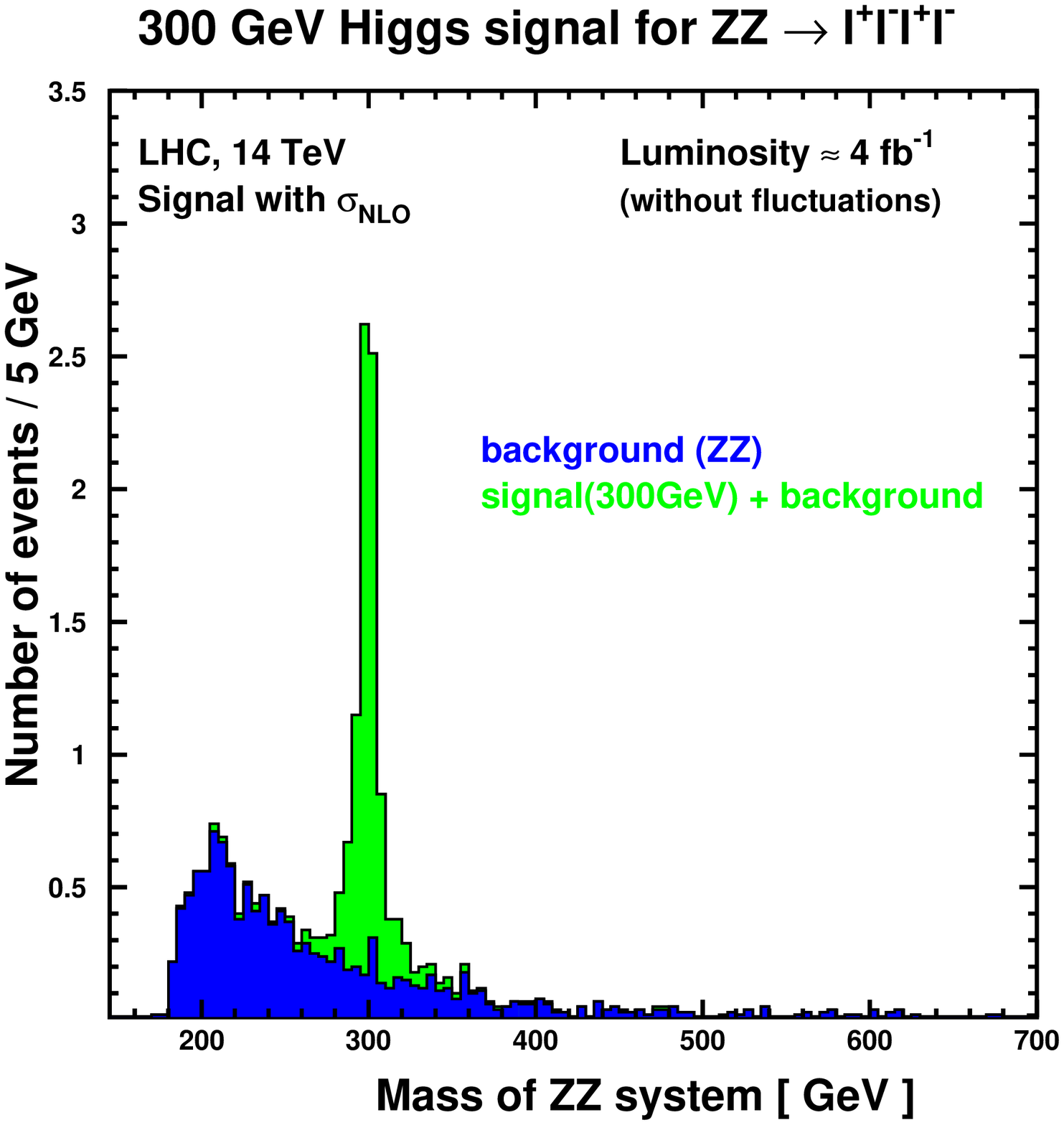}
\caption{Higgs mass peak for a Higgs mass of 300~GeV, when no jet tagging is
done. The signal and background were generated in a way to reduce
as much as possible the statistical fluctuations.}
\label{masspeak}
\end{center}
\end{figure}

\begin{figure}[htb]
\begin{center}
\mbox{
\subfigure{\includegraphics*[width=.5\textwidth]{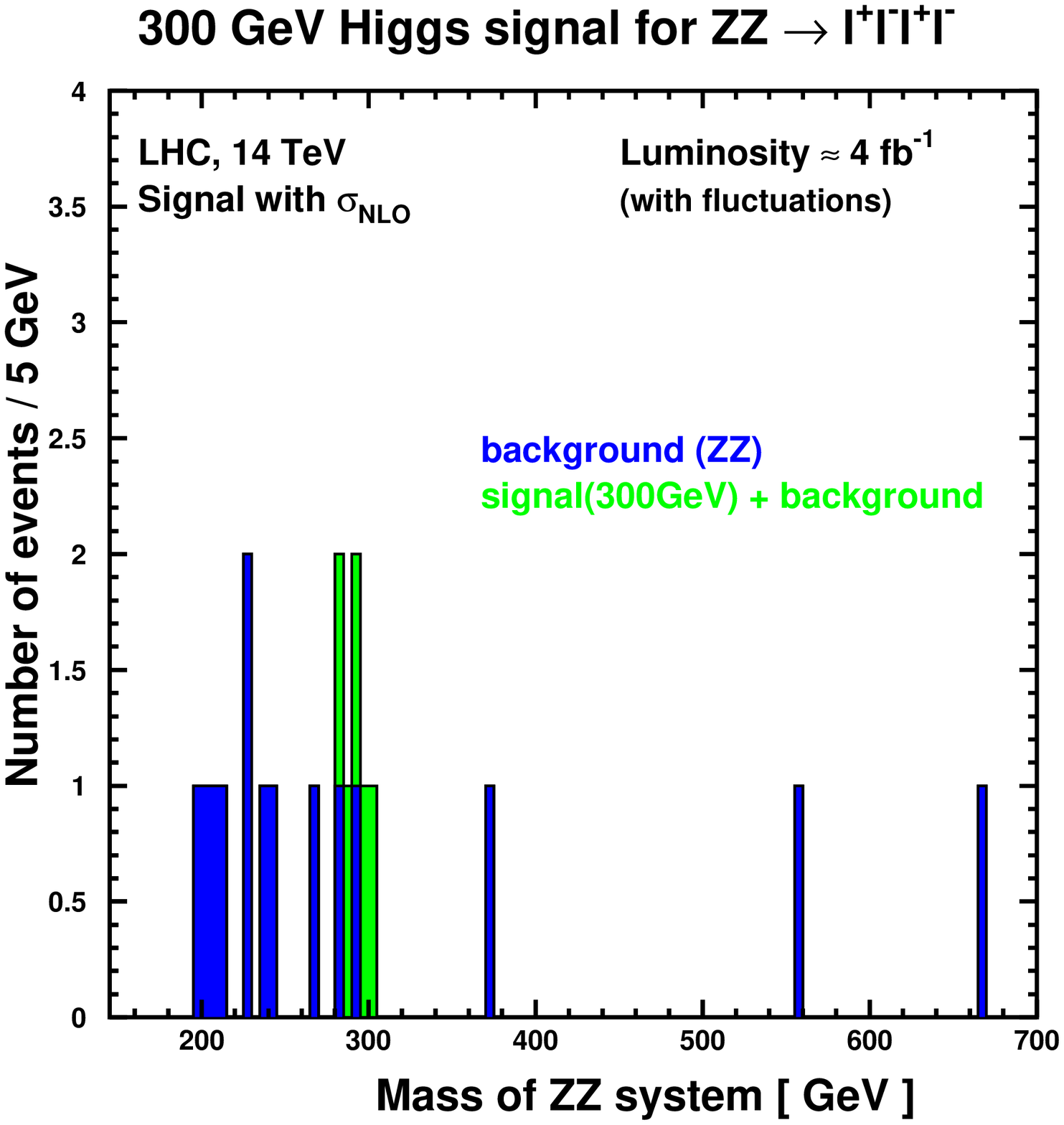} \quad }
\subfigure{
\includegraphics*[width=.5\textwidth]{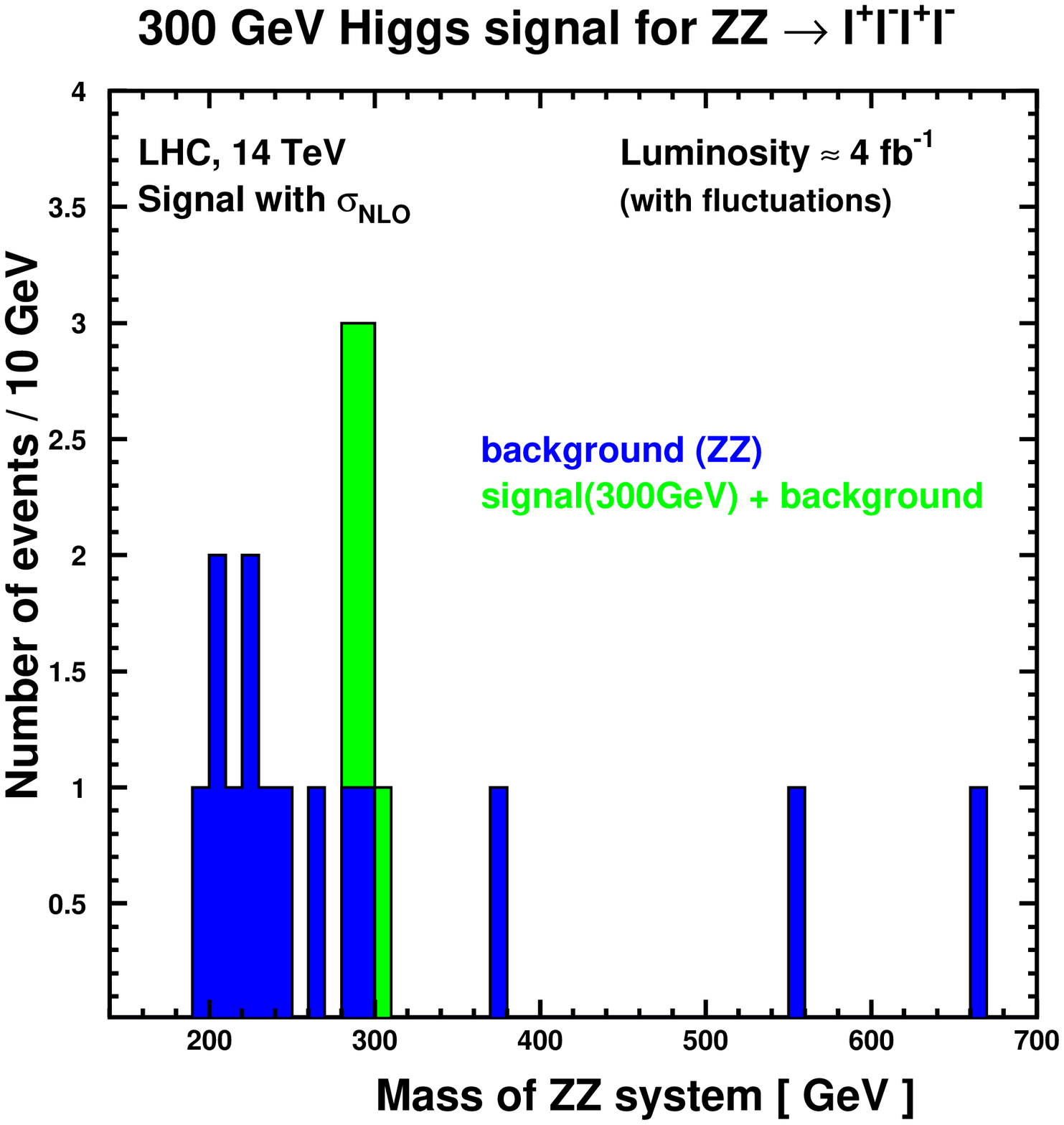}
}
}
\caption{Higgs mass peak for a Higgs mass of 300GeV, for a 'real' experiment and
two different binnings, the 5~GeV bins are optimized 
according to our high luminosity plot
while the 10~GeV bins are more for the small statistic samples.}
\label{masspeakreal}
\end{center}
\end{figure}

In Figure \ref{masspeak}, we can see how a mass peak for a 300~GeV Higgs 
would look like in that channel and for
a number of events corresponding to the discovery luminosity. 
However, we would like to draw attention 
to the y-axis, which represents
a small and non-integer number of events.
Actually, in order to get rid of the 
statistical fluctuations, the
signal was simulated for a number of events corresponding to a luminosity
$10'000 \,fb^{-1}$ and the background for a number of events corresponding
to a luminosity of $1'000 \,fb^{-1}$.
However, in a real 
experiment, we would rather get something like in Figure \ref{masspeakreal},
where only the number of events corresponding to the discovery luminosity
were generated. For a 300~GeV Higgs, as the discovery luminosity is $4 fb^{-1}$, it
represented 106 signal events and 652 background events.
We get 5 signal events in the good mass region against 2 background event.
The expected values are of 5.6 signal events and 0.16 background
events. We see than
that we were not really
lucky with our 'experiment', as the background is a bit high~!
The plots on the left and on the right on Figure \ref{masspeakreal} 
represent exactly the same events, 
but with a binning of 5~GeV and 10~GeV. 
While a signal has to be guessed on the plot on the left,
a Higgs peak becomes visible on the plot on the right. 

\begin{figure}[htb]
\begin{center}
\subfigure{\includegraphics*[scale=0.45]{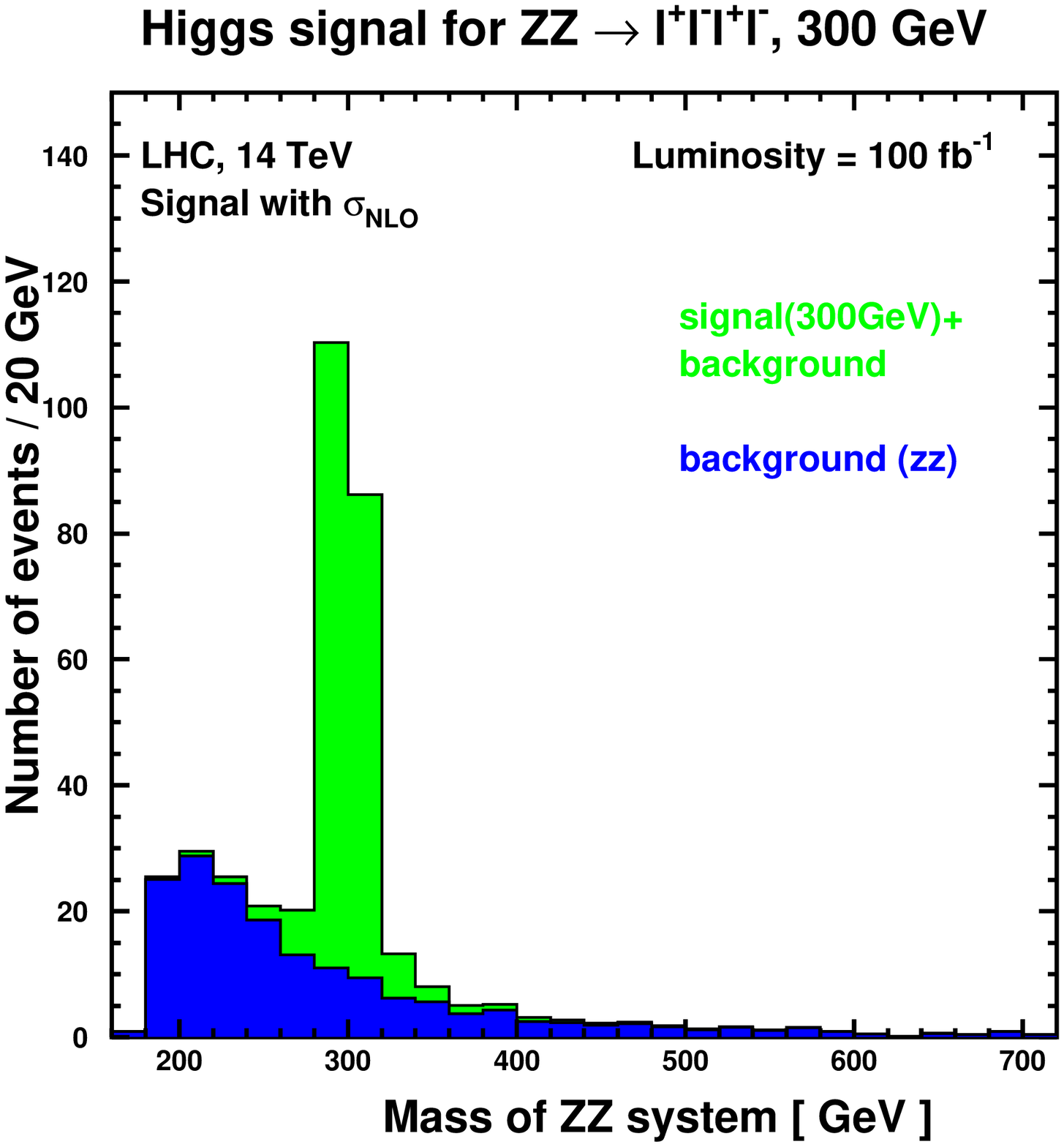}}
\\
\subfigure{\includegraphics*[scale=0.45]{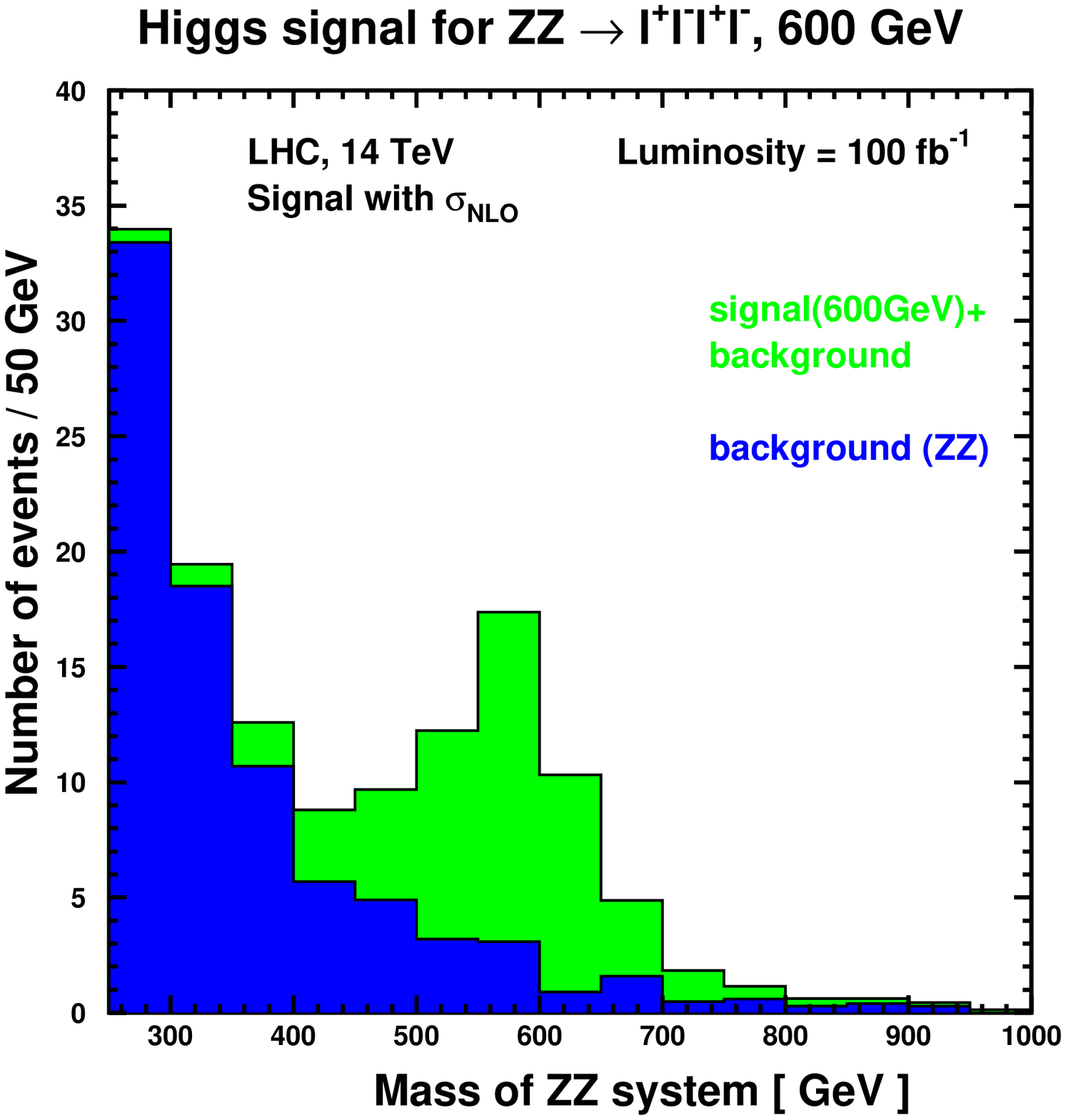}
}
\caption{Mass spectrum for a 300 (Up) and 600 (Down)~GeV Higgs for 
a luminosity of $100 fb^{-1}$, without jet tagging.}
\label{mnotag}
\end{center}
\end{figure}

Finally Figure \ref{mnotag} shows the mass peak we can get with
a luminosity of $100\,fb^{-1}$ without jet tagging for a 300 and a 600~GeV
Higgs.

When the jet tagging technique is used, all the
background events are removed, as continuum $ZZ$ production contains few jets.
However, not enough 
events are left to draw any significant conclusion:
17, 7 and 5 events are left for a 300, 450 and 600~GeV Higgs respectively.
\\[0.5cm]

For this channel, the Higgs search without tagging is 
giving excellent results, with
a good signal to background ratio.
The four leptons channel is thus a good candidate for
a discovery channel. On the other hand, the jet tagging technique do not give here
significant results, due to a too small statistic.

	\section{$H \fl ZZ \fl \ell^+ \ell^- \nu \bar{\nu} $}
This signature was already studied in detail 
(for instance see Ref. \cite{barca}) 
and it is known to give good signals
for Higgs masses between 500 and 800~GeV.
For such high masses, it seems to be superior
to the four leptons channel.
Notice that the branching faction for this channel
is six times larger than the one
for the four leptons channel.
Even if the Higgs mass peak cannot be totally reconstructed, the presence of the 
two leptons and a large missing energy gives this signature enough
specifications to be isolated for large Higgs masses.

The signal cross sections times branching ratio for this signature
are given in Table \ref{cssigzzllnunu} 
for the 
two most significant production mechanisms of the Higgs.

\begin{table}[htb]
\begin{center}
\begin{tabular}{|c|c|c|}
\hline
\multicolumn{3}{|c|}{\emph{Signal}} \\
\hline
Channel: &$qq \fl qqH \fl ZZ \fl \ell \ell \nu \nu$ &
$gg \fl H \fl ZZ \fl \ell \ell \nu \nu$ \\
\hline
Mass of Higgs &$\sigma \times BR$, NLO 
& $\sigma \times BR$, NLO \\
(GeV) & ($fb$) & ($fb$) \\
\hline
 300 &  16  \hspace*{0.3cm} & 
  90  \hspace*{0.3cm}\\
\hline
 450 & 7  \hspace*{0.3cm} & 
  60  \hspace*{0.3cm} \\ 
\hline
 600 & 4  \hspace*{0.3cm}  &
  23  \hspace*{0.3cm} \\ 
\hline
\end{tabular}
\caption{Cross sections times branching ratio for the signal in the channel 
$H \fl ZZ \fl \ell^+ \ell^- \nu \bar{\nu} $, given at the NLO.}
\label{cssigzzllnunu}
\end{center}
\end{table}

The main feature of this signature is thus two isolated leptons, coming from a
$Z$ decay and a large missing $p_t$.
The potential backgrounds are then all the processes which
can produce two isolated leptons and missing $p_t$ as given in
Table \ref{csbakzzllnunu}.

\begin{table}[htb]
\begin{center}
\begin{tabular}{|l|rc|}
\hline
\multicolumn{3}{|c|}{\emph{Backgrounds}} \\
\hline
Channel & $\sigma \times BR$ ($fb$) &\\
\hline
$qq \fl ZZ \fl \ell \ell \nu \nu$ & 502 &\\ 
\hline
$qq \fl WW \fl \ell \nu \ell \nu$ & 8'050 &\\
\hline
$qq \fl WZ \fl \ell \nu \ell \ell$ & 950 & \\
\hline
$qq \fl Z + jets \quad (Z \fl \ell\ell)$ & 486'000 &
($\widehat{p_t}>50 \,\mathrm{GeV}$) \\
\hline
$qq \fl t \bar{t}\fl WbWb \fl \ell \nu b\; \ell \nu b $ 
& 65'400 &\\
\hline
\end{tabular}
\caption{Cross sections times branching ratio
for the potential backgrounds in the channel 
$H \fl ZZ \fl \ell^+ \ell^- \nu \bar{\nu}$.}
\label{csbakzzllnunu}
\end{center}
\end{table}

When there are jets in the event, to get rid of the $ZZ$ continuum production,
a cut on the $p_t$ of the reconstructed $ZZ$ system was made,
which reduces this background by a factor 3 while the signal is only
reduced by a factor 1.5
(if there are no jets in the event, the reconstructed Higgs has 
obviously $p_t=0$).

The $Z$+ jets background can be reduced strongly with cuts on the 
reconstructed $Z$ $p_t$'s.

The $t\bar{t}$ background is
not really a problem here, as the isolated leptons do not give
the $Z$ mass.

The Higgs signal events are isolated with the following cuts:
\begin{itemize}
    \item We ask for two isolated leptons from the same flavor in event ($e$, $\mu$ 
with $p_t>10\,\mathrm{GeV}$ and $|\eta|<2.5$), whose invariant mass has to be
within an interval of 10~GeV centered in 91~GeV.
    \item Then we ask, as a pre-cut,
the missing $p_t$ in the event to be higher than 75~GeV. 
The $p_t$ of the reconstructed $Z$ decaying into leptons has to be higher than 75~GeV.
    \item As explained before,
when there are jets in the event, the $p_t$ of the reconstructed Higgs
has to be larger than 50~GeV.
    \item A harder cut on the missing $p_t$ is finally done:
$p_{t}^{miss}>150(175,\,225)\,\mathrm{GeV}$, 
for a 300(450, 600)~GeV Higgs.
\end{itemize}

A background which remains problematic when the jet tagging is done, is the 
single $Z$ production with jets.
It definitely looks like a signal event if the $Z$ decays into leptons and the
jets emitted together with the $Z$ play the role
of the tagging jets. The missing $p_t$ comes
from badly reconstructed jets. 
However the jets identified as the tagging jets are 
more centrally emitted than the ones coming from the signal events
and the cut on their invariant mass reduces this background
by a factor of 40 against a factor of 2 for the signal. Furthermore, the missing $p_t$
spectrum for the main background
is falling much more steeply than the one for the signal, 
as there are no neutrinos in the 
event. A final cut on the missing $p_t$ will then be lethal for that
background.
Cuts on the $p_t$'s of the reconstructed $ZZ$ system and the reconstructed $Z$'s
allows finally to get a better efficiency.
The cuts used to isolate weak boson fusion are the following:
\begin{itemize}		
     \item The events must survive the above cuts,
except for the final $p_t$ cuts.
     \item Exactly 2 jets must be in the event with $p_t>20\,\mathrm{GeV}$
and $|\eta|<4.5$. Their invariant mass has to be higher than 800~GeV.
    \item A last cut on the missing $p_t$ is done:
$p_{t}^{miss}>100\,(125,\,150)\,\mathrm{GeV}$, for a 300 (450, 600)~GeV Higgs.
\end{itemize}

The expected results for $\mathcal{L}=100\,fb^{-1}$ and for different Higgs masses
are given in Table \ref{reszzllnunu}. We see that the main background 
is $qq \fl Z + jets$. The
cut which will be efficient against it will be the
final cut on the missing $p_t$ which reduces this background from a factor
25.

\begin{table}
\begin{center}
\scalebox{0.9}{
\begin{tabular}{|l||r|r|r||r|}
\hline
\multicolumn{5}{|c|}{$H \fl ZZ \fl \ell^+ \ell^- \nu \bar{\nu} $}\\
\hline
\hspace*{1.5cm} Channel & \multicolumn{4}{c|}{Number of events} \\
& Generated & Min. rec. cuts    
& $p_t$ cuts & Jet tag. \\
\hline
\multicolumn{5}{|c|}{} \\
\multicolumn{5}{|c|}{$m_{Higgs}=300\,\mathrm{GeV}$} \\ 
\multicolumn{5}{|c|}{} \\
\hline
$qq \fl qqH \fl ZZ \fl \ell \ell \nu \nu$ &1'600 & 260 & 55 & 55 \\
\hline
$gg \fl H \fl ZZ \fl \ell \ell \nu \nu$ & 9'000 & 1'500 & 90 & 7 \\
\hline
Sum of all backgrounds & 56'090200 & 37'960 & 2'115 & 66 \\
\hline
\hline
\multicolumn{5}{|l|}{Detailed backgrounds} \\
\hline
$qq \fl ZZ \fl \ell \ell \nu \nu$ & 50'200 & 2'600 & 500 & 0 \\
\hline
$qq \fl WW \fl \ell \nu \ell \nu$ & 805'000 & 300 & 10 & 0 \\
\hline
$qq \fl WZ \fl \ell \nu \ell \ell $ & 95'000 & 1'000 & 155 & 0 \\
\hline
$qq, gg \fl Z + jets \quad (Z \fl \ell\ell) $ & 48'600'000 & 30'000 & 1'100 & 60 \\
\hline
$qq \fl t \bar{t}\fl WbWb \fl \ell \nu b\; \ell \nu b $ 
& 6'540'000 & 4'060 & 350 & 6 \\
\hline
\hline
\multicolumn{5}{|c|}{} \\
\multicolumn{5}{|c|}{$m_{Higgs}=450\,\mathrm{GeV}$} \\ 
\multicolumn{5}{|c|}{} \\
\hline
$qq \fl qqH \fl ZZ \fl \ell \ell \nu \nu$ & 700 & 190 & 110 & 55 \\
\hline
$gg \fl H \fl ZZ \fl \ell \ell \nu \nu$ & 6'000 & 1'300 & 700 & 10 \\
\hline
Sum of all backgrounds & 56'090'200 & 37'960 & 1'313 & 33 \\
\hline
\hline
\multicolumn{5}{|l|}{Detailed backgrounds} \\
\hline
$qq \fl ZZ \fl \ell \ell \nu \nu $ & 50'200 & 2'600 & 350 & 0 \\
\hline
$qq \fl WW \fl \ell \nu \ell \nu $ & 805'000 & 300 & 3 & 0 \\
\hline
$qq \fl WZ \fl \ell \nu \ell \ell $ & 95'000 & 1'000 & 110 & 0 \\
\hline
$qq, gg \fl Z + jets \quad (Z \fl \ell\ell)$ & 48'600'000 & 30'000 & 670 & 30 \\
\hline
$qq \fl t \bar{t}\fl WbWb \fl \ell \nu b\; \ell \nu b $
& 6'540'000 & 4'060 & 180 & 3\\
\hline
\hline
\multicolumn{5}{|c|}{} \\
\multicolumn{5}{|c|}{$m_{Higgs}=600\,\mathrm{GeV}$} \\ 
\multicolumn{5}{|c|}{} \\
\hline
$qq \fl qqH \fl ZZ \fl \ell \ell \nu \nu$ & 400 & 120 & 80 & 40 \\
\hline
$gg \fl H \fl ZZ \fl \ell \ell \nu \nu$ & 2'300 & 710 & 330 & 7 \\
\hline
Sum of all backgrounds & 56'090'200 & 37'960 & 401 & 12 \\
\hline
\hline
\multicolumn{5}{|l|}{Detailed backgrounds} \\
\hline
$qq \fl ZZ \fl \ell \ell \nu \nu $ & 50'200 & 2'600 & 150 & 0 \\
\hline
$qq \fl WW \fl \ell \nu \ell \nu $ & 805'000 & 300 & 1 & 0 \\
\hline
$qq \fl WZ \fl \ell \nu \ell \ell $ & 95'000 & 1'000 & 40 & 0 \\
\hline
$qq, gg \fl Z + jets \quad (Z \fl \ell\ell)$ & 48'600'000 & 30'000 & 170 & 10 \\
\hline
$qq \fl t \bar{t}\fl WbWb \fl \ell \nu b\; \ell \nu b $
& 6'540'000 & 4'150 & 40 & 2\\
\hline
\end{tabular}}
\caption{Expected
results for $\mathcal{L}=100\,fb^{-1}$ and for different Higgs masses
in the $H \fl ZZ \fl \ell^+ \ell^- \nu \bar{\nu} $, 
channel, NLO cross sections. 
After the total number of events generated for each process,
we give the number of events left after the minimal reconstruction
cuts, which means all cuts before the cut on the missing $p_{t}$ and the
number of events left without and with the jet tagging technique.
Notice than looser $p_t$ cuts are done if the jet tagging technique is used.}
\label{reszzllnunu}
\end{center}
\end{table}

Is this possible to discover the Higgs in
that channel~? It is interesting to calculate the
necessary luminosity for a 5 standard deviation with and without the
use of the jet tagging technique.
For the search without tagging,
as we have here more statistic, we can
approximate the Poisson distribution with a Gaussian distribution and the
formula for the discovery luminosity simplifies to:
$$\mathcal{L}_{disc}=2'500\frac{B}{S^2}$$
where $B$ and $S$ are the number of background and signal events
corresponding to a luminosity of $100\,fb^{-1}$.
For the results obtained with the jet tagging technique, 
the Poisson distribution has to be used.
These results are given in Table \ref{ldiscllnunu}.
For Higgs masses higher than 450~GeV, 
interesting signals with good signal to background ratios are obtained. \\

\begin{table}[htb]
\begin{center}
\begin{tabular}{|c|r|r|r|r|}
\hline
\multicolumn{5}{|c|}{$H \fl ZZ \fl \ell^+ \ell^- \nu \bar{\nu} $}\\
\hline
& Signal & Background
& S/B & $\mathcal{L}_{disc}$ ($fb^{-1}$)\\
\hline
\multicolumn{5}{|c|}{$m_{Higgs}=300\,\mathrm{GeV}$} \\ 
\hline
no tagging & 145 & 2115 & 0.07 & 251 \\
tagging & 62 & 66 & 0.94 & 43 \\
\hline
\multicolumn{5}{|c|}{$m_{Higgs}=450\,\mathrm{GeV}$} \\ 
\hline
no tagging & 810 & 1'313 & 0.62 & 5 \\
tagging & 65 & 33 & 1.97 & 27 \\
\hline
\multicolumn{5}{|c|}{$m_{Higgs}=600\,\mathrm{GeV}$} \\ 
\hline
no tagging & 410 & 400 & 1.03 & 6 \\
tagging & 47 & 12 & 3.92 & 30 \\
\hline
\end{tabular}
\caption{Expected discovery luminosities for the channel 
$H \fl ZZ \fl \ell^+ \ell^- \nu \bar{\nu} $, without systematic errors taken into
account. The number of events for signal and background 
corresponds to a luminosity of $100\,fb^{-1}$.}
\label{ldiscllnunu}
\end{center}
\end{table}

We conclude from this that it would be interesting
to combine this channel with the four leptons channel 
in the search of the Higgs.

\begin{figure}[htb]
\begin{center}
\mbox{
\subfigure{\includegraphics*[width=.5\textwidth]{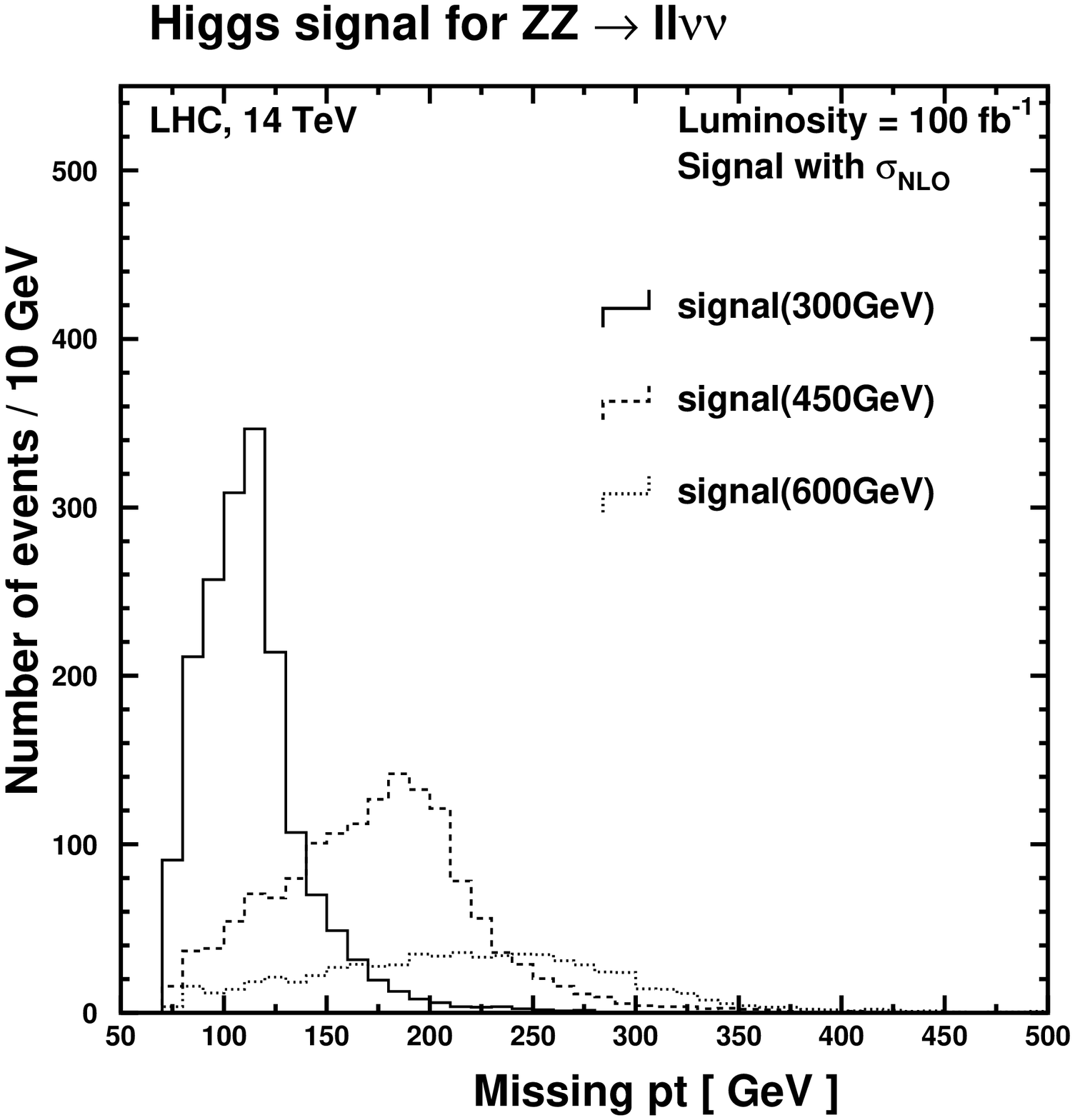}\quad
}
\subfigure{\includegraphics*[width=.5\textwidth]{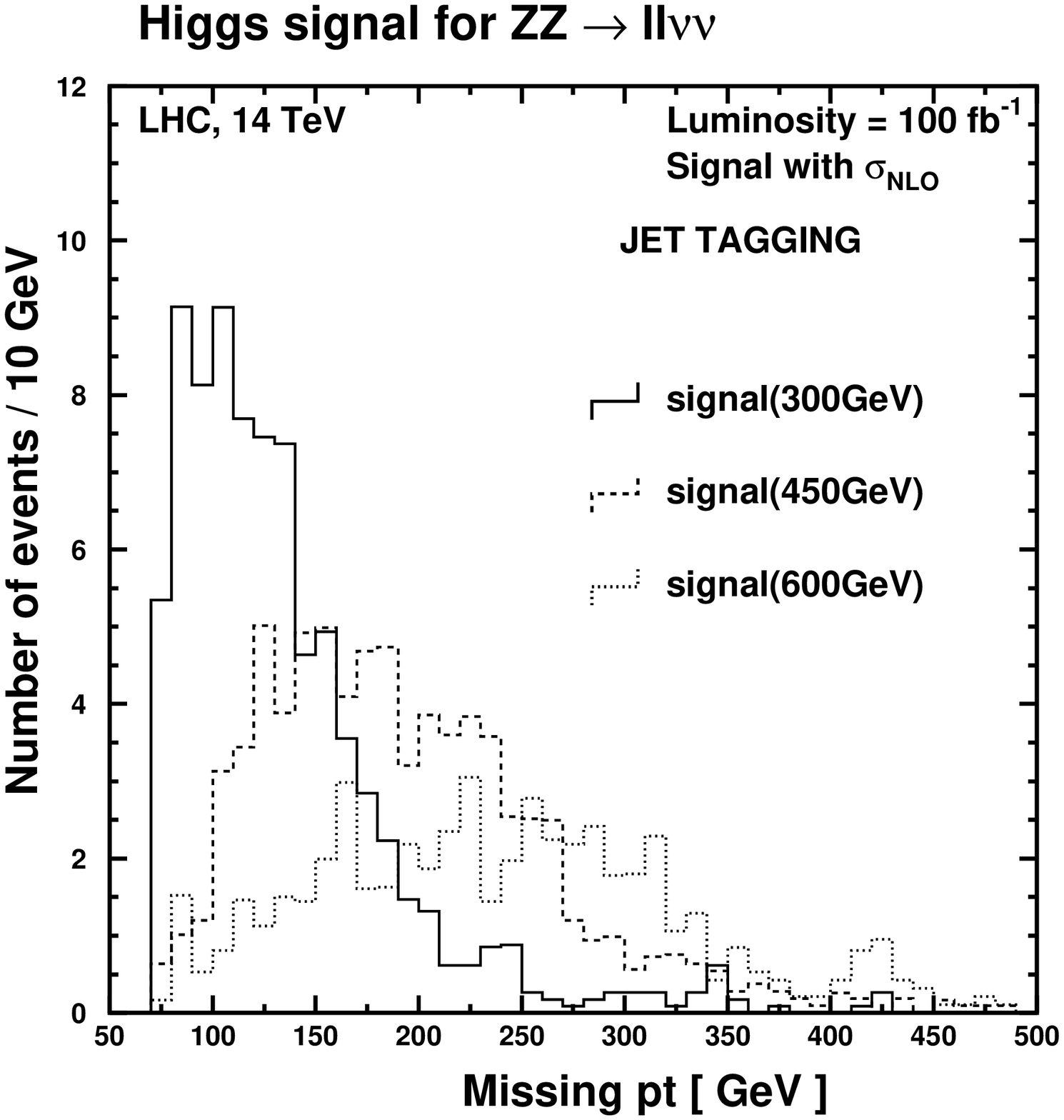}
}
}
\caption{Missing $p_t$ spectrum for a 300, 450 and 600~GeV Higgs without the
backgrounds.
(Left) without tagging (Right) with tagging. We can clearly see in all cases
a peak in the distribution.}
\label{ptsig}
\end{center}
\end{figure}

\begin{figure}[htb]
\begin{center}
\mbox{
\subfigure{\includegraphics*[width=.5\textwidth]{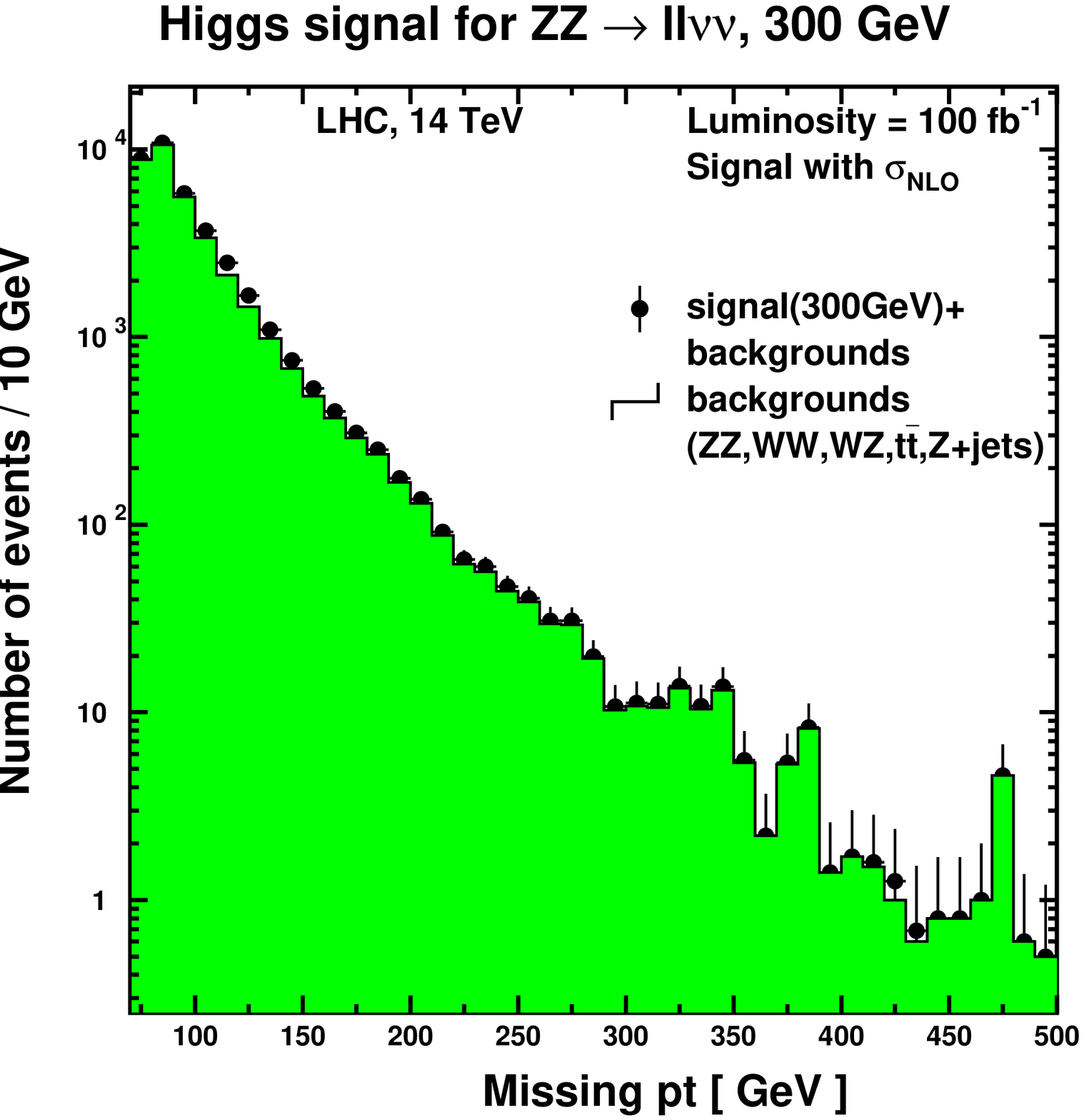}\quad
}
\subfigure{\includegraphics*[width=.5\textwidth]{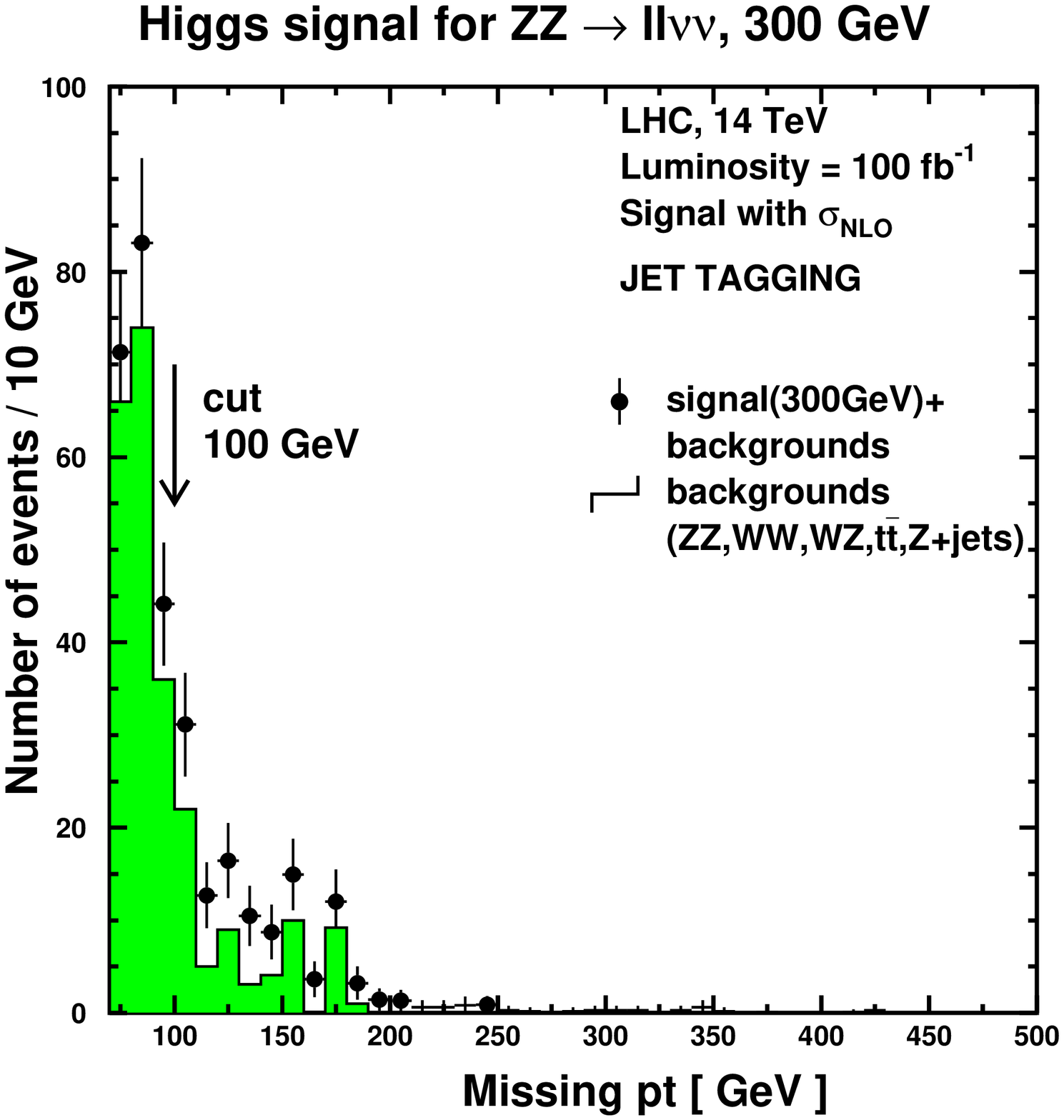}
}
}
\caption{Missing $p_t$ spectrum for a 300~GeV Higgs signal with the backgrounds.
(Left) without tagging (Right) with tagging.}
\label{ptllnunu300}
\end{center}
\end{figure}

\begin{figure}[htb]
\begin{center}
\mbox{
\subfigure{\includegraphics*[width=.5\textwidth]{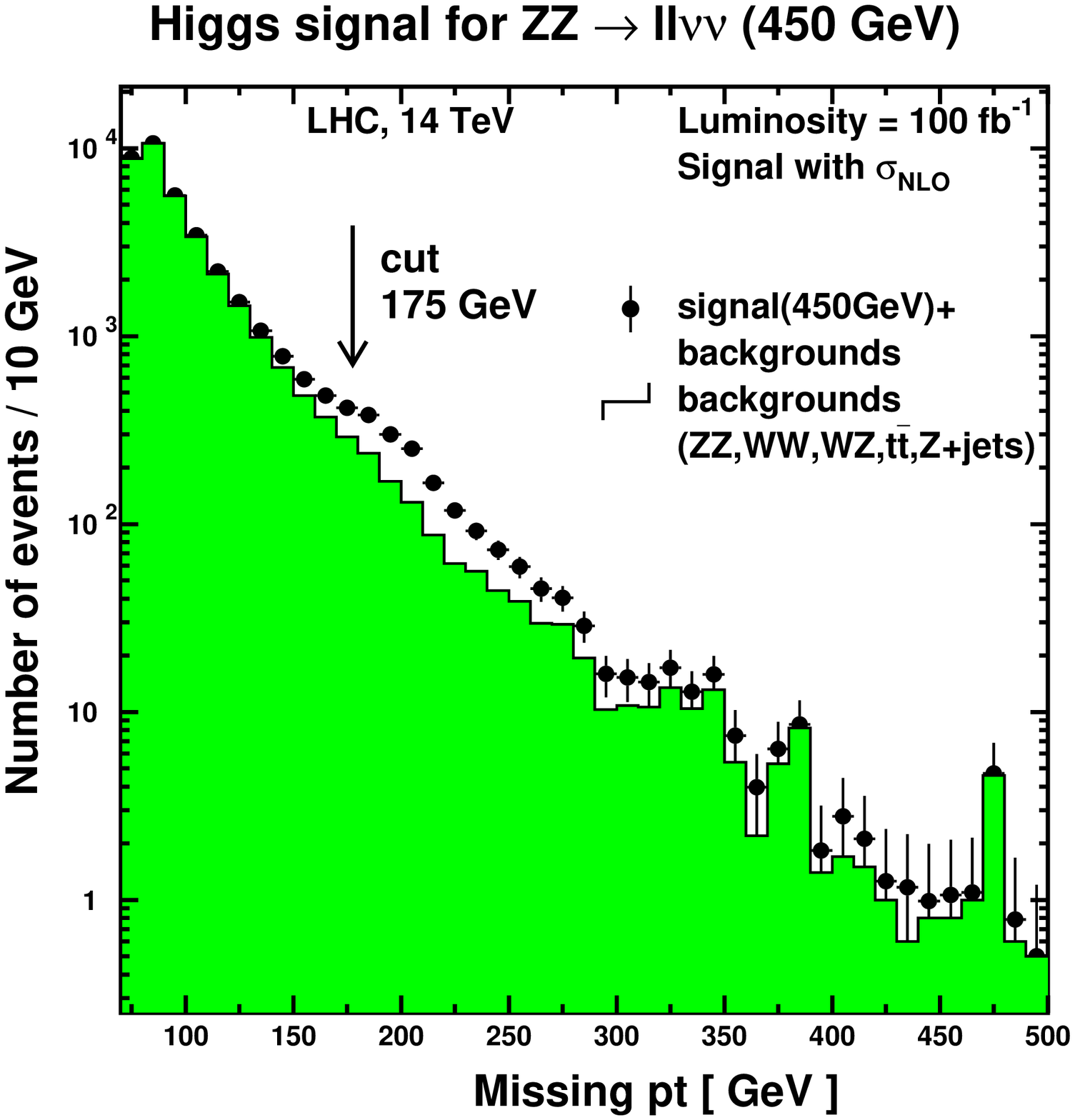}\quad
}
\subfigure{\includegraphics*[width=.5\textwidth]{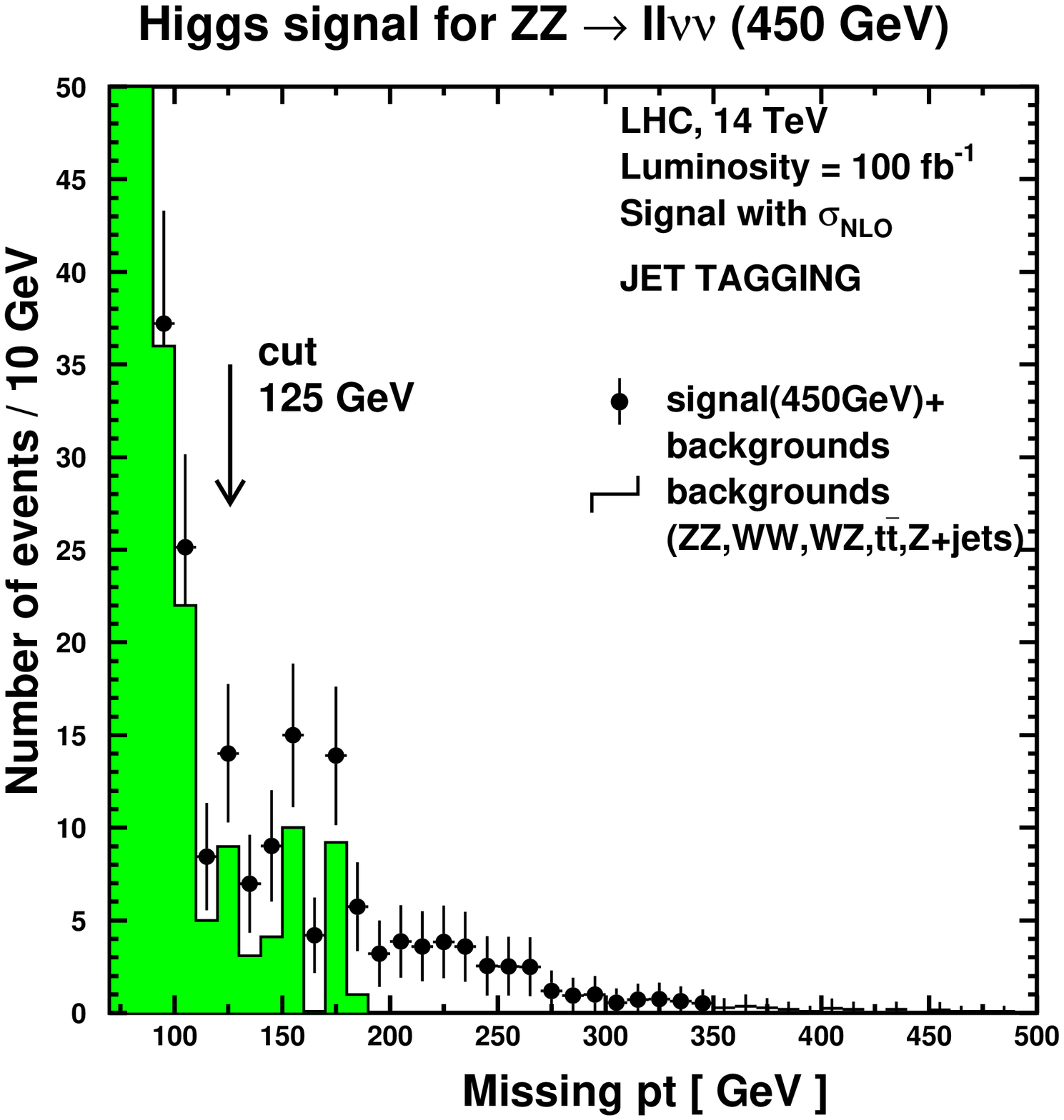}
}
}
\caption{Missing $p_t$ spectrum for a 450~GeV Higgs signal with the backgrounds.
(Left) without tagging (Right) with tagging.}
\label{ptllnunu450}
\end{center}
\end{figure}

\begin{figure}[p]
\begin{center}
\includegraphics[width=.6\textwidth]{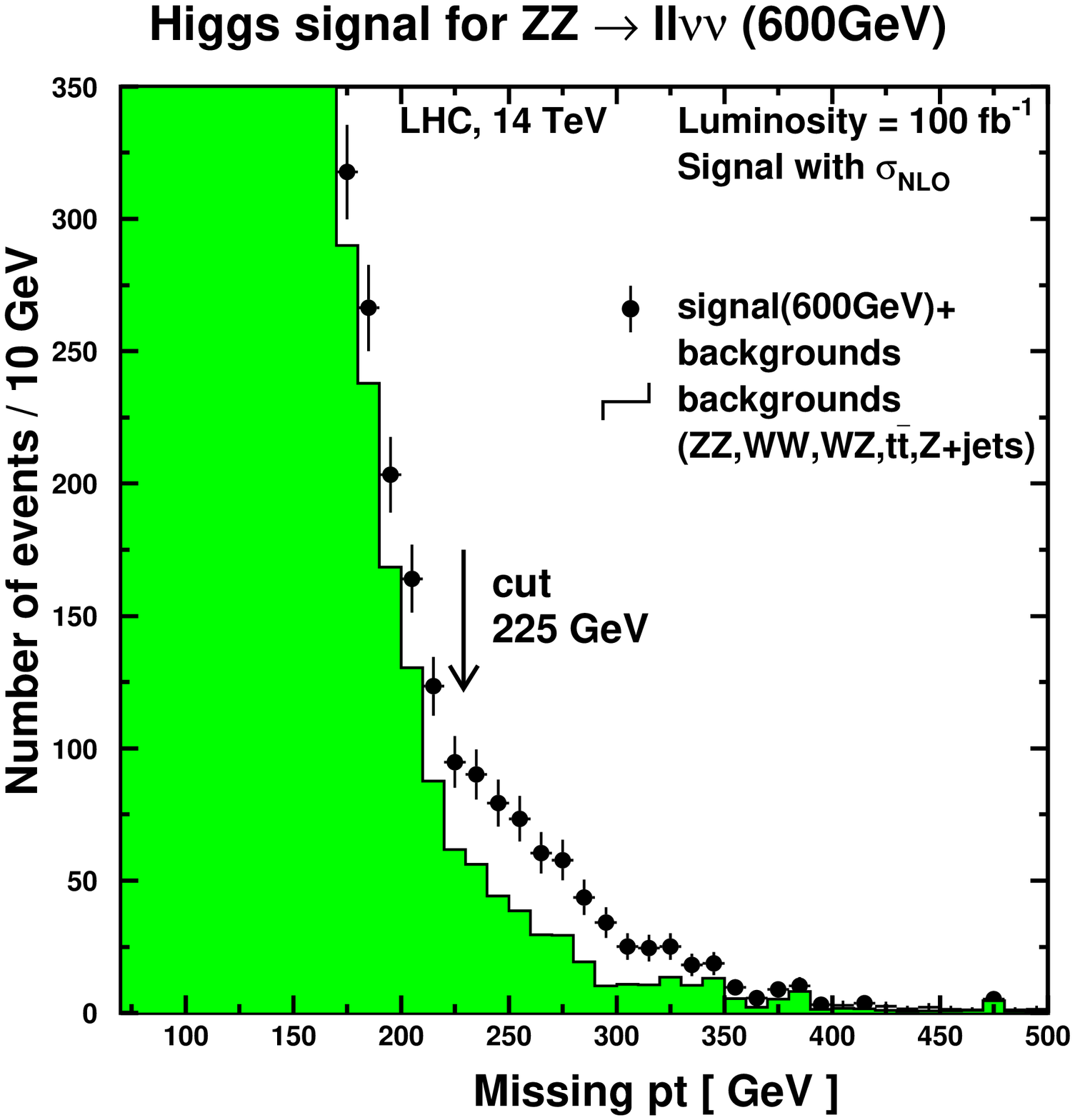}
\\[0.5cm]
\includegraphics[width=.6\textwidth]{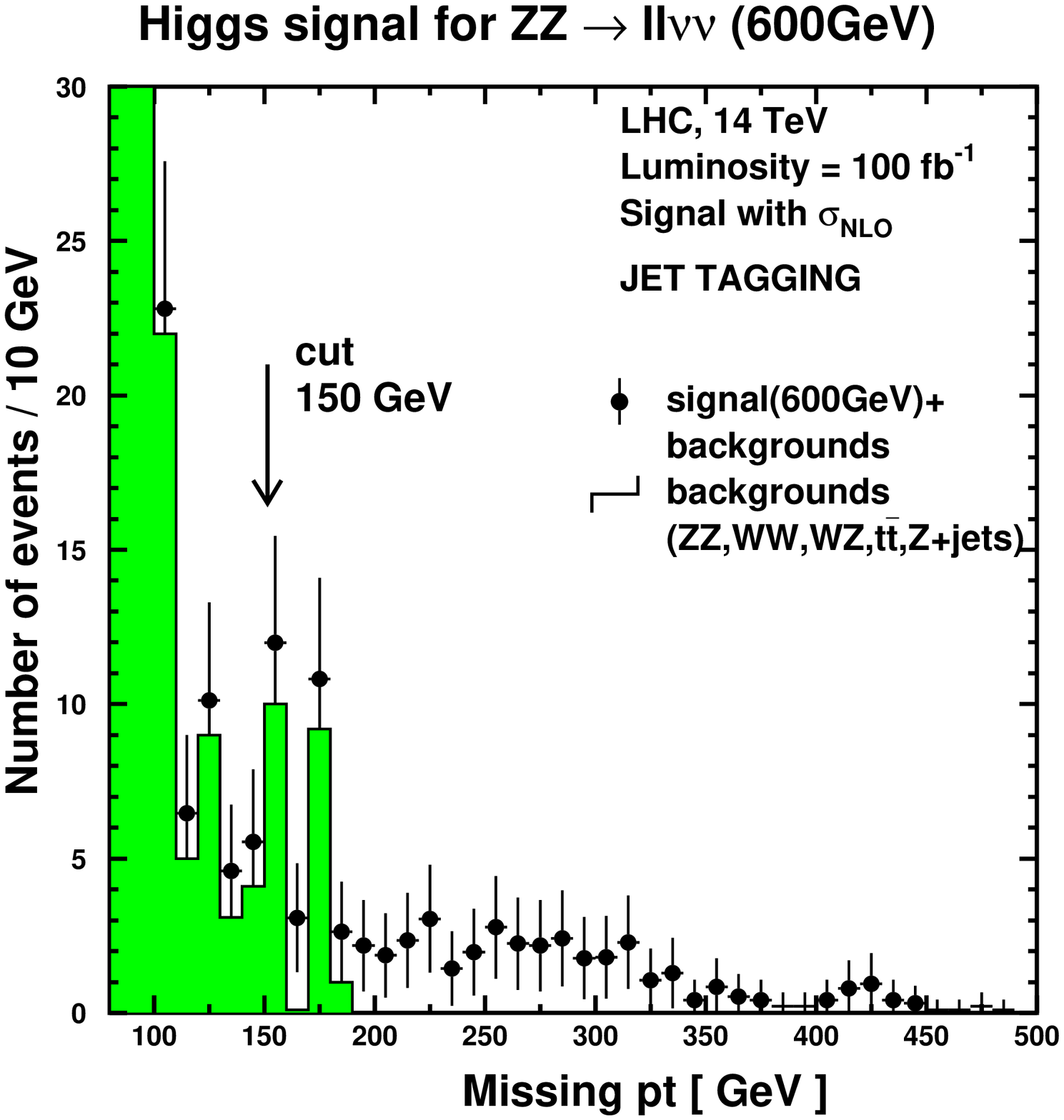}
\caption{Missing $p_t$ spectrum for a 600~GeV Higgs signal with the backgrounds.
(Up) without tagging (Down) with tagging.}
\label{ptllnunu600}
\end{center}
\end{figure}

In Figures \ref{ptllnunu300}, \ref{ptllnunu450} and  \ref{ptllnunu600}
we can see the missing transverse momentum in the event for
a 300, 450 and respectively 600~GeV Higgs signal,
plotted above the different backgrounds.
The plot on the left is made
before jet tagging and the plot on the right
after the use of the jet tagging technique. 

Notice that without tagging, no excess can be seen for a 300~GeV Higgs.
The problem is that for low Higgs masses, the peak in the missing transverse
energy is just sitting where the background is huge. 
That means that it will be very difficult to
measure an excess in that region. For higher Higgs masses the peak gets shifted
and coincide with a region where the background in not that important any more 
and an excess can be seen. 
To get an idea of where the peak in the signal is sitting, see 
Figure \ref{ptsig}, which shows the $p_t$ spectrum
of the signal.
Actually the missing energy is 
correlated to the mass of the system.
This can be explained in the following way: In the center of mass referential,
the two vector bosons are emitted back to back.
Then the $W$'s get a boost along the z-axis and their $p_t$ 
remain unchanged.
Thus the higher their transverse momentum is, the
larger the mass of the system will be. 
Making a cut on the missing transverse momentum will then automatically
make a cut on the mass of the system.\\
\label{norm}
Notice also (on Figures \ref{ptllnunu300}, \ref{ptllnunu450} and 
\ref{ptllnunu450})
that the low missing $p_t$ region could be used to normalize the
background. It is good to keep that region
in order to
check if the background is indeed well described by the simulation.
This was the reason to use first some softer 
cuts on the $p_t$'s, in our case asking them to be higher than 75~GeV, 
and subsequently 
make harder cuts to get a good efficiency.\\[0.5cm]
This channel is thus giving interesting results for Higgs masses higher than 
300~GeV.
The jet tagging technique allows to get better signal to background ratios
and for a 300~GeV Higgs, we have a smaller discovery luminosities when
the jet tagging is used.

	\section{$H \fl WW \fl \ell \nu \ell \nu $}

This channel is a bit different from the channels studied so far, since 
with that signature,
it is impossible to reconstruct the Higgs mass and the cuts to isolate 
the signal have to rely only on the kinematic properties of the signal.
Note that in spite of this, it was shown \cite{lnulnu} that it gives a 
very good signal for 
Higgs masses between 150
and 200~GeV. 
To get a signal in that mass region, the spin correlations properties of the 
$W$'s coming from the Higgs were used as well as the differences
in the $p_{t}$ spectrum of the leptons coming from the decays of the $W$'s.
However if we go to higher Higgs masses, the strong angular 
correlations that the $W$'s had for masses around 160 GeV are not 
present any more, due to 
the large boost that the $W$ are receiving. New kinematical 
cuts have then to be found.

The cross section times branching ratio for the 
two most significant production mechanisms of the Higgs 
for this channel are given in Table \ref{cssigwwlnulnu}.

\begin{table}[htb]
\begin{center}
\begin{tabular}{|c|c|c|}
\hline
\multicolumn{3}{|c|}{\emph{Signal}} \\
\hline
Channel: & $qq \fl qqH \fl WW \fl \ell \nu \ell \nu$ &
$gg \fl H \fl WW \fl \ell \nu \ell \nu$ \\
\hline
Mass of Higgs &$\sigma \times BR$, NLO 
& $\sigma \times BR$, NLO \\
(GeV) & ($fb$) & ($fb$) \\
\hline
 300 &  88  \hspace*{0.3cm} & 
  533  \hspace*{0.3cm}\\
\hline
 450 & 40  \hspace*{0.3cm} & 
  332  \hspace*{0.3cm} \\ 
\hline
 600 & 22  \hspace*{0.3cm} &
  130 \hspace*{0.3cm} \\ 
\hline
\end{tabular}
\caption{Cross sections times branching ratio for the signal in the channel 
$H \fl WW \fl \ell \nu \ell \nu $, given at the NLO.}
\label{cssigwwlnulnu}
\end{center}
\end{table}

The potential backgrounds for this signatures are all the processes which
can produce two isolated leptons and possibly missing $p_t$. 
They are given in
Table \ref{csbakwwlnulnu}. 
We demand, for the signature when no jet tagging is applied, 
no jets in event to reduce the 
top-antitop background. 

To get rid of most of the single $Z$
production, the invariant mass of the leptons is asked not to be
the $Z$ mass.

Finally note that, as a jet veto
in the selection cuts without jet tagging is made, two different processes 
for the single vector boson production have to be generated:
Either $qq, gg \fl Z$ or $qq, gg \fl Z + jets$, corresponding to two
different PYTHIA processes and depending on the signature we want
to study (with or without tagging).
The $qq, gg \fl Z + jets$ background was generated with 
generation cuts as explained on page \pageref{cutbak}.

\begin{table}[htb]
\begin{center}
\begin{tabular}{|l|rc|}
\hline
\multicolumn{3}{|c|}{\emph{Backgrounds}} \\
\hline
Channel & $\sigma \times BR$ ($fb$) &\\
\hline
$qq \fl WW \fl \ell \nu \ell \nu$ & 8'050 & \\ 
\hline
$qq, gg \fl Z \fl \ell \ell$ 
 & 230'000 & \\
\hline
$qq, gg \fl Z + jets \fl \ell \ell +jets$ & 52'000 & 
($\widehat{p_t}>100\,\mathrm{GeV}$, \\ 
& & $\:\widehat{m}>300\,\mathrm{GeV}$) \\
\hline
$qq \fl t \bar{t} \fl WbWb \fl \ell \nu b \; \ell \nu b$ & 65'400 & \\
\hline
\end{tabular}
\caption{Cross sections times branching ratio 
for the  potential backgrounds in the channel 
$H \fl WW \fl \ell \nu \ell \nu $.}
\label{csbakwwlnulnu}
\end{center}
\end{table}

\begin{figure}[htb]
\begin{center}
\includegraphics*[scale=.4]{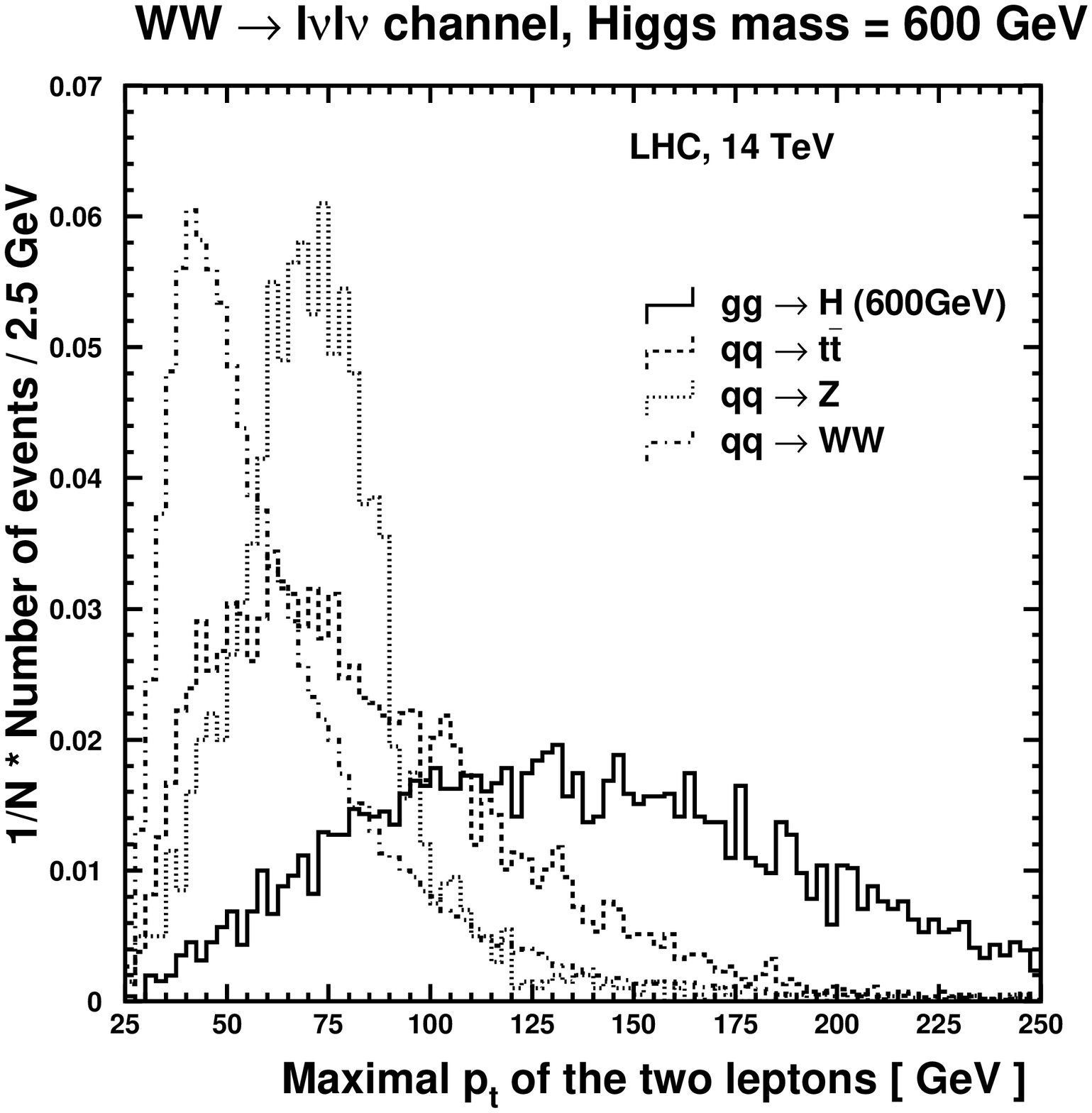}
\caption{$p_t$ spectrum of the lepton which has 
the highest $p_t$ from the two, for a 600~GeV Higgs produced through weak boson
fusion and for the backgrounds just before the cut on that $p_t$ is applied
in the $WW \fl \ell \nu \ell \nu$ channel.
The curves are normalized to 1.}
\label{ptlmax}
\end{center}
\end{figure}

The Higgs events are isolated with the following cuts:
\begin{itemize}
    \item We ask for two isolated leptons in event ($e$, $\mu$ 
with $p_t>10\,\mathrm{GeV}$ and $|\eta|<2.5$), whose invariant mass 
should not to be
within an interval of 20~GeV centered in 91~GeV, 
in order to reduce the single $Z$ production background.
    \item This channel is also characterized by a missing $p_t$, 
that is asked to be higher than 50~GeV. 
    \item A cut on the $p_t$ of the lepton which has the highest $p_t$ 
    from the two is also done, asking  
$p_{t}^{\ell}(max)> 100(150) \,\mathrm{GeV}$ for Higgs masses of
300 and 600~GeV respectively.
    \item We finally cut on the angle that make the leptons
in the transversal plane, $\varphi_{\ell\ell}$, asking them not being 
emitted back to back: $\varphi_{\ell\ell}< 140^{\mathrm{o}} (175^{\mathrm{o}})$, 
for Higgs masses of 300 and 600~GeV respectively.
\end{itemize}
The cuts used for a 450~GeV Higgs mass are the same than the ones used for
a 600~GeV mass.
Figure \ref{ptlmax} shows the $p_t$ spectrum of the lepton which has 
the highest $p_t$ from the two for a 600~GeV Higgs produced through weak boson
fusion and for the backgrounds just before the cut on that $p_t$ is applied.

The cuts with the jet tagging are a bit different. This comes from the fact
that the main
signal process will be then weak boson fusion which have other kinematic
properties than the main signal process when no tagging is done, the
gluon fusion.
When the jet tagging is done, we do not use the cut on $\varphi_{\ell\ell}$ 
but rather make a cut on the angle between the two leptons, $\theta_{\ell \ell}$.

A cut against the top-antitop background is also made, as explained on page
\pageref{cuttop}.
In that case, we combine one jet with one lepton and keep the
combination, whose mass, $m_{\ell jet}$,
is closest to the mass of the top to make the cut.

The cuts used for the $H \fl WW \fl \ell \nu \ell \nu$ channel
when the jet tagging is done are the following:
\begin{itemize}
    \item Exactly 2 jets must be in the event with $p_t>20\,\mathrm{GeV}$
and $|\eta|<4.5$. Their invariant mass has to be higher than 800~GeV.
    \item A explained above, we ask $m_{\ell jet}$ to be higher than
200~GeV, to reduce the $t\bar{t}$ background.
    \item The missing $p_{t}$ in event should be higher than 50 GeV.
    \item The $p_{t}$ of the lepton with the highest $p_{t}$
    from the two must be higher than $50 \,\mathrm{GeV}$.
    \item We make a cut on the angle between the two leptons
asking its cosine to be negative, ie. that the leptons are flying 
more back to back: $cos\theta_{\ell \ell} < 0$.
\end{itemize}

Some other cuts were tried, like a cut on the rapidity difference 
between the leptons or on the invariant mass of the two leptons.
But with these other cuts, we did not manage to improve the 
significance of the signal
in any significant way.

The expected results for $\mathcal{L}=100\,fb^{-1}$ and for different Higgs masses
are given in Table \ref{reswwlnulnu}. We see that the jet tagging technique
works very well here and manage to rise the signal to background ratio from
about 0.1 up to a factor 1.

\begin{table}
\begin{center}
\scalebox{0.9}{
\begin{tabular}{|l||r|r|r||r|}
\hline
\multicolumn{5}{|c|}{$H \fl WW \fl \ell \nu \ell \nu $}\\
\hline
\hspace*{1.5cm} Channel & \multicolumn{4}{c|}{Number of events} \\
& Generated & Min. rec. cuts   
& Full cuts & Jet tag. \\
&&& (No tag.) &\\
\hline
\multicolumn{5}{|c|}{} \\
\multicolumn{5}{|c|}{$m_{Higgs}=300\,\mathrm{GeV}$} \\ 
\multicolumn{5}{|c|}{} \\
\hline
$qq \fl qqH \fl \ell \nu \ell \nu$ & 8'800 & 25 & 7 & 80 \\
\hline
$gg \fl H \fl \ell \nu \ell \nu$ & 53'300 & 980 & 460 & 10 \\
\hline
Sum of all backgrounds & 35'545'000 & 9'860 & 2'800 & 90 \\
\hline
\hline
\multicolumn{5}{|l|}{Detailed backgrounds} \\
\hline
$qq \fl WW \fl \ell \nu \ell \nu $ & 805'000 & 5'120 & 1'400 & 0 \\
\hline
$qq, gg \fl Z \fl \ell\ell$ & 23'000'000 & 1'400 & 100 & - \\
\hline
$qq, gg \fl Z + jets \fl \ell\ell +jets$ & 5'200'000 & - & - & 10 \\
\hline
$qq \fl t \bar{t} \fl WbWb \fl \ell \nu b \; \ell \nu b $ & 6'540'000 
& 3'340 & 1300 & 80 \\
\hline
\hline
\multicolumn{5}{|c|}{} \\
\multicolumn{5}{|c|}{$m_{Higgs}=450\,\mathrm{GeV}$} \\ 
\multicolumn{5}{|c|}{} \\
\hline
$qq \fl qqH \fl \ell \nu \ell \nu$ & 4'000 & 30 & 5 & 140 \\
\hline
$gg \fl H \fl \ell \nu \ell \nu$ & 33'200 & 1'570 & 530 & 10 \\
\hline
Sum of all backgrounds & 35'545'000 & 9'860 & 2'550 & 90 \\
\hline
\hline
\multicolumn{5}{|l|}{Detailed backgrounds} \\
\hline
$qq \fl WW \fl \ell \nu \ell \nu $ & 805'000 & 5'120 & 1500 & 0 \\
\hline
$qq, gg \fl Z \fl \ell\ell$ & 23'000'000 & 1'400 & 300 & - \\
\hline
$qq, gg \fl Z + jets \fl \ell\ell+jets$ & 5'200'000 & - & - & 10 \\
\hline
$qq \fl t \bar{t} \fl WbWb \fl \ell \nu b \; \ell \nu b $ & 6'540'000 
& 3'340 & 750 & 80 \\
\hline
\hline
\multicolumn{5}{|c|}{} \\
\multicolumn{5}{|c|}{$m_{Higgs}=600\,\mathrm{GeV}$} \\ 
\multicolumn{5}{|c|}{} \\
\hline
$qq \fl qqH \fl \ell \nu \ell \nu$ & 2'200 & 20 & 10 & 110 \\
\hline
$gg \fl H \fl \ell \nu \ell \nu$ & 13'000 & 550 & 310 & 20 \\
\hline
Sum of all backgrounds & 35'545'000 & 9'860 & 2'550 & 90 \\
\hline
\hline
\multicolumn{5}{|l|}{Detailed backgrounds} \\
\hline
$qq \fl WW \fl \ell \nu \ell \nu $ & 805'000 & 5'120 & 1500 & 0 \\
\hline
$qq, gg \fl Z \fl \ell\ell$ & 23'000'000 & 1'400 & 300 & - \\
\hline
$qq, gg \fl Z + jets \fl \ell\ell+jets$ & 5'200'000 & - & - & 10 \\
\hline
$qq \fl t \bar{t} \fl WbWb \fl \ell \nu b \; \ell \nu b $ & 6'540'000 
& 3'340 & 750 & 80 \\
\hline
\end{tabular}}
\caption{Expected results for $\mathcal{L}=100\,fb^{-1}$ 
and for different Higgs masses
in the $H \fl WW \fl \ell \nu \ell \nu $
channel, NLO cross sections. 
After the total number of events generated for each process,
we give the number of events left after the minimal reconstruction
cuts, which means $p_{t}^{miss}>50\,\mathrm{GeV}$,
$p_{t}^{\ell}(max)> 100\,\mathrm{GeV}$
and $\varphi_{\ell \ell}<175^{\mathrm{o}}$. The
number of events left without and with the jet tagging
are given in the two last columns.}
\label{reswwlnulnu}
\end{center}
\end{table}

Now let's calculate the
luminosity for a 5 standard deviation with and without the jet
tagging.
The results are given in Table \ref{ldisclnulnu}.

\begin{table}[htb]
\begin{center}
\begin{tabular}{|c|r|r|r|r|}
\hline
\multicolumn{5}{|c|}{$H \fl WW \fl \ell \nu \ell \nu $}\\
\hline
& Signal & Background 
& S/B & $\mathcal{L}_{disc}$ ($fb^{-1}$)\\
\hline
\multicolumn{5}{|c|}{$m_{Higgs}=300\,\mathrm{GeV}$} \\ 
\hline
no tagging & 467 & 2'800 & 0.19 & 32 \\
tagging & 90 & 90 & 1 & 28 \\
\hline
\multicolumn{5}{|c|}{$m_{Higgs}=450\,\mathrm{GeV}$} \\ 
\hline
no tagging & 535 & 2'550 & 0.22 & 20 \\
tagging & 150 & 90 & 1.67 & 12 \\
\hline
\multicolumn{5}{|c|}{$m_{Higgs}=600\,\mathrm{GeV}$} \\ 
\hline
no tagging & 320 & 2'550 & 0.13 & 62 \\
tagging & 130 & 90 & 1.44 & 13 \\
\hline
\end{tabular}
\caption{Expected discovery luminosities for the channel 
$H \fl WW \fl \ell \nu \ell \nu$, without systematic errors taken into
account. The number of events for signal and background 
corresponds to a luminosity of $100\,fb^{-1}$.}
\label{ldisclnulnu}
\end{center}
\end{table}

Once again, we see that it would be interesting to combine the results
obtained with the jet tagging with the results of the other
channels for the Higgs search. However, notice that it is hard to 
find the Higgs with this channel as no mass peak can be reconstructed. 
In order to determine the Higgs mass, we could
fit for instance the transversal mass\footnote{The 
transversal mass of a particle is the mass is would have if its
$p_z$ is set to zero.} distribution. This lead to a 
supplementary uncertainty on 
the simulation used to make the fit.

\begin{figure}[htb]
\begin{center}
\mbox{
\subfigure{\includegraphics*[width=.5\textwidth]{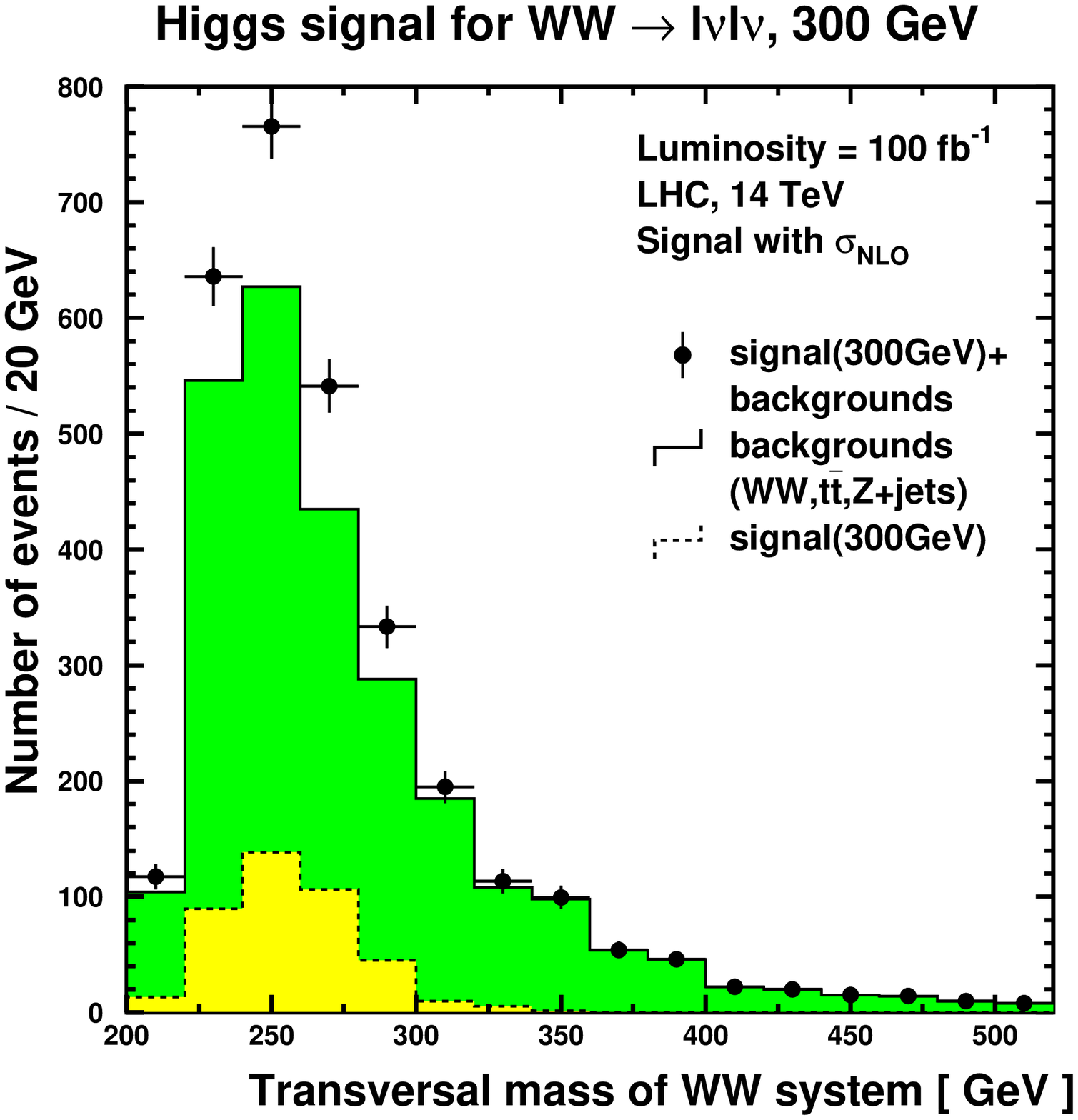} \quad }
\subfigure{
\includegraphics*[width=.5\textwidth]{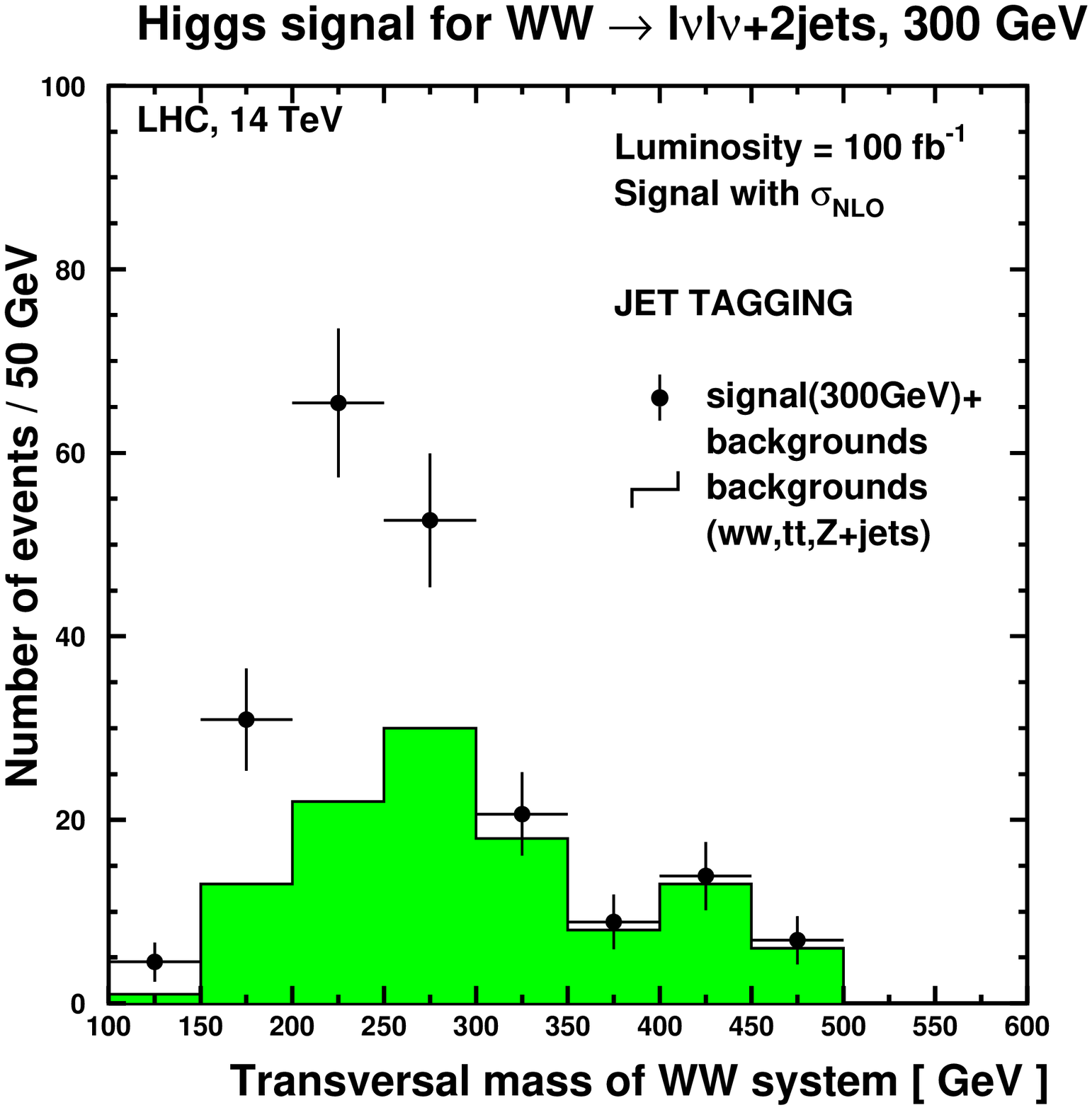}
}
}
\caption{Transverse mass of the $WW$ system for a 300~GeV
Higgs, before tagging (Left) and after tagging (Right).}
\label{masspeak1}
\end{center}
\end{figure}

\begin{figure}[htb]
\begin{center}
\mbox{
\subfigure{\includegraphics*[width=.5\textwidth]{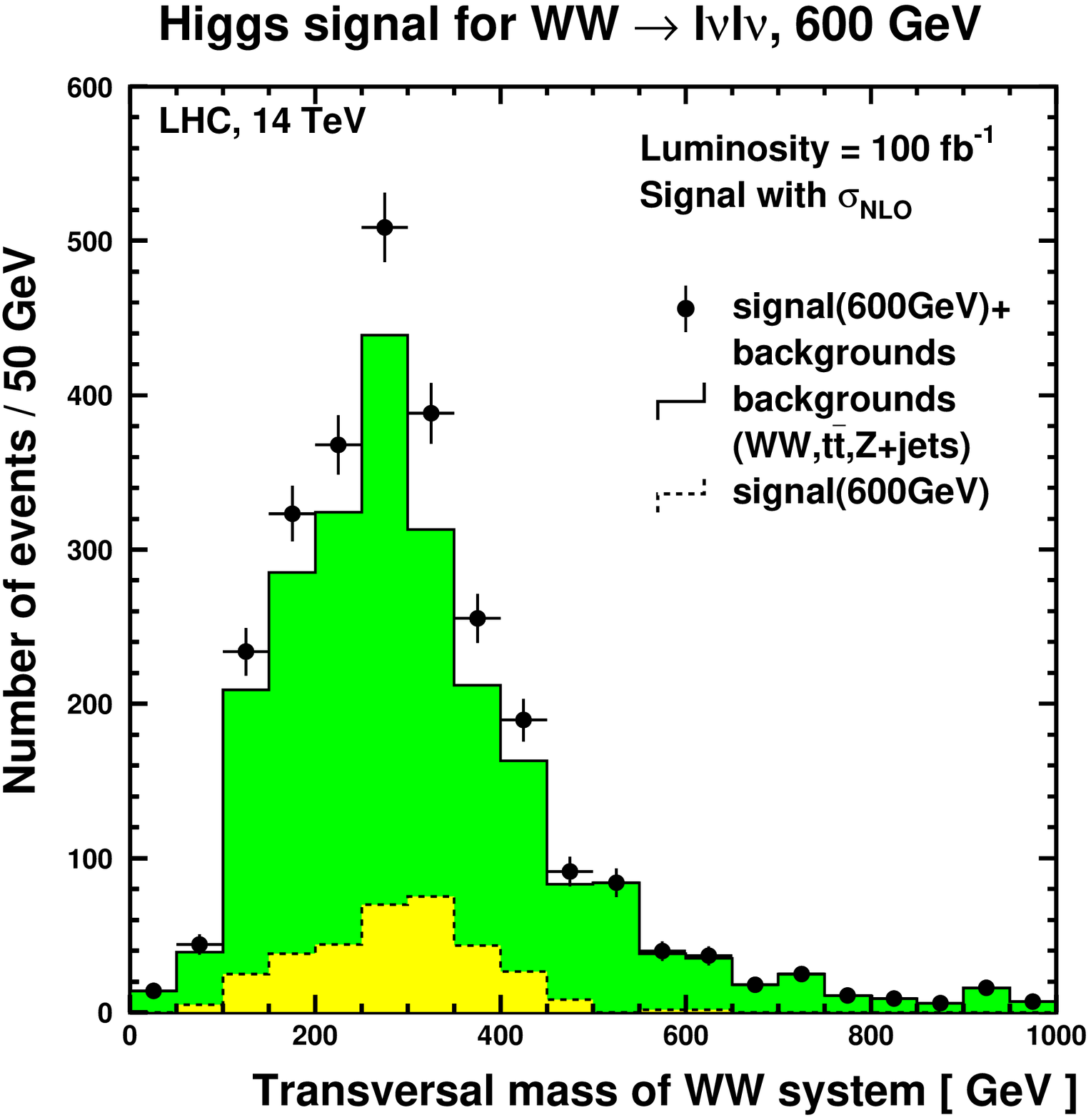} \quad }
\subfigure{
\includegraphics*[width=.5\textwidth]{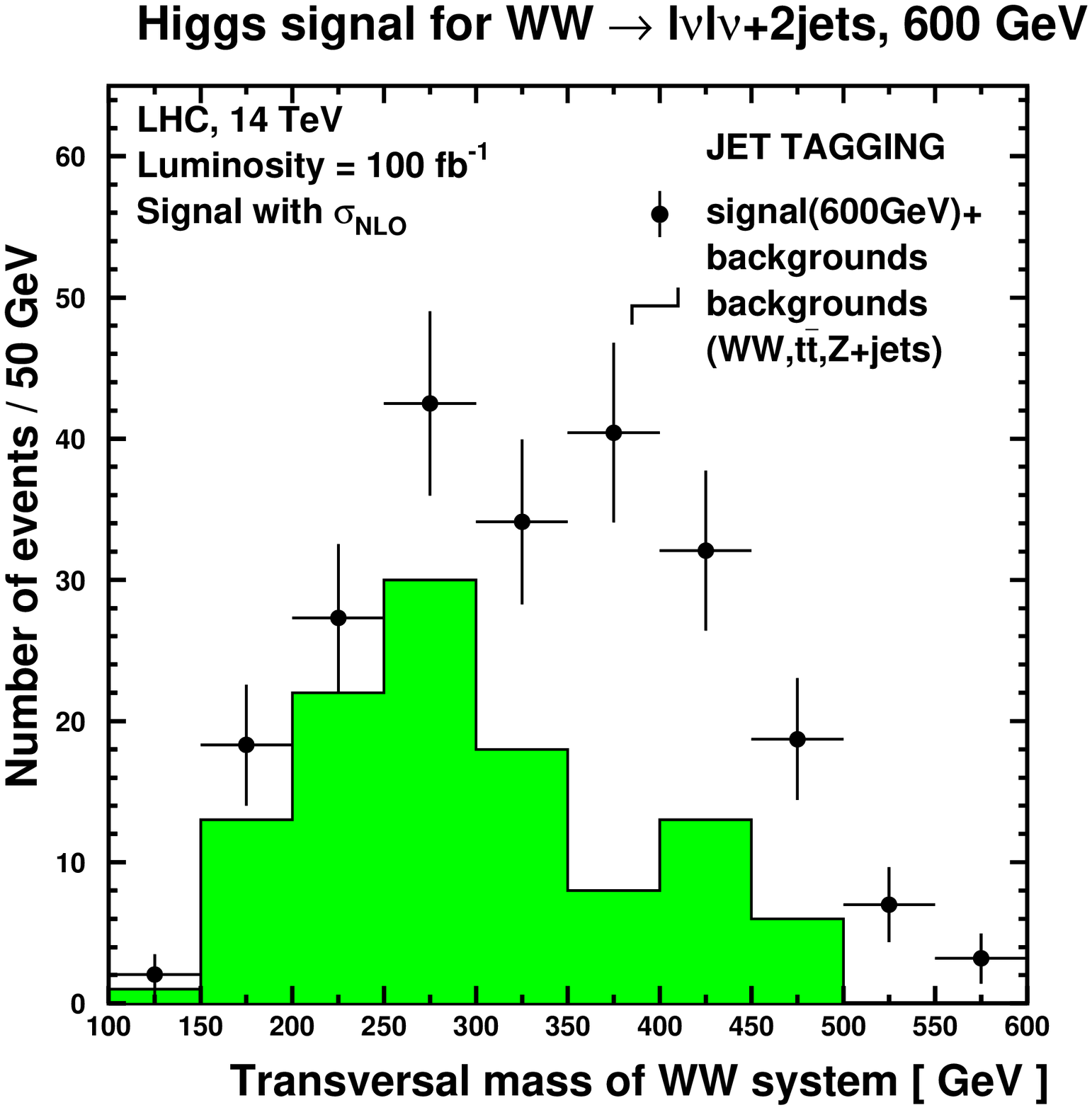}
}
}
\caption{Transverse mass of the $WW$ system for a 600~GeV
Higgs. (Left) before tagging (Right) after tagging.}
\label{masspeak2}
\end{center}
\end{figure}

Figures \ref{masspeak1} and \ref{masspeak2} show how the transversal 
mass of
the 'leptons plus missing energy' system 
for a 300 and a 600~GeV Higgs would look like, after the cuts,
with and without forward jet tagging. 
A signal can clearly be seen when the tagging is made. The observation
of a signal when no tagging is done implies a good
knowledge of the background, as the signal peak stands right where the
background has a maximum.
\\[0.5cm]
In the $H \fl WW \fl \ell \nu \ell \nu$ channel, 
when the jet tagging is added in the selection cuts, 
the signal is reduced by a factor 3 and the background by
a factor 30, thus allowing
a Higgs signal to be seen.
The jet tagging improves the signal to background ratio 
of this signature from about a factor 10.
As we will see later, this assumption will
be true also for other channels studied.

	\section{$H \fl WW \fl \ell \nu j j $}
This channel is characterized by a large
branching fraction (135 times larger than for
$H \fl ZZ \fl 4\ell^{\pm}$).
It is unfortunately also characterized by sources of backgrounds
with huge cross sections
due to the presence of jets in the signature.
However thanks to the large number of 
signal events, 
harder cuts to isolate the signal can be made.

The cross sections times branching ratio for this channel
are given in the Table \ref{cssigwwlnujj}.

\begin{table}[htb]
\begin{center}
\begin{tabular}{|c|c|c|}
\hline
\multicolumn{3}{|c|}{\emph{Signal}} \\
\hline
Channel: &$qq \fl qqH \fl WW \fl \ell \nu jj$ &
$gg \fl H \fl WW \fl \ell \nu jj$ \\
\hline
Mass of Higgs &$\sigma \times BR$, NLO
& $\sigma \times BR$, NLO \\
(GeV) & ($fb$) & ($fb$) \\
\hline
 300 &  370 \hspace*{0.3cm} &
  2126 \hspace*{0.3cm} \\
\hline
 450 & 158  \hspace*{0.3cm} &
  1350  \hspace*{0.3cm} \\ 
\hline
 600 & 90 \hspace*{0.3cm} & 
  518 \hspace*{0.3cm} \\ 
\hline
\end{tabular}
\caption{Cross sections times branching ratio for the signal in the channel 
$H \fl WW \fl \ell \nu j j $, given at the NLO.}
\label{cssigwwlnujj}
\end{center}
\end{table}

This signature is characterized by one isolated lepton and at least 
two jets.
The potential backgrounds are then all the processes which
can produce one isolated lepton, jets and missing $p_t$. They are listed in
Table \ref{csbakwwlnujj}. The $qq, gg \fl W + jets \fl \ell \nu +jets$
background can be generated with generation 
cuts, like explained on page \pageref{cutbak}. This is a main difference
between the study we did and the study 
previously made by the Atlas group \cite{atlas}.

\begin{table}[htb]
\begin{center}
\begin{tabular}{|l|rc|}
\hline
\multicolumn{3}{|c|}{\emph{Backgrounds}} \\
\hline
Channel & $\sigma \times BR$ ($fb$) &\\
\hline
$qq \fl WW \fl \ell \nu j j$ & 32'340 & \\
\hline
$qq, gg \fl W + jets \fl \ell \nu +jets$ 
& 358'000 & ($\widehat{p_t}>100\, \mathrm{GeV}$, \\
&  & $\:\widehat{m}>300\,\mathrm{GeV}$) \\
\hline
$qq \fl t \bar{t} \fl WbWb \fl \ell\nu b\; \ell \nu b$ & 623'000 & \\
\hline
\end{tabular}
\caption{Cross sections times branching ratio
for the  potential backgrounds in the channel 
$H \fl WW \fl \ell \nu jj $.}
\label{csbakwwlnujj}
\end{center}
\end{table}

Here is a summary of the cuts to get a signal without jet tagging:
\begin{itemize}
    \item We ask for one isolated lepton ($e$ or $\mu$ 
with $p_t>10\,\mathrm{GeV}$ and $|\eta|<2.5$) and at least two jets (with 
$p_t>20\,\mathrm{GeV}$ and $|\eta|<4.5$).
    \item We reconstruct the first $W$ by asking the two jets to have a mass
within a interval of 40~GeV centered in the shifted $W$ mass (here 73~GeV).
    \item Cuts are done on the $p_t$ of the lepton and 
on the missing $p_t$:
$p_{t}(\ell)>25\,\mathrm{GeV}$ and $p_{t}^{miss}>25\,\mathrm{GeV}$. 
The $W$ decaying into a lepton and a neutrino is reconstructed as
explained on page \pageref{wlnu}.
We make a loose cut on the $W$'s transversal mass,
asking it to be between 20 and 100~GeV.
    \item Only events where the invariant mass of the $WW$ system is 
within an interval 
of 70 (100, 200)~GeV centered in 300 (450, 600)~GeV were kept.
    \item Finally, cuts on the $p_t$'s of the reconstructed 
$W$'s are done:\\
$p_{t}(W_{jj},W_{\ell \nu})>100\, (150,200)\,\mathrm{GeV}$, 
for a 300 (450, 600)~GeV Higgs.
\end{itemize}

The cuts applied to isolate weak boson fusion, ie. when the jet tagging technique 
is used, are the following:
\begin{itemize}
    \item The events must survive the same cuts as before.
    \item Exactly four jets with $p_t>20\,\mathrm{GeV}$
and $|\eta|<4.5$
are asked in event (two jets to reconstruct the $Z$
and two jets to be the tagging jets). The invariant mass 
of the tagging jets has to be higher than 800~GeV.
    \item As explained on page \pageref{cuttop}, a cut against 
the $t\bar{t}$ background
is added, namely the mass of the $W$ plus jet system, $m_{Wjet}$, has to be 
higher than 300~GeV.
\end{itemize}

The expected results for $\mathcal{L}=100\,fb^{-1}$ and for different Higgs masses
are given in Table \ref{reswwlnujj}. 
We do not get very interesting results when no jet tagging technique is used.
This is mainly caused by the presence of the huge $t\bar{t}$ background.

Like in the 
$H \fl WW \fl \ell \nu \ell \nu$ channel, we see that the jet
tagging technique improves a lot the signal to background ratio, making it
rise from 1/20 to about 1. One main source of background, the 
top antitop production, is strongly reduced with the cut on the
$m_{Wjet}$. 
We find that when it is applied just after asking the
cut on the invariant mass of the two tagging jets and before the cuts on the 
$p_t$'s, it makes us loose a factor 1.5 (for 600~GeV) 
and 2 (for 300~GeV) in the signal and
5 in the top-antitop background. 
To visualize the effect of that cut, see Figure \ref{top} on
page \pageref{top}. 

The other important background is the single $W$
production, which is also strongly reduced thanks to the 
cuts on the $m_{Wjet}$ and on the mass of the tagging jets requirement.
The reason is that the jets
for this background are emitted rather centrally and cannot lead to so high
masses like the forward going jets coming from the weak boson fusion process.

\begin{table}[p]
\begin{center}
\scalebox{0.9}{
\begin{tabular}{|l||r|r|r||r|}
\hline
\multicolumn{5}{|c|}{$H \fl WW \fl \ell \nu jj $}\\
\hline
\hspace*{1.5cm} Channel & \multicolumn{4}{c|}{Number of events} \\
& Generated & Min. rec. cuts   
& $p_t$ cuts  & Jet tag. \\
\hline
\multicolumn{5}{|c|}{} \\
\multicolumn{5}{|c|}{$m_{Higgs}=300\,\mathrm{GeV}$} \\ 
\multicolumn{5}{|c|}{} \\
\hline
$qq \fl qqH \fl \ell \nu jj$ & 37'000 & 5'000 & 1'900 & 290 \\
\hline
$gg \fl H \fl \ell \nu jj$ & 212'600 & 28'000 & 14'700 & 40 \\
\hline 
Sum of all backgrounds & 65'234'000 & 1'750'000 & 530'000 & 280 \\
\hline
\hline
\multicolumn{5}{|l|}{Detailed backgrounds} \\
\hline
$qq \fl WW  \fl \ell \nu jj$ & 3'234'000 & 60'000 & 10'000 & 0 \\
\hline
$qq, gg \fl W + jets\fl \ell \nu + jets$ & 35'800'000 & 290'000 & 160'000 & 130 \\
\hline
$qq \fl t \bar{t} \fl WbWb \fl \ell \nu b \; \ell\nu b$ & 26'200'000 & 
1'400'000 & 360'000 & 150 \\
\hline
\hline
\multicolumn{5}{|c|}{} \\
\multicolumn{5}{|c|}{$m_{Higgs}=450\,\mathrm{GeV}$} \\ 
\multicolumn{5}{|c|}{} \\
\hline
$qq \fl qqH \fl \ell \nu jj$ & 15'800 & 2'400 & 1'650 & 330 \\
\hline
$gg \fl H \fl \ell \nu jj$ & 135'000 & 19'000 & 14'000 & 90 \\
\hline
Sum of all backgrounds & 65'234'000 & 760'000 & 304'000 & 200 \\
\hline
\hline
\multicolumn{5}{|l|}{Detailed backgrounds} \\
\hline
$qq \fl WW  \fl \ell \nu jj$ & 3'234'000 & 20'000 & 4'000 & 0 \\ 
\hline
$qq, gg \fl W + jets \fl \ell \nu + jets$ & 35'800'000 
& 270'000 & 150'000 & 120 \\ 
\hline
$qq \fl t \bar{t} \fl WbWb \fl \ell \nu b \; \ell\nu b$ & 26'200'000 & 
470'000 & 150'000 & 80 \\
\hline
\hline
\multicolumn{5}{|c|}{} \\
\multicolumn{5}{|c|}{$m_{Higgs}=600\,\mathrm{GeV}$} \\ 
\multicolumn{5}{|c|}{} \\
\hline
$qq \fl qqH \fl \ell \nu jj$ & 9'000 & 1'500 & 900 & 220 \\
\hline
$gg \fl H \fl \ell \nu jj$ & 51'800 & 6'400 & 4'700 & 40 \\
\hline
Sum of all backgrounds & 65'234'000 & 462'000 & 153'000 & 150 \\
\hline
\hline
\multicolumn{5}{|l|}{Detailed backgrounds} \\
\hline
$qq \fl WW  \fl \ell \nu jj$ & 3'234'000 & 12'000 & 3'000 & 0 \\
\hline
$qq, gg \fl W + jets \fl \ell \nu + jets$ & 35'800'000 & 190'000 & 85'000 & 100 \\
\hline
$qq \fl t \bar{t} \fl WbWb \fl \ell \nu b \; \ell\nu b$ & 26'200'000 
& 260'000 & 65'000 & 50 \\
\hline
\end{tabular}}
\caption{Expected results for $\mathcal{L}=100\,fb^{-1}$ and for different Higgs masses
in the $H \fl WW \fl \ell \nu jj $
channel, NLO cross sections. 
After the total number of events generated for each process,
we give the number of events left after the minimal reconstruction
cuts, which means all cuts except the ones
on the $p_{t}$'s 
of the reconstructed $W$'s. In the two last columns, the expected
number of events left without and with the jet tagging technique is given.}
\label{reswwlnujj}
\end{center}
\end{table}

As usual, let's calculate
necessary luminosity for a 5 standard deviation, with and without the jet
tagging technique. These results are given in Table \ref{ldisclnujj}.

\begin{table}[htb]
\begin{center}
\begin{tabular}{|c|r|r|r|r|}
\hline
\multicolumn{5}{|c|}{$H \fl WW \fl \ell \nu jj $}\\
\hline
& Signal
& Background 
& S/B &  $\mathcal{L}_{disc}$ ($fb^{-1}$)\\
\hline
\multicolumn{5}{|c|}{$m_{Higgs}=300\,\mathrm{GeV}$} \\ 
\hline
no tagging & 16'600 & 530'000 & 0.03 & 5 \hspace*{0.3cm}\\
tagging & 330 & 280 & 1.18 & 6 \hspace*{0.3cm} \\
\hline
\hline
\multicolumn{5}{|c|}{$m_{Higgs}=450\,\mathrm{GeV}$} \\ 
\hline
no tagging & 15'650 & 304'000 & 0.05 & 3 \hspace*{0.3cm} \\
tagging & 420 & 200 & 2.10 & 3 \hspace*{0.3cm}\\
\hline
\hline
\multicolumn{5}{|c|}{$m_{Higgs}=600\,\mathrm{GeV}$} \\ 
\hline
no tagging & 5'600 & 153'000 & 0.04 & 12 \hspace*{0.3cm}\\
tagging & 260 & 150 & 1.73 & 6 \hspace*{0.3cm}\\
\hline
\end{tabular}
\caption{Expected discovery luminosities for the channel 
$H \fl WW \fl \ell \nu jj$, without systematic errors taken into
account. The results without tagging give too small signal to background
ratios and cannot be taken into account. 
The number of events for signal and background 
corresponds to a luminosity of $100\,fb^{-1}$.}
\label{ldisclnujj}
\end{center}
\end{table}

\begin{figure}[htb]
\begin{center}
\mbox{
\subfigure{\includegraphics*[width=.45\textwidth]{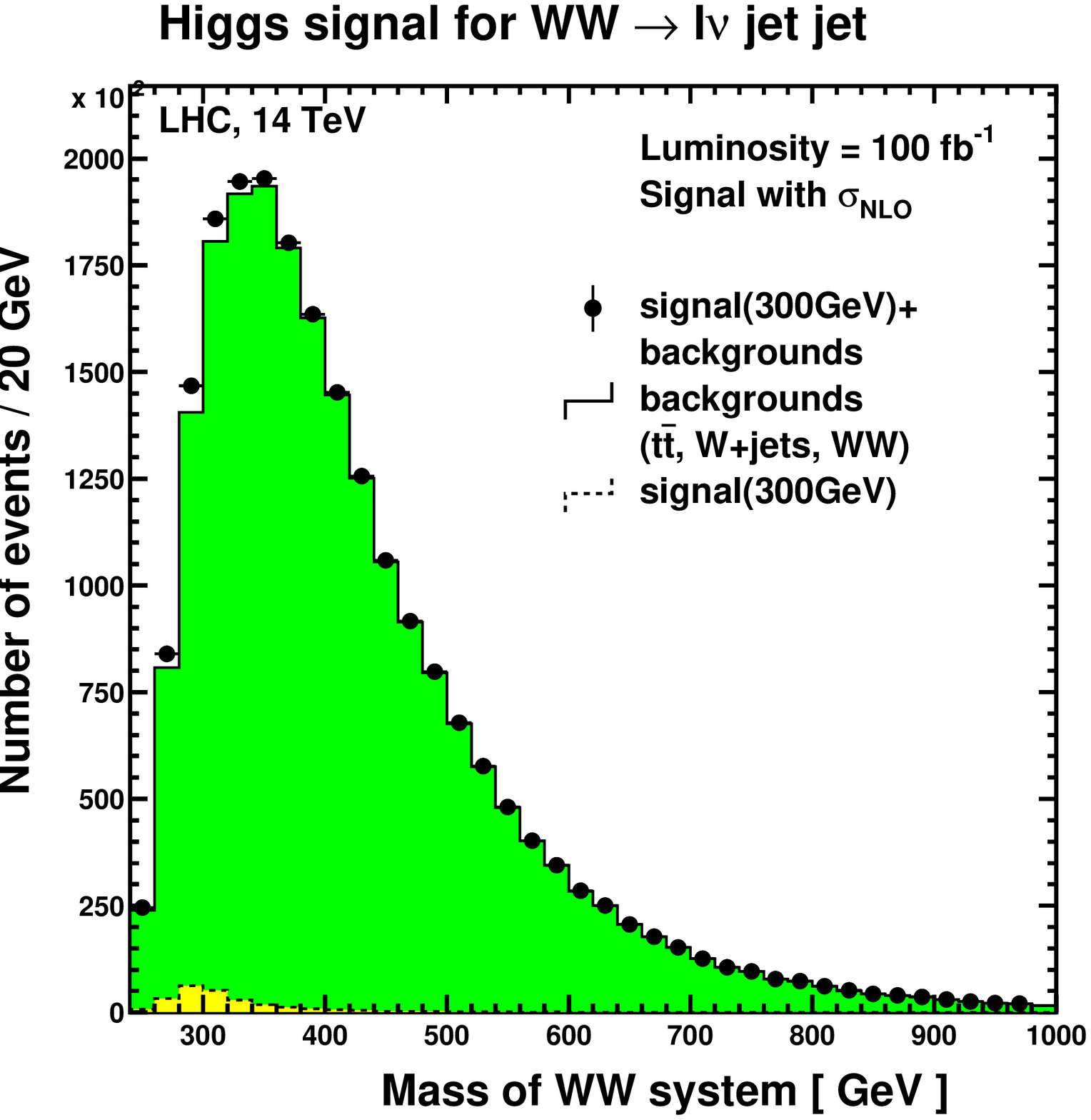}\quad
}
\subfigure{\includegraphics*[width=.45\textwidth]{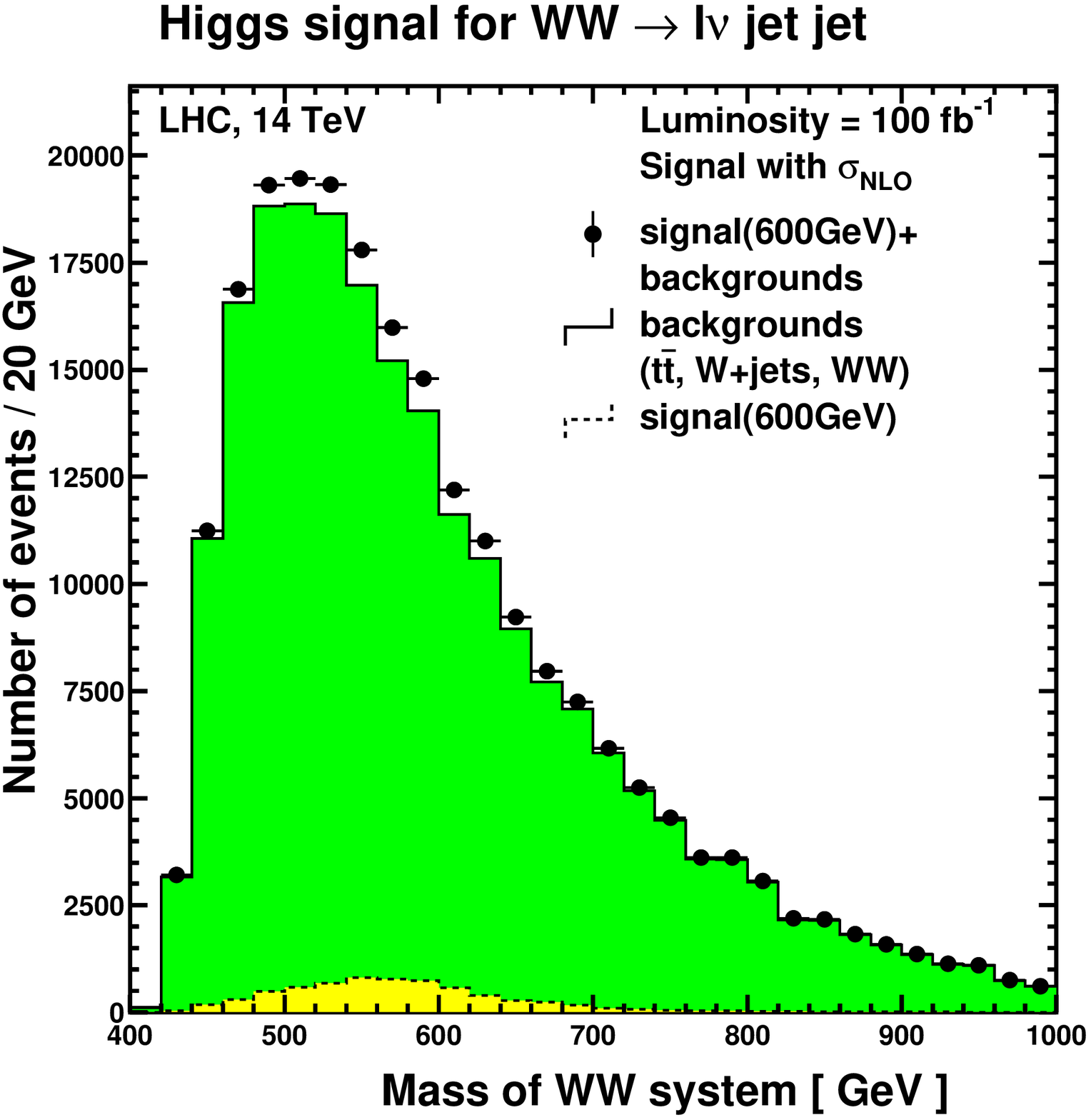} 
}
}
\caption{Mass spectrum of a 300 (Left) and a 600~GeV Higgs (Right) in the 
$H \fl WW \fl \ell \nu q \bar{q} $ channel, when no jet tagging is applied.}
\label{notagjets}
\end{center}
\end{figure}

It is interesting to discuss the results obtained without jet tagging. 
That this channel seems
to lead to small discovery luminosities,
at least as good as
the ones obtained in the four leptons channels.
Actually one thing need to
be pointed out here. Even if the signal significance without tagging
is very good, due to the large amount of events, the signal to
background ratio is really low. 
That means that if the background is not known with
a precision of 3\%, the error on the number of events
in the background will be bigger than the 
expected signal itself !
Since the signal to background ratio is small, systematic
errors have to be taken into account to get relevant results.
Moreover if we look at the mass plots for a 300 and a 600~GeV 
Higgs without tagging (see Figure \ref{notagjets}), 
no big excess is observable, which confirms this last remark~!

\begin{figure}[htb]
\begin{center}
\includegraphics*[width=.5\textwidth]{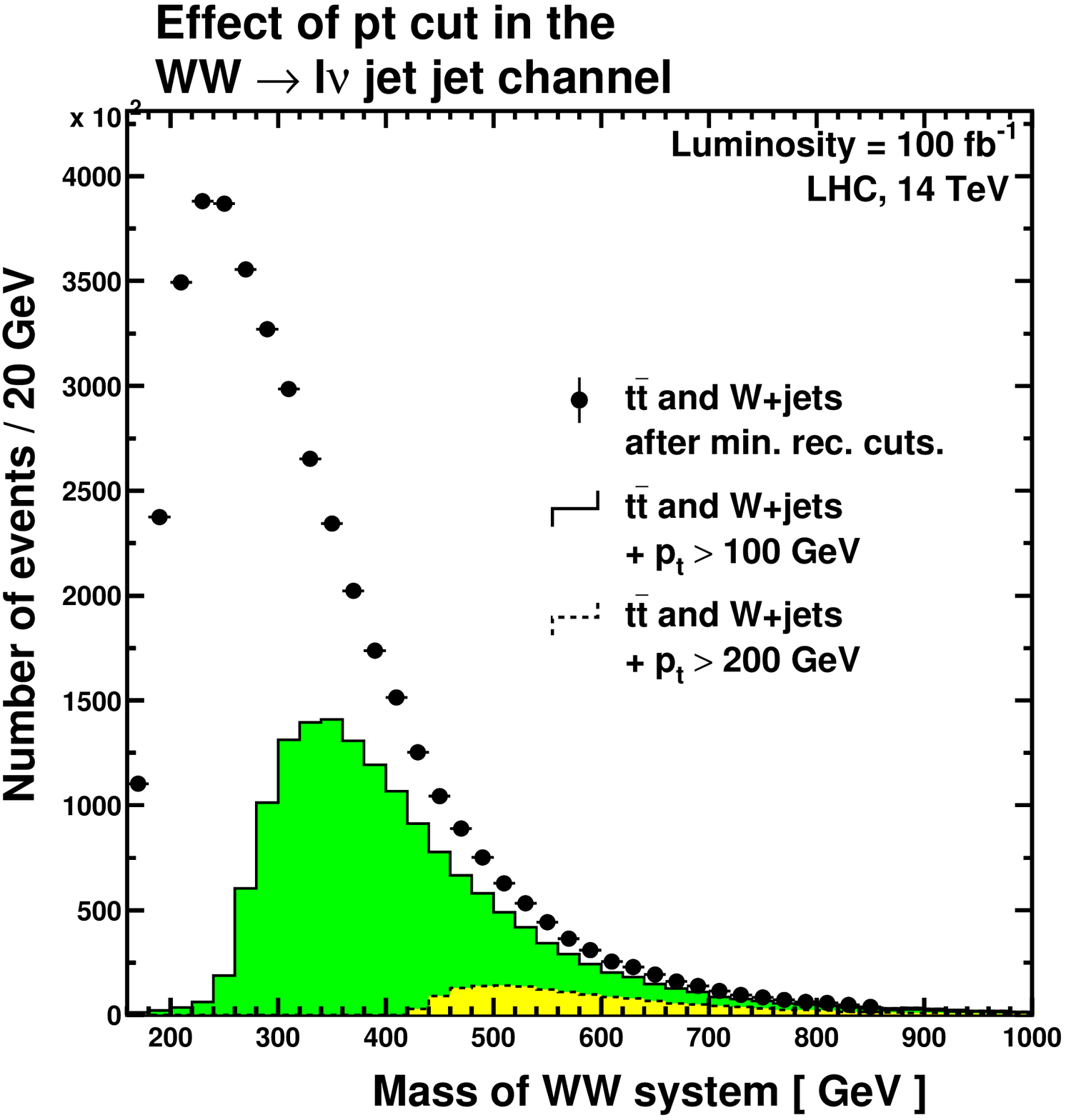}
\caption{Effect of the cuts on the reconstructed $WW$ mass spectrum.
It is shown for $W+jets$ and $t \bar{t}$ backgrounds, with a number
of events corresponding to a luminosity of $100 fb^{-1}$.
A first plot is done with the events that survived the minimal reconstruction cuts.
As $p_t$ cuts of 100 and 200~GeV were done to isolate a 300 and 
600~GeV Higgs respectively, we show how the
$WW$ mass spectrum for
background looks like after these two cuts.} 
\label{pteff}
\end{center}
\end{figure}

One thing which is interesting to notice on the mass plots
is that the background is not peaking around 200~GeV any more.
It is a shifted. Actually, as soon as the high $p_t$ cuts are made
for the selection,
the low mass region is much more strongly 
reduced than the high mass region.
Figure \ref{pteff} shows that the low-mass region in the $WW$ mass spectrum
for the backgrounds is much more reduced than the high mass region as soon as the
cuts on the $p_t$s are applied.
For a 300~GeV Higgs, the
background maximum is then shifted to $\approx 350\,\mathrm{GeV}$.
However this is not the case for the signal, which still peaks around 300~GeV.
The signal will then not stand where the background has a maximum
(see Figure \ref{notagjets}).
Unfortunately, this is not the case any more for a 600~GeV Higgs.

\begin{figure}[htb]
\begin{center}
\mbox{
\subfigure{\includegraphics*[width=.5\textwidth]{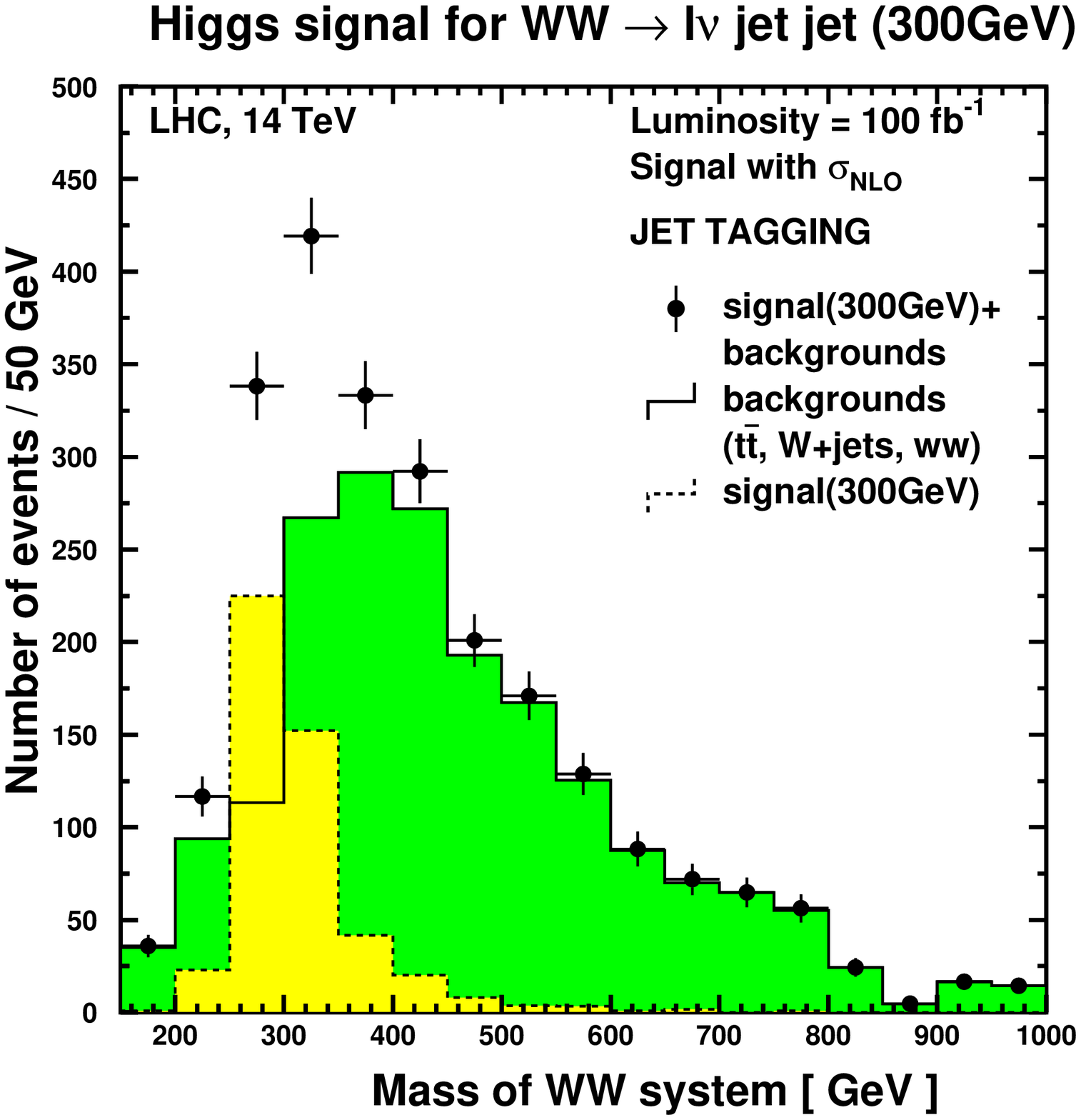}\quad
}
\subfigure{\includegraphics*[width=.5\textwidth]{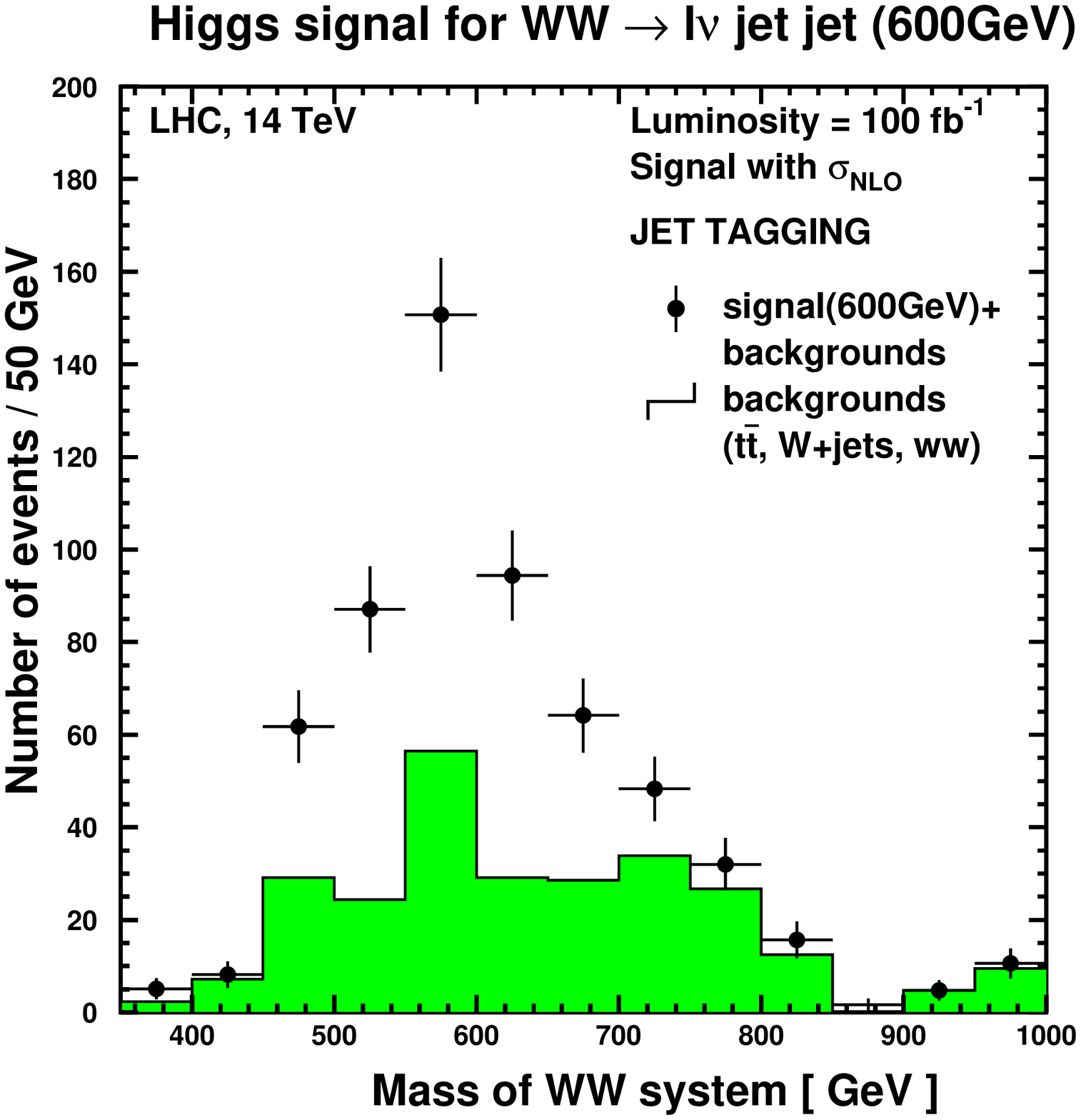} 
}
}
\caption{Mass spectrum of a 300 (Left) and a 600~GeV Higgs (Right) 
in the 
$H \fl WW \fl \ell \nu jj $ channel, 
with forward jet tagging.}
\label{tagjets}
\end{center}
\end{figure}

As soon as the jet tagging technique
is applied, the number of events
is reduced, but a signal to background ratio larger than one is obtained
with still a sizable number of events left.
Figure \ref{tagjets} shows the reconstructed Higgs mass spectrum
after the jet tagging.
That means that even if the background is not well known,
it should be possible to observe a signal.
In that case, fluctuations in the background of for instance
10\% allows nevertheless
a signal to be observed.
Moreover the mass plots show clearly a signal
(see Figure \ref{tagjets}). This channel might thus be very interesting
if the tagging technique is used in the selection cuts.

\begin{figure}[htb]
\begin{center}
\mbox{
\subfigure{\includegraphics*[width=.45\textwidth]{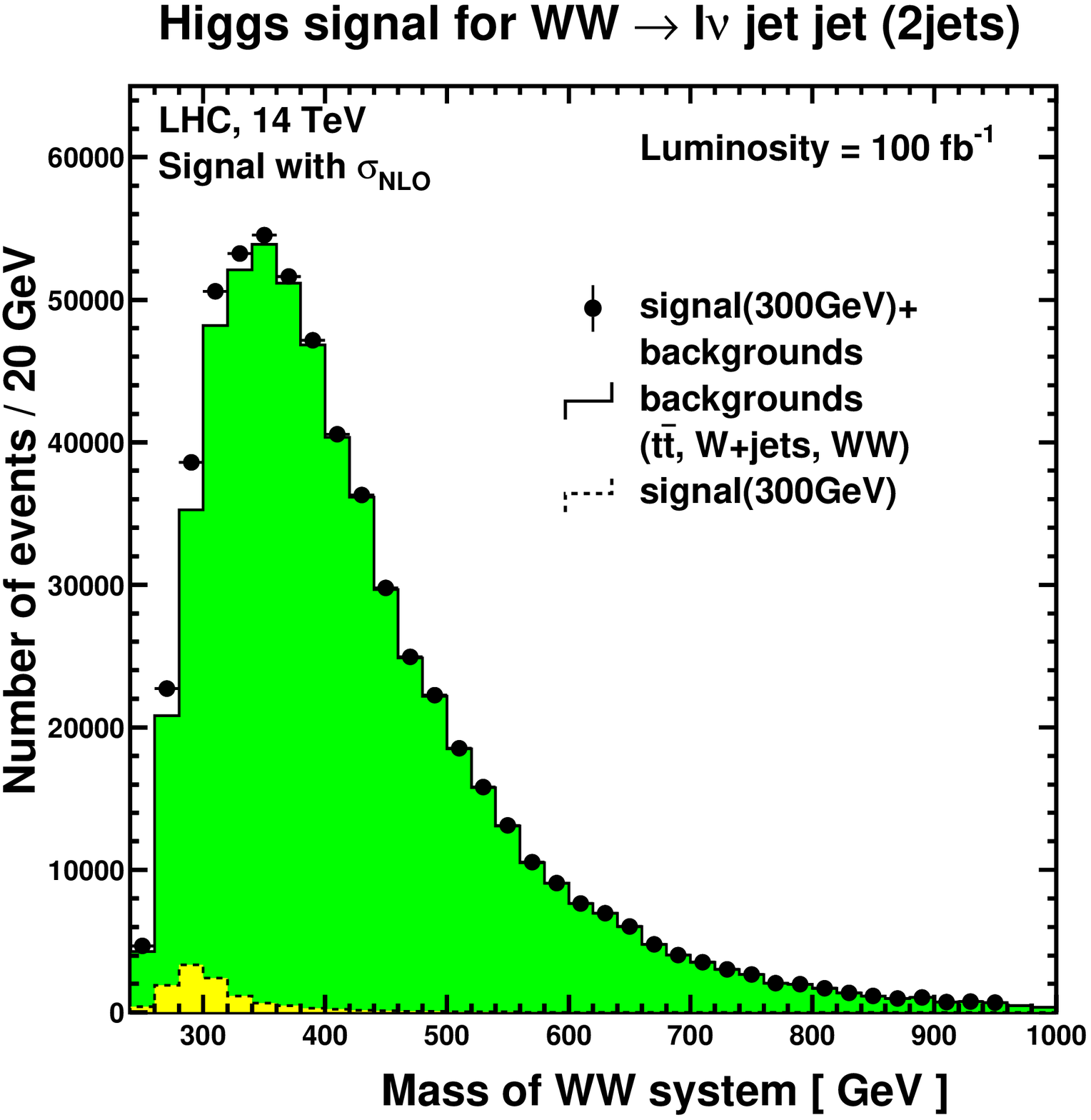}\quad
}
\subfigure{\includegraphics*[width=.45\textwidth]{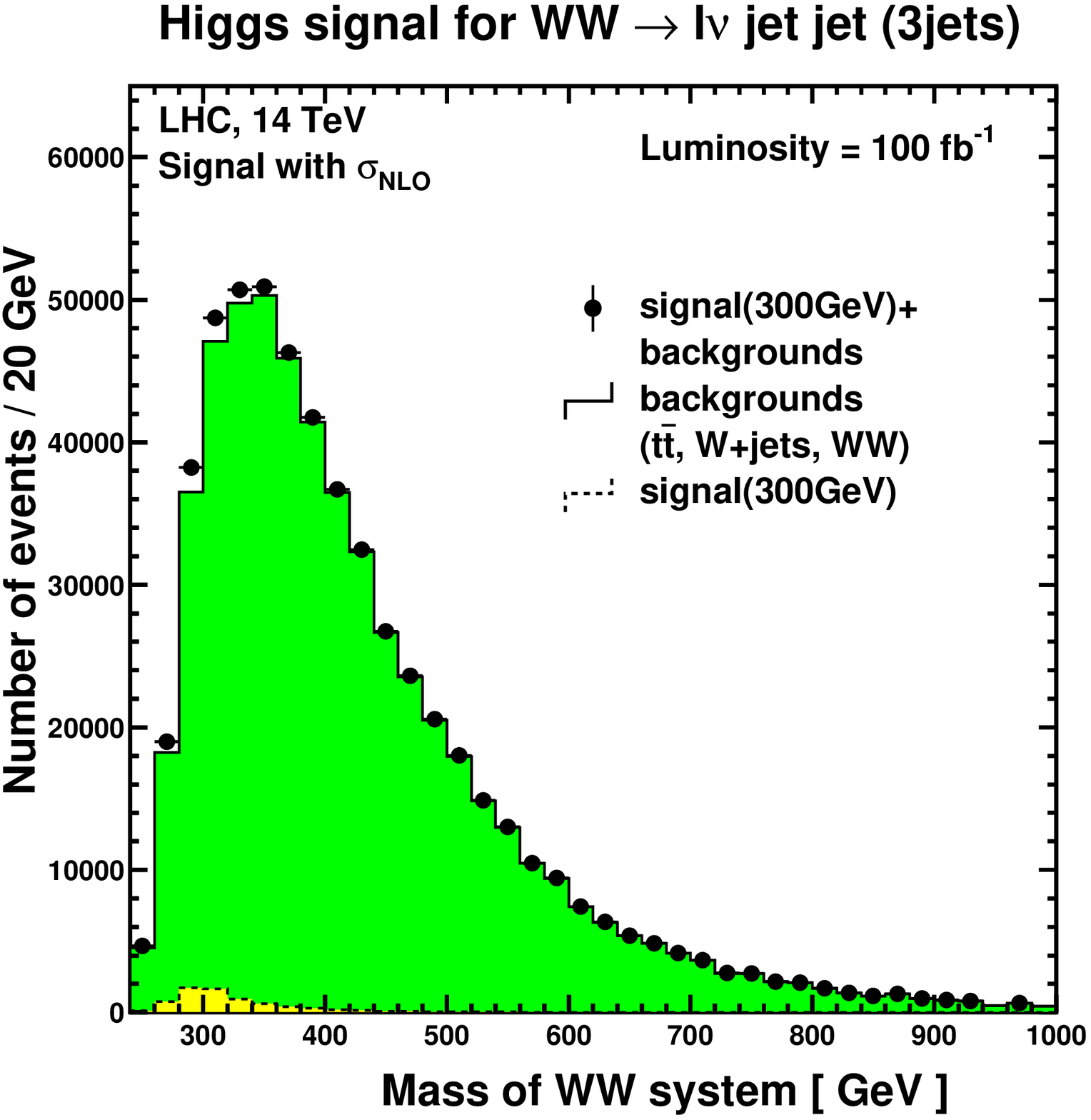} 
}
}
\\
\mbox{
\subfigure{\includegraphics*[width=.45\textwidth]{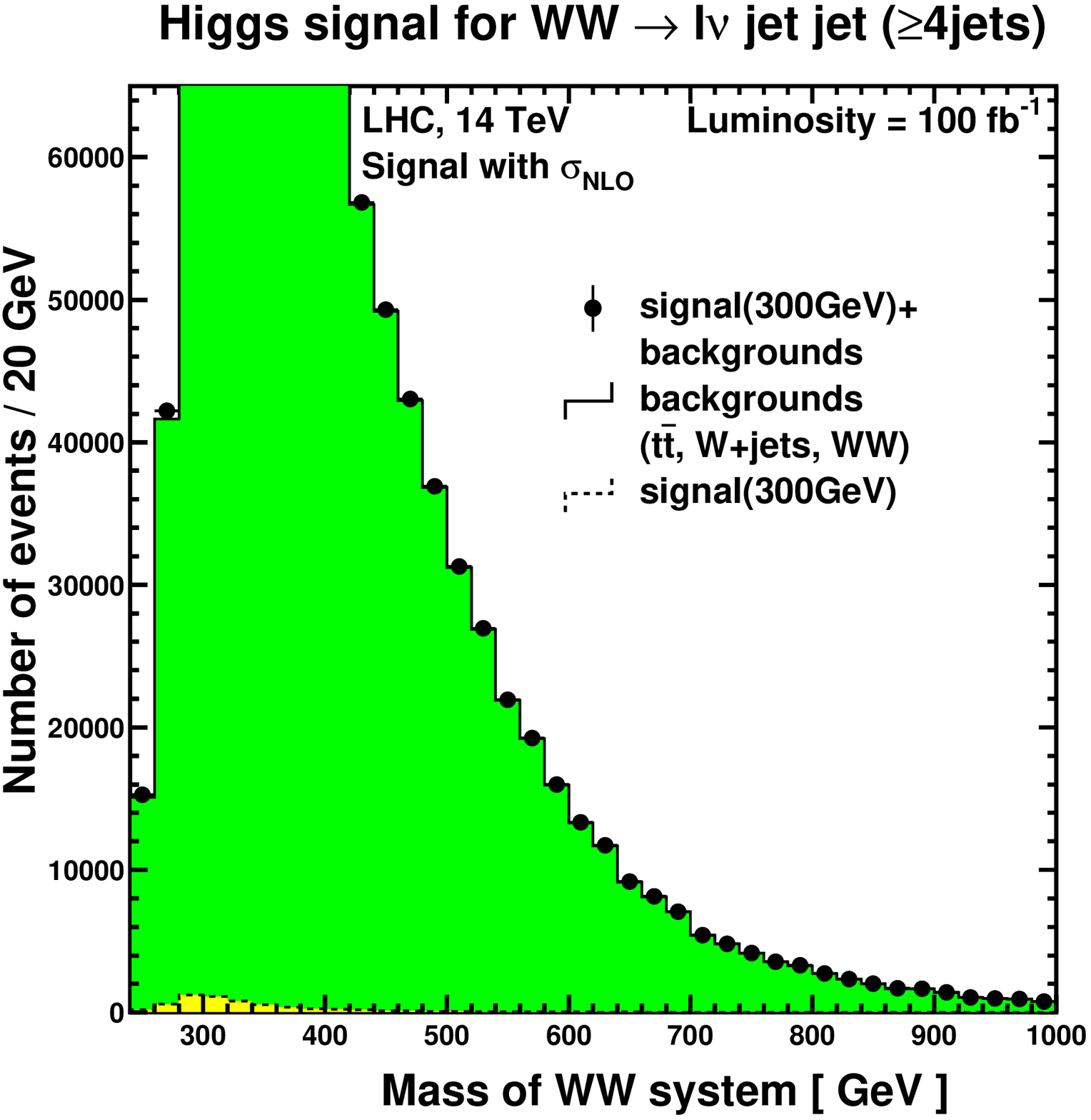}\quad}
\subfigure{\includegraphics*[width=.45\textwidth]{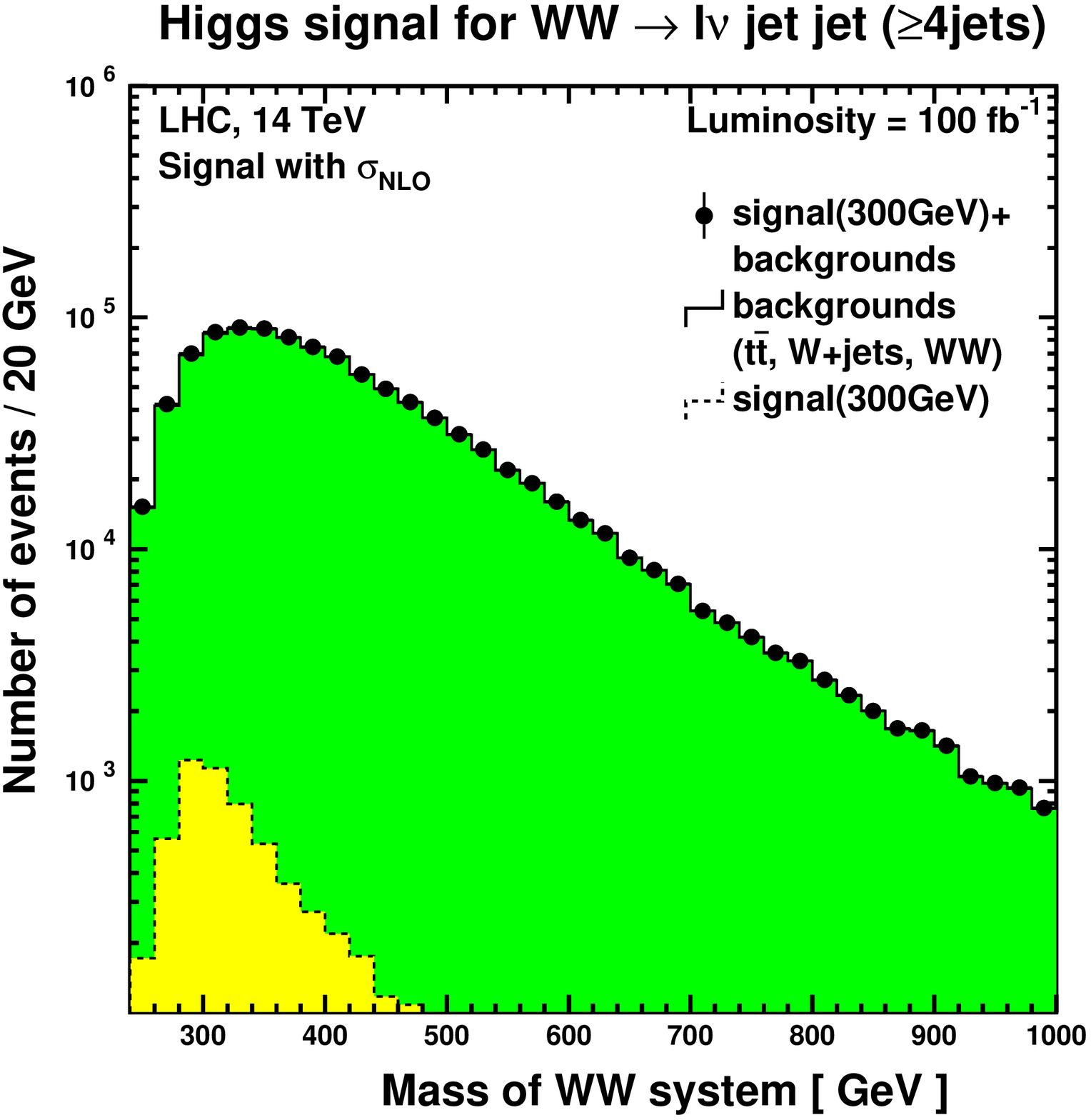}
}
}
\caption{Mass spectrum of a 300~GeV Higgs in the 
$H \fl WW \fl \ell \nu jj $ channel, for different number of jets
in event. (Up left) exactly 2 jets (Up right) exactly 3 jets 
(Down) more than 3 jets. (Left) with the same scale used than for the other plots
(Right) with a logarithmic scale.
As soon as there are more than 3 jets in events, the $t\bar{t}$ background
begins to be important and overwhelm a possible signal. Furthermore the number
of signal events produced through gluon fusion
is strongly reduced as soon as the number of required jets 
is higher than 2.}
\label{notagjet}
\end{center}
\end{figure}

It is finally interesting to point out that the number of jets
in event influences the signal to background ratio, especially when
no jet tagging is done.
A  signal to background ratio of about 0.1 is obtained when there 
are two jets in event, while it falls to about 0.01
than where more jets are present.
Actually, as soon as there are more than two jets in the event, the $t\bar{t}$
background begins to be very important. 
This can be seen on the plots of the
Figure \ref{notagjet}, which shows the invariant mass distribution of the
reconstructed $WW$ system for a 300~GeV Higgs and the corresponding backgrounds, 
plotted for different 
numbers of jets in the event.\\

In the $H\fl WW \fl \ell \nu jj$ channel,
very good results were obtained
thanks to the jet tagging technique. We manage to get
a signal to background ratio of about 2 together with a good
statistic (about 300 events left for a luminosity of $100 fb^{-1}$).
Moreover, 
somewhat better results were obtained
than the ones obtained previously
(for instance \cite{atlas}). The reason appears to be the introduction of 
the cut on $m_{Wjet}$.

	\section{$H \fl ZZ \fl \ell^+ \ell^- j j$}

This signature is not really different
from the four leptons one,
except for one $Z$ which will decay in jets instead of leptons.
The cuts to isolate
this signal are then similar to the ones defined
for the four leptons signature.
However, problems arise from the fact that
it is difficult to reconstruct the $Z$ decaying
into two jets (see page \pageref{zlljj}) and that new backgrounds
with huge cross sections
become dangerous due to the jets in the signature. Even if the branching fraction 
for this channel is 21 times bigger
than the one of the four leptons channel, it is then hard to get a signal.

For this channel, the signal cross sections times branching ratio are given 
in the Table \ref{cssigzzlljj}.

\begin{table}[htb]
\begin{center}
\begin{tabular}{|c|c|c|}
\hline
\multicolumn{3}{|c|}{\emph{Signal}} \\
\hline
Channel: &$qq \fl qqH \fl ZZ \fl \ell \ell j j$ &
$gg \fl H \fl ZZ \fl \ell \ell j j$ \\
\hline
Mass of Higgs &$\sigma \times BR$, NLO  
& $\sigma \times BR$, NLO \\
(GeV) & ($fb$) & ($fb$) \\
\hline
 300 &  54  \hspace*{0.3cm} &
  310  \hspace*{0.3cm} \\
\hline
 450 & 25  \hspace*{0.3cm} &
  203  \hspace*{0.3cm}  \\ 
\hline
 600 & 15  \hspace*{0.3cm} &
  81  \hspace*{0.3cm}  \\ 
\hline
\end{tabular}
\caption{Cross sections times branching ratio for the signal in the channel 
$H \fl ZZ \fl \ell^+ \ell^- jj $, given at the NLO.}
\label{cssigzzlljj}
\end{center}
\end{table}

The signature is here two isolated leptons of the same flavor
and at least two jets for the signature without jet tagging.
The potential backgrounds are then all the processes which
can produce two isolated leptons together with jets and are given in
Table \ref{csbakzzlljj}. Notice that for the $qq \fl Z + jets$ background
we did not let the $Z$ decay into $\tau$'s, 
because we checked that
these events did not survive the selection cuts. This reduces the
cross section
for this background by a factor 2/3, allowing less events to be generated.

\begin{table}[htb]
\begin{center}
\begin{tabular}{|l|rc|}
\hline
\multicolumn{3}{|c|}{\emph{Backgrounds}} \\
\hline
Channel & $\sigma \times BR$ ($fb$) &\\
\hline
$qq \fl ZZ \fl \ell \ell j j$ & 1'700 & \\ 
\hline
$qq \fl Z + jets$ , ($Z \fl e^{\pm},\,
\mu^{\pm}$) & 320'000 & 
 ($\widehat{p_t}>50 \,\mathrm{GeV}$) \\
\hline
$qq \fl t \bar{t\fl WbWb \fl \ell \nu b \; \ell \nu b} $ & 65'400 &\\ 
\hline
\end{tabular}
\caption{Cross sections times branching ratio 
for the  potential backgrounds in the channel 
$H \fl ZZ \fl \ell^+ \ell^- jj $.}
\label{csbakzzlljj}
\end{center}
\end{table}
The cuts used for the Higgs search without jet tagging in that channel are then
the followings:
\begin{itemize}
    \item We ask for two isolated leptons from the same flavor in the event ($e$, $\mu$ 
with $p_t>10\,\mathrm{GeV}$ and $|\eta|<2.5$), whose invariant mass has to be
within 5~GeV around 91~GeV and at least 2 jets with ($p_t>20\,\mathrm{GeV}$
and $|\eta|<4.5$).
    \item We then take all jets combinations which gave a mass within an interval of 
40~GeV centered in the shifted Z mass (85~GeV) and from these good combinations,
the one with the highest $p_t$ was chosen.
    \item As the mass of the $Z$ 
decaying into jets is 
shifted, the reconstructed $ZZ$ system mass is also slightly shifted
and the Higgs peak is around
285~GeV for a 300~GeV Higgs (for a 600
and a 450~GeV Higgs, as the peak
is wider, the effect does not need to be taken into account). 
Only events where the invariant mass of
the $ZZ$ system is within an interval of 70 (100, 200)~GeV 
centered in 285 (450, 600)~GeV are kept.
    \item A cut on the $p_t$'s of the reconstructed vector bosons and of the
reconstructed Higgs is finally done: $p_{t}(ZZ)>50 \,\mathrm{GeV}$,
$p_{t}(Z_{\ell\ell}$ or $Z_{jj})>100\, (150,\,200)\,\mathrm{GeV}$ 
for a 300 (450, 600)~GeV Higgs
\end{itemize}

If the jet tagging is added, the following selection cuts are applied:
\begin{itemize}
    \item The events must survive the above cuts.
    \item Exactly four jets are asked in event (two jets to reconstruct the $Z$
and two jets to be the tagging jets). The invariant mass 
of the tagging jets has to be higher than 800~GeV.
\end{itemize}

The expected results for $\mathcal{L}=100\,fb^{-1}$ and for 
different Higgs masses
are given in Table \ref{reszzlljj}. We give the expected number of events
without the jet tagging technique before and after 
the application of the different cuts on the 
$p_t$'s. We see that the main background in that case is the 
single $Z$ production. Because of that, it will be hard to see a signal
if no jet tagging is made. We see that the cuts on the different $p_t$'s
allows us to double the signal to background ratio.
Since the jet tagging is done, we get a good
signal to background ratio (more than 1) and a clear excess over the
background is seen.

\begin{table}
\begin{center}
\scalebox{0.95}{
\begin{tabular}{|l||r|r|r||r|}
\hline
\multicolumn{5}{|c|}{$H \fl ZZ \fl \ell^+ \ell^- jj $}\\
\hline
\hspace*{1.5cm} Channel & \multicolumn{4}{c|}{Number of events} \\
& Generated & Min. rec. cuts   
& $p_t$ cuts & Jet tag. \\
\hline
\multicolumn{5}{|c|}{} \\
\multicolumn{5}{|c|}{$m_{Higgs}=300\,\mathrm{GeV}$} \\ 
\multicolumn{5}{|c|}{} \\
\hline
$qq \fl qqH \fl \ell\ell jj$ & 5'400 & 980 & 220 & 50 \\ 
\hline
$gg \fl H \fl \ell\ell jj$ & 31'000 & 4'490 & 1'350 & 3 \\
\hline
Sum of all backgrounds & 38'710'000 & 193'280 & 27'750 & 46 \\
\hline
\hline
\multicolumn{5}{|l|}{Detailed backgrounds} \\
\hline
$qq \fl ZZ \fl \ell\ell jj$ & 170'000 & 3'400 & 400 & 0 \\
\hline
$qq, gg \fl Z + jets \fl \ell\ell +jets$ & 32'000'000 & 186'000 & 27'000 & 40 \\ 
\hline
$qq \fl t \bar{t}\fl WbWb \fl \ell \nu b \; \ell \nu b $ & 6'540'000 & 
3'880 & 350 & 6 \\
\hline
\hline
\multicolumn{5}{|c|}{} \\
\multicolumn{5}{|c|}{$m_{Higgs}=450\,\mathrm{GeV}$} \\
\multicolumn{5}{|c|}{} \\
\hline
$qq \fl qqH \fl \ell\ell jj$ & 2'500 & 460 & 260 & 60 \\
\hline
$gg \fl H \fl \ell\ell jj$ & 20'300 & 3'070 & 2'030 & 15 \\
\hline
Sum of all backgrounds & 38'710'000 & 60'320 & 18'760 & 50 \\
\hline
\hline
\multicolumn{5}{|l|}{Detailed backgrounds} \\
\hline
$qq \fl ZZ\fl \ell\ell jj$ & 170'000 & 420 & 320 & 0 \\
\hline
$qq, gg \fl Z + jets \fl \ell\ell +jets$ &  32'000'000 & 59'500 & 18'300 & 50\\
\hline
$qq \fl t \bar{t}\fl WbWb \fl \ell \nu b \; \ell \nu b $ & 6'540'000 &
400 & 140 & 0 \\ 
\hline
\multicolumn{5}{|c|}{} \\
\multicolumn{5}{|c|}{$m_{Higgs}=600\,\mathrm{GeV}$} \\
\multicolumn{5}{|c|}{} \\
\hline
$qq \fl qqH \fl \ell\ell jj$ & 1'500 & 320 & 210 & 70 \\ 
\hline
$gg \fl H \fl \ell\ell jj$ & 8'100 & 1'090 & 680 & 5 \\
\hline
Sum of all backgrounds & 38'710'000 & 43'970 & 11'211 & 25 \\
\hline
\hline
\multicolumn{5}{|l|}{Detailed backgrounds} \\
\hline
$qq \fl ZZ\fl \ell\ell jj$ & 170'000 & 760 & 210 & 0 \\
\hline
$qq, gg \fl Z + jets \fl \ell\ell +jets$ & 32'000'000 & 43'000 & 11'000 & 25 \\
\hline
$qq \fl t \bar{t}\fl WbWb \fl \ell \nu b \; \ell \nu b $ & 6'540'000 & 
210 & 1 & 0 \\
\hline
\end{tabular}}
\caption{Expected results for $\mathcal{L}=100\,fb^{-1}$ and for different Higgs masses
in the $H \fl ZZ \fl \ell^+ \ell^- j j $
channel, NLO cross sections. 
After the total number of events generated for each process,
we give the number of events left after the minimal reconstruction
cuts, which means all cuts before the cut on the 
different $p_{t}$'s. The two last columns show the expected 
number of events left after all cuts
without and with the jet tagging technique.}
\label{reszzlljj}
\end{center}
\end{table}

The luminosities required for a five standard deviation in that channel are
given in Table \ref{ldisclljj}.
\begin{table}[htb]
\begin{center}
\begin{tabular}{|c|r|r|r|r|}
\hline
\multicolumn{5}{|c|}{$H \fl ZZ \fl \ell^+ \ell^- jj $}\\
\hline
& Signal & Background
& S/B &  $\mathcal{L}_{disc}$ ($fb^{-1}$)\\
\hline
\multicolumn{5}{|c|}{$m_{Higgs}=300\,\mathrm{GeV}$} \\ 
\hline
no tagging & 1'570 & 27'750 & 0.06 & 28 \hspace*{0.3cm} \\
tagging & 53 & 46 & 1.15 & 41 \hspace*{0.3cm} \\
\hline
\hline
\multicolumn{5}{|c|}{$m_{Higgs}=450\,\mathrm{GeV}$} \\ 
\hline
no tagging & 2'290 & 18'760 & 0.12 & 9 \hspace*{0.3cm} \\
tagging & 75 & 50 & 1.50 & 22 \hspace*{0.3cm} \\
\hline
\hline
\multicolumn{5}{|c|}{$m_{Higgs}=600\,\mathrm{GeV}$} \\ 
\hline
no tagging & 890 & 11'211 & 0.08 & 35 \hspace*{0.3cm} \\
tagging & 75 & 25 & 3.00 & 20 \hspace*{0.3cm} \\
\hline
\end{tabular}
\caption{Expected discovery luminosities for the channel 
$H \fl ZZ \fl \ell^+\ell^- jj$, without systematic errors taken into
account.The number of events for signal and background 
corresponds to a luminosity of $100\,fb^{-1}$.}
\label{ldisclljj}
\end{center}
\end{table}

Like for the $H \fl WW\fl \ell \nu jj$ channel, the results without jet tagging
have to be taken with care, as the signal to background ratio is 
a bit too small to give relevant results when one does not take into
account the systematic errors.

\begin{figure}[htb]
\begin{center}
\mbox{
\subfigure{\includegraphics*[width=.5\textwidth]{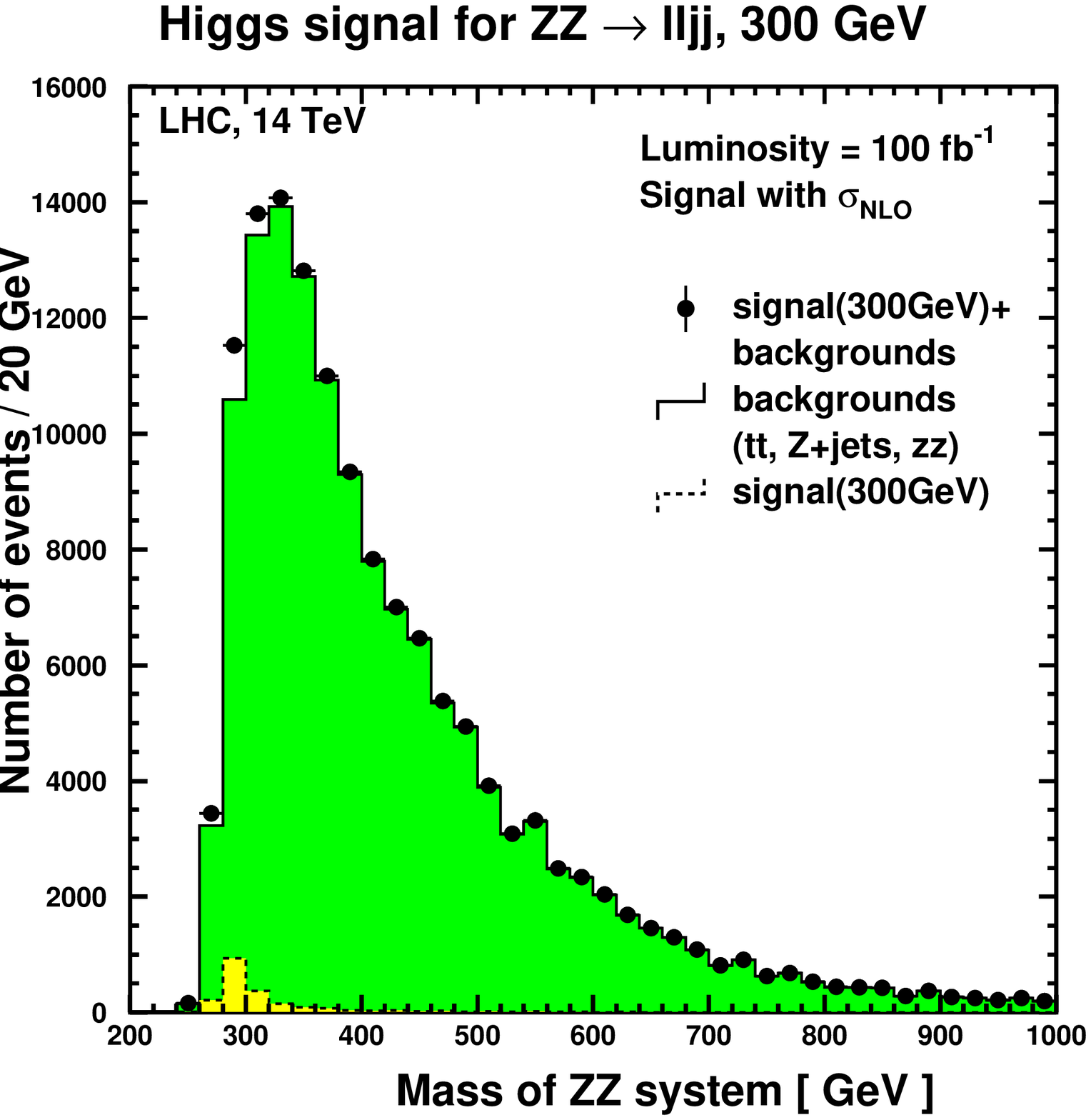}\quad
}
\subfigure{\includegraphics*[width=.5\textwidth]{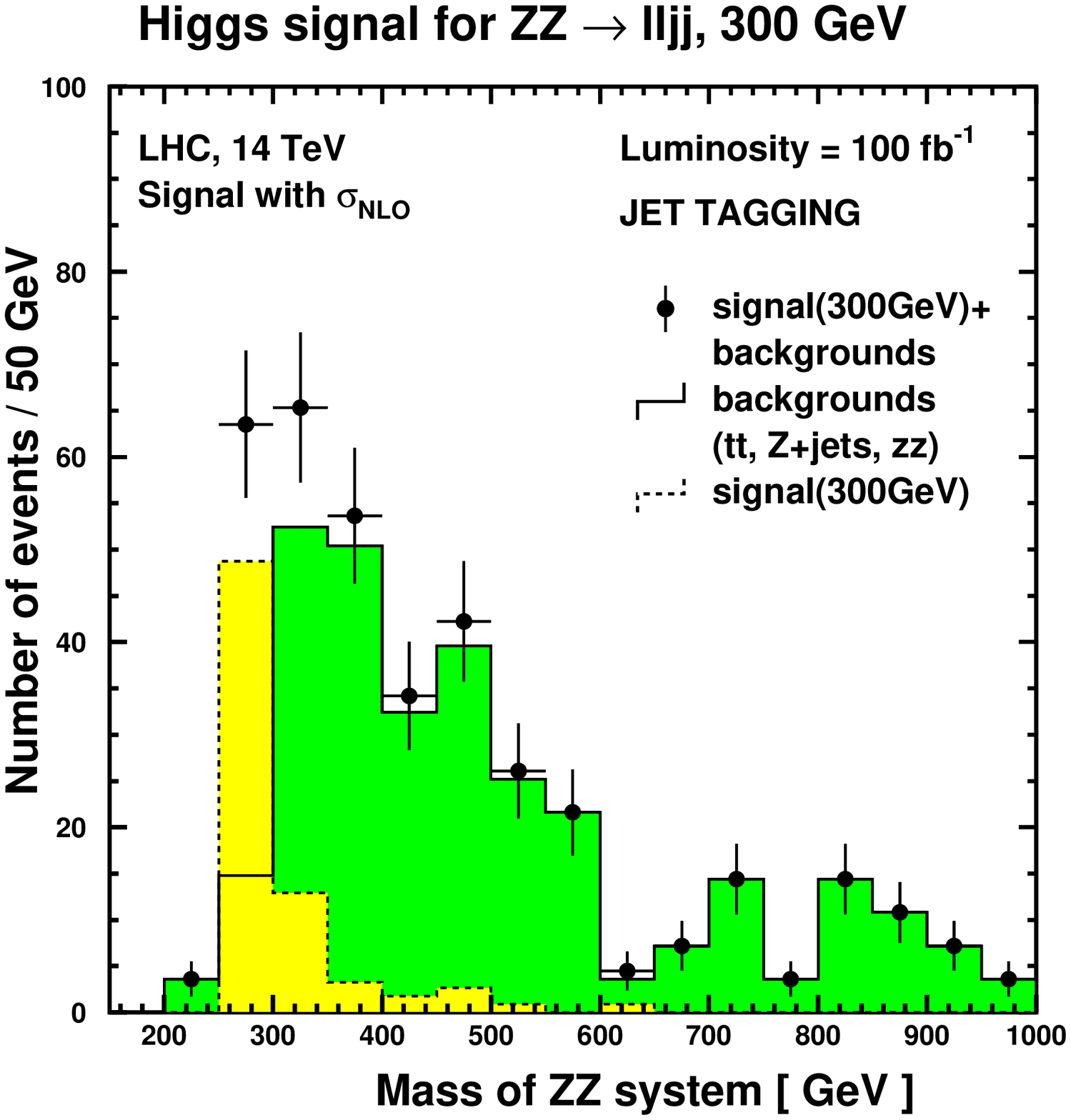}
}
}
\caption{Mass spectrum of the reconstructed $ZZ$ system
for a 300~GeV Higgs. (Left) without jet tagging
(Right) with jet tagging.}
\label{mpeak300}
\end{center}
\end{figure}

\begin{figure}[htb]
\begin{center}
\mbox{
\subfigure{\includegraphics*[width=.5\textwidth]{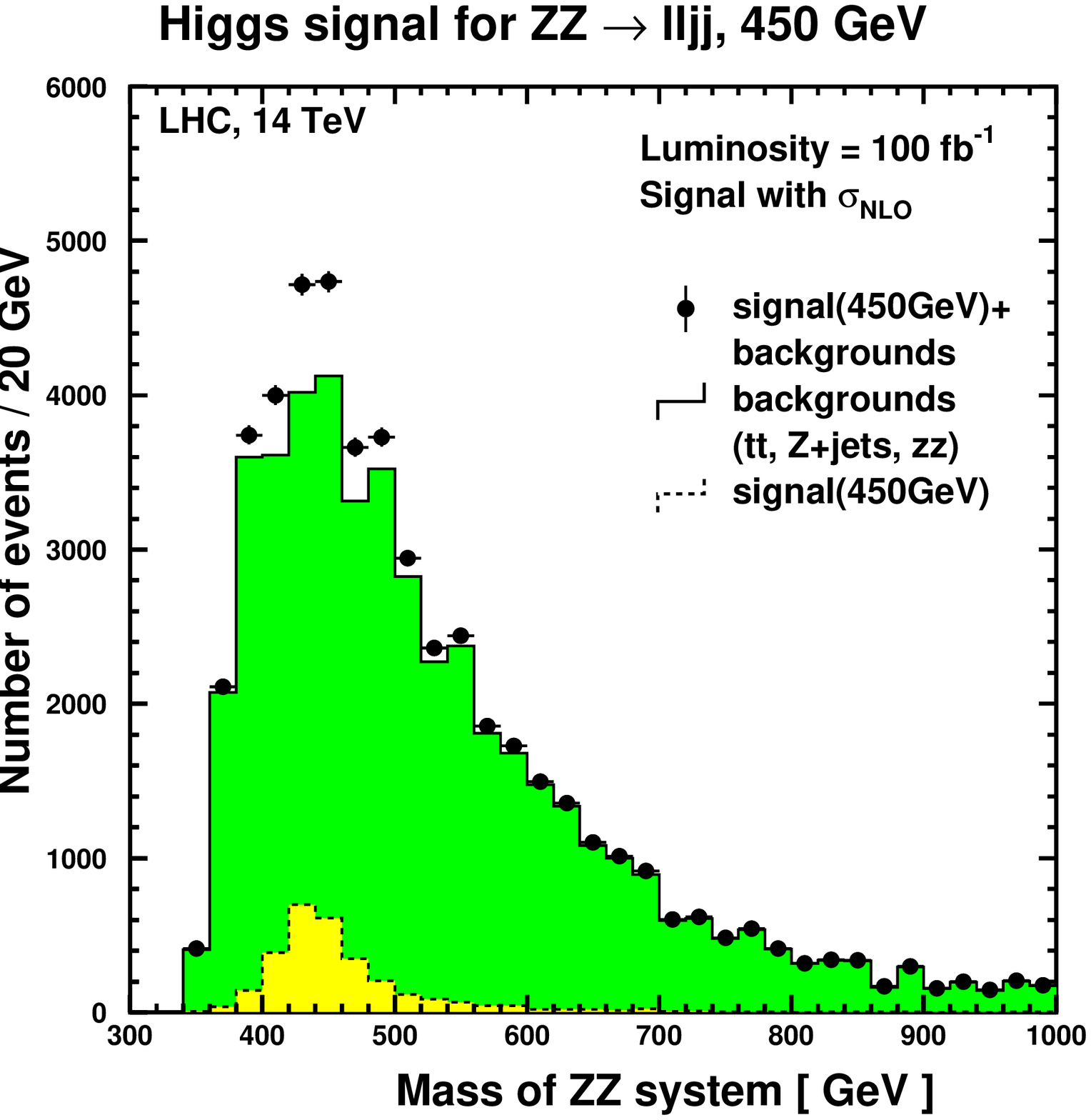}\quad
}
\subfigure{\includegraphics*[width=.5\textwidth]{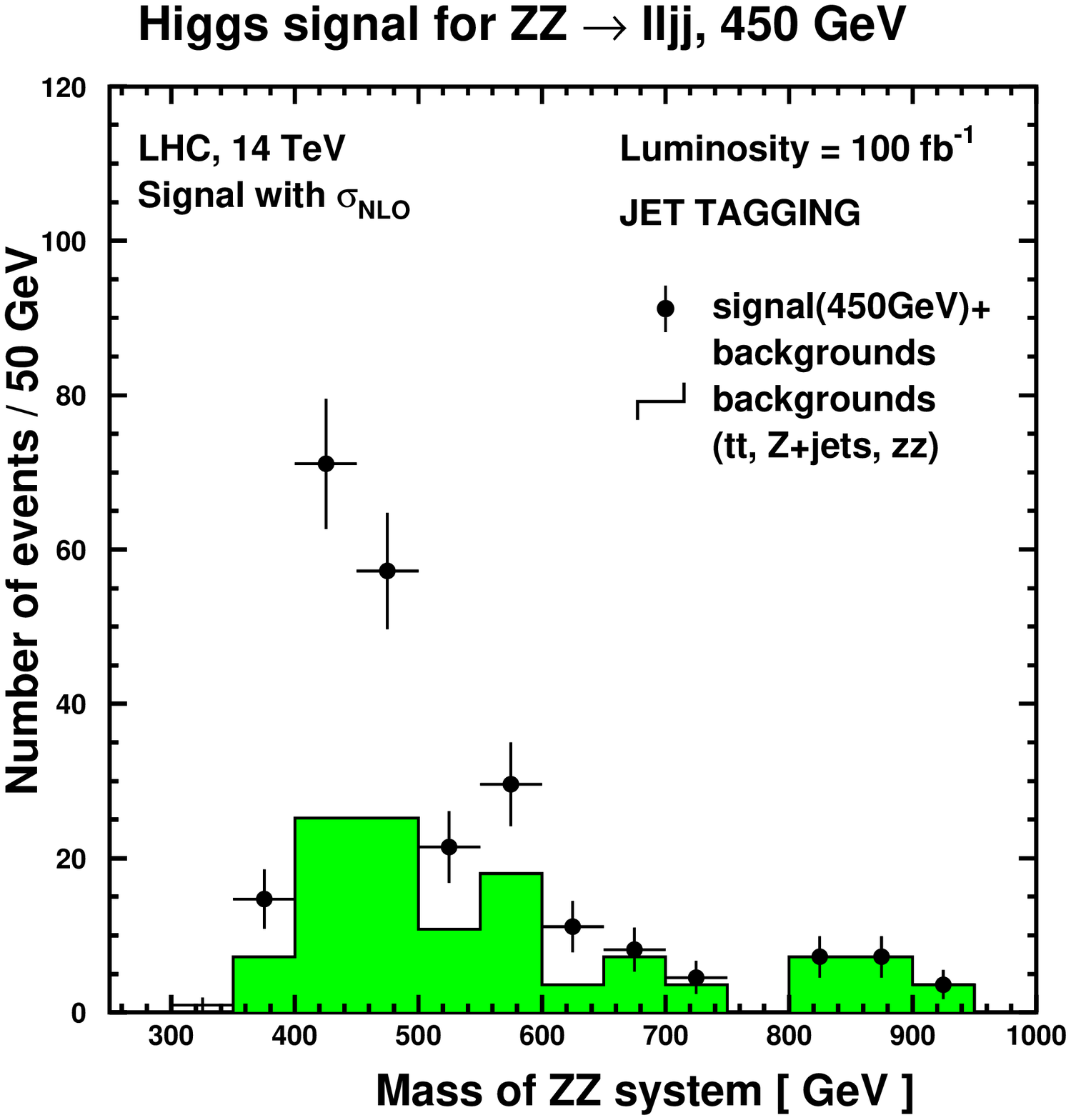}
}
}
\caption{Mass spectrum of the reconstructed $ZZ$ system
for a 450~GeV Higgs. (Left) without jet tagging
(Right) with jet tagging.}
\label{mpeak450}

\end{center}
\end{figure}

\begin{figure}[htb]
\begin{center}
\includegraphics[width=.65\textwidth]{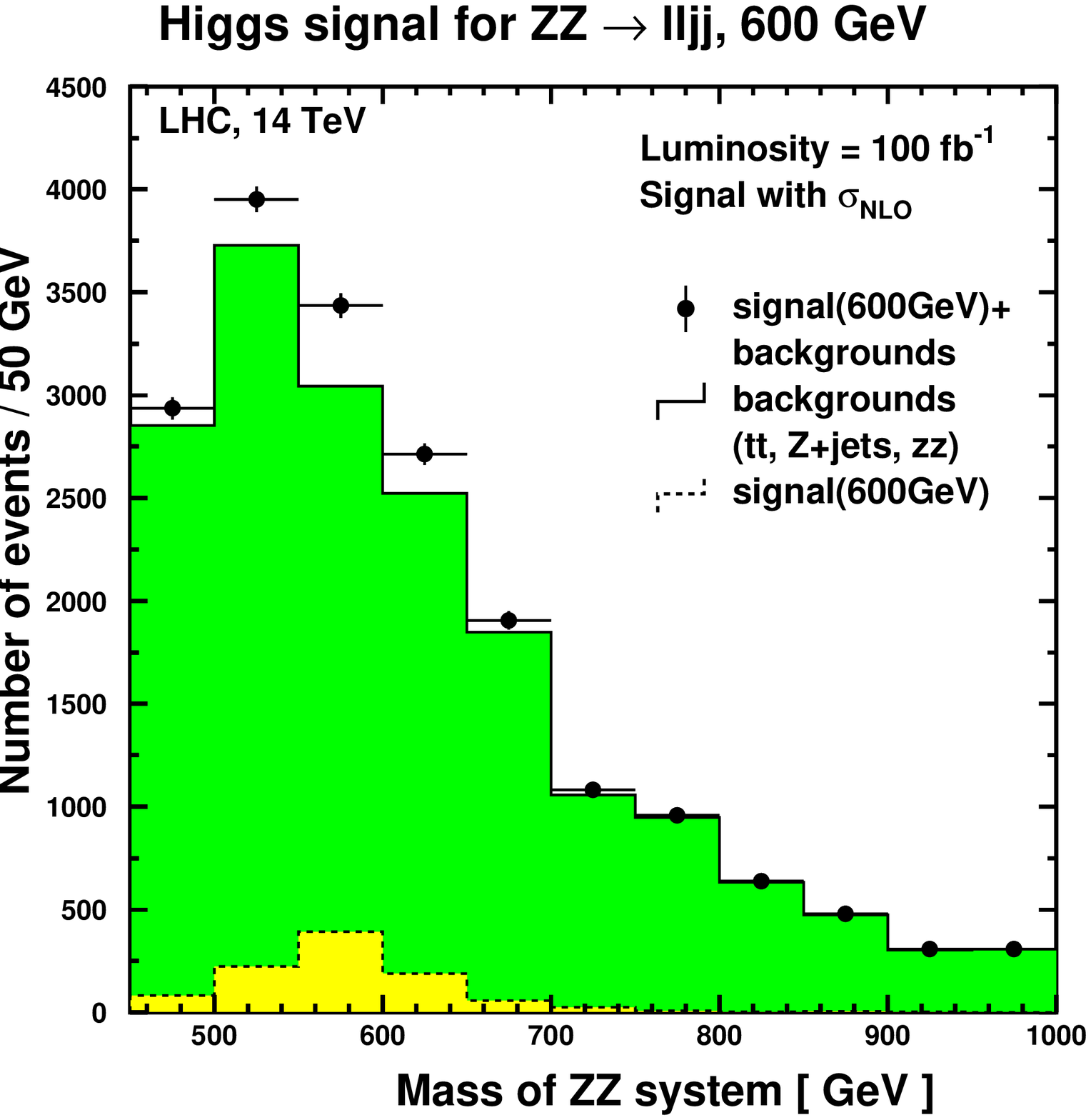}
\\
\includegraphics[width=.65\textwidth]{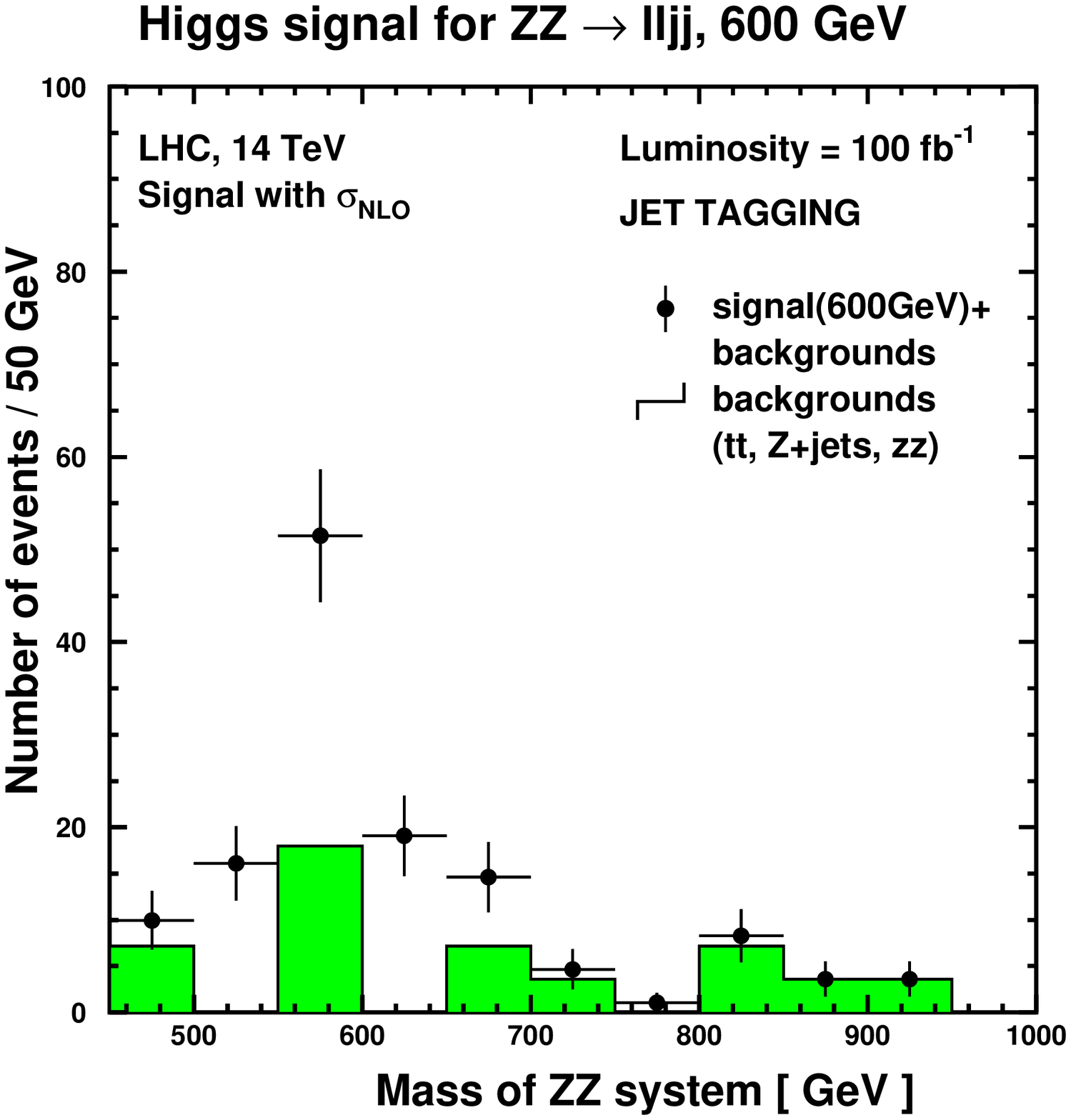}
\caption{Mass spectrum of the reconstructed $ZZ$ system
for a 600~GeV Higgs. (Up) without jet tagging
(Down) with jet tagging.}
\label{mpeak600}
\end{center}
\end{figure}

Figures \ref{mpeak300}, \ref{mpeak450} and \ref{mpeak600},
show the mass of the $ZZ$ system.
When no tagging is done, for a 300 and a 600~GeV Higgs,
it is clear that only a signal might be seen if the background
is very well known. 
Nevertheless, a good peak can be obtained if the
signal alone is considered.
For a 450~GeV Higgs, a bigger excess can be seen. It confirms the good
discovery luminosity we find there: $9fb^{-1}$. 
However the signal to background ratio is 
a bit low, 0.12.

As soon as the tagging is done, clear excesses become visible
for every masses, as shown on Figures 
\ref{mpeak300}, \ref{mpeak450} and \ref{mpeak600}. 

Once more, only the number of events corresponding to a 
lower luminosity
were generated for the background,
which explains the big fluctuations on the plots
in the distributions of the background.
\\[0.5cm]
This channel can lead an interesting contribution
if it is combined to the results of the other channels.
Due to the presence of high backgrounds, it does not lead alone
to very useful results.

        \section{$H \fl ZZ \fl \nu \bar{\nu} j j$}

This signature has not been really studied yet.
Among the channels which contain Higgs decays into $Z$,
it has the larger branching fraction:
64 times higher than the four leptons channels one and 10 times 
higher than the
$H\fl ZZ \fl \ell\ell \nu\nu$ one. 
However,
this signature is really hard to isolate due to the presence of the large QCD 
background. 
As no isolated leptons are present here, all the processes like
$q_i q_j \fl q_i q_j$ might be important background sources.
However, as we find with the previous signatures that the jet tagging
technique managed to rise considerably the signal to background ratios,
it is interesting to study 
if, with the help of the jet tagging cut, one could nevertheless
have a signal with that signature.
We chose to consider only a 600~GeV Higgs, as
harder cuts can be done than with a 300~GeV Higgs.

The cross sections times branching ratio at the next to leading order
are given in the Table \ref{cssigzzjjnunu}.

\begin{table}[htb]
\begin{center}
\begin{tabular}{|c|c|c|}
\hline
\multicolumn{3}{|c|}{\emph{Signal}} \\
\hline
Channel: &$qq \fl qqH \fl ZZ \fl \nu \bar{\nu} jj $ &
$gg \fl H \fl ZZ \fl \nu \bar{\nu} jj $ \\
\hline
Mass of Higgs &$\sigma \times BR$, NLO
& $\sigma \times BR$, NLO \\
(GeV) & ($fb$) & ($fb$) \\
\hline
 600 & 30  \hspace*{0.3cm} &
  164 \hspace*{0.3cm} \\ 
\hline
\end{tabular}
\caption{Cross sections times branching ratio for the signal in the channel 
$H \fl ZZ \fl \nu \bar{\nu} j j$, given at the NLO.}
\label{cssigzzjjnunu}
\end{center}
\end{table}

\begin{table}[htb]
\begin{center}
\begin{tabular}{|l|rl|}
\hline
\multicolumn{3}{|c|}{\emph{Backgrounds}} \\
\hline
Channel & \multicolumn{2}{|l|}{$\sigma \times BR$ ($fb$)}\\
\hline
$qq \fl ZZ \fl \nu \bar{\nu} j j $ & 3'200 & \\ 
\hline
$qq \fl Z + jets$ , ($Z \fl \nu \bar{\nu}$) & 811'000 & 
 ($\widehat{p_t}>50 \,GeV$) \\
\hline
$qq \fl t \bar{t}\fl WbWb \fl \mathrm{anything}$ & 622'000 &\\ 
\hline
QCD backgrounds & $7 \cdot 10^7$ & 
($\widehat{m}>1000 \,GeV$, $\widehat{p_t}>100 \,GeV$) \\
\hline
\end{tabular}
\caption{Cross sections  times branching ratio 
for the  potential backgrounds in the channel 
$H \fl ZZ \fl \nu\bar{\nu} j j$.}
\label{csbakzzjjnunu}
\end{center}
\end{table}

The backgrounds for this signature, which is characterized by two jets, 
which have to give the $Z$
mass, a large missing $p_t$ and no isolated leptons, are given in Table
\ref{csbakzzjjnunu}. What should be understood under \emph{QCD backgrounds} are the
following processes:
\begin{itemize}
\item $q_i q_j \fl q_i q_j$
\item $q_i \bar{q}_i \fl q_k \bar{q}_k$
\item $q_i \bar{q}_i \fl g g$
\item $q_i g \fl q_i g$
\item $g g \fl q_i \bar{q}_i$
\item $g g \fl g g$
\end{itemize}
Actually these processes can be generated asking the mass of the total system,
$\widehat{m}$,
being higher than 1000~GeV. This can be done, as a high $\widehat{m}$
is anyway asked by the selection cuts:
The tagging jet system to have a mass higher than 800~GeV
and the two other jets a mass close to 90~GeV.

\begin{figure}[htb]
\begin{center}
\includegraphics*[scale=0.5]{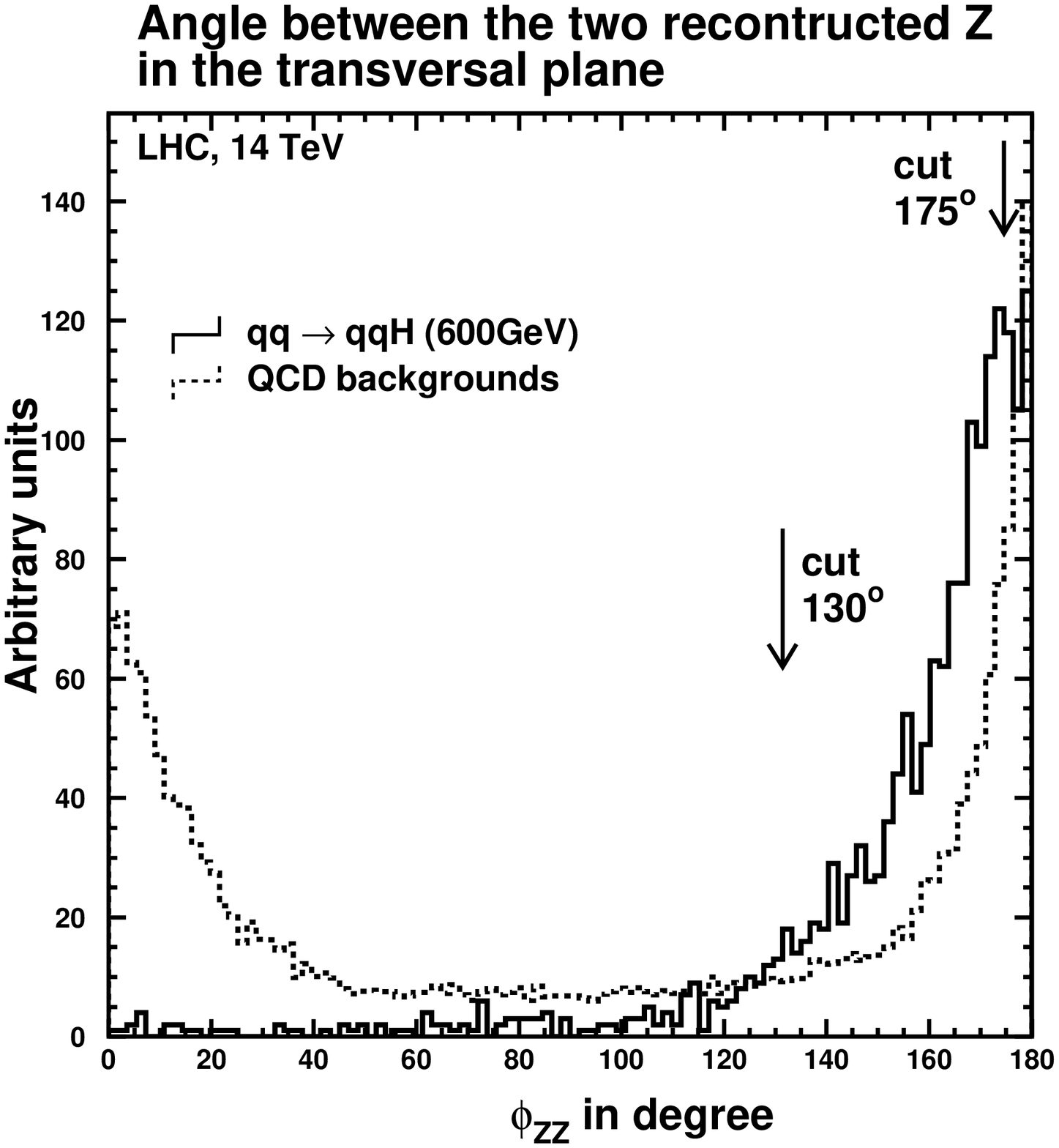}
\caption{Angle between the two reconstructed Z, for the signal
and the QCD background.
The two histograms are not made for the same luminosity
in order to compare the
shapes.
To have a number of events corresponding to the same luminosity for the
signal and the background, the number of background events should be 
multiplied by $2 \cdot 10^{6}$.}
\label{backtoback}
\end{center}
\end{figure}

Here are the cuts that were tried to isolate a signal with that signature:
\begin{itemize}
    \item First we ask for four jets with $p_t>20\,\mathrm{GeV}$
and $|\eta|<4.5$ and no isolated leptons in the event.
    \item We then took all jets combinations which gave a mass within an interval of 
40~GeV centered in the shifted $Z$ mass (85~GeV) and from these good combinations,
the one with the highest $p_t$ was chosen.
The missing $p_t$ vector is identified with the $Z$
decaying into neutrinos.
    \item A cut on the angle in the transverse plane, $\phi$,
between the two reconstructed $Z$, 
was also made, asking it to be between $130^{\mathrm{o}}$ 
and $170^{\mathrm{o}}$. This allows us 
to get rid of the half of the QCD background,
as it is shown on Figure \ref{backtoback}.
    \item A cut on the $p_t$'s of the two reconstructed $Z$'s is also done, asking
$p_{t}(Z_{jj})>200\,GeV$, $p_{t}^{miss}>200\,GeV$
    \item Finally, the mass of the tagging jets system has to be higher than 
800~GeV or 1200~GeV.
\end{itemize}

The expected results for $\mathcal{L}=100\,fb^{-1}$ 
and for a 600~GeV Higgs are given in Table \ref{reszzjjnunu}.
There is a big uncertainty on the QCD background results
as its cross section is too high to generate the number of events
corresponding to a luminosity of $100fb^{-1}$, even with hard cuts 
in the generation.
We then had to multiply the number of QCD background events with 700~!

\begin{table}[htb]
\begin{center}
\scalebox{0.9}{
\begin{tabular}{|l||r|r|r|}
\hline
\multicolumn{4}{|c|}{$H \fl ZZ \fl \nu \bar{\nu} jj $}\\
\hline
 & \multicolumn{3}{c|}{Number of events} \\
\hspace*{1.5cm}Channel& Generated 
& \multicolumn{2}{c|}{With tagging} \\
 & &$m_{jj}>800\,\mathrm{GeV}$ & $m_{jj}>1200\,\mathrm{GeV}$ \\
\hline
$qq \fl qqH \fl \nu \bar{\nu} jj$ & 3'000 & 190 & 110 \\
\hline
$gg \fl H \fl \nu \bar{\nu} jj$ & 16'400 & 30 & 20 \\
\hline
Sum of all backgrounds & $7 \cdot 10^9$ & 4'700 & 1'600 \\
\hline
\hline
\multicolumn{4}{|l|}{Detailed backgrounds} \\
\hline
$qq \fl ZZ \fl \nu \bar{\nu} jj$ & 320'000 & 0 & 0 \\
\hline
$qq, gg \fl Z + jets \fl \nu \nu +jets$ & 81'100'000 & 300 & 100 \\
\hline
$qq \fl t \bar{t}\fl WbWb \fl \mathrm{anything} $ & 62'200'000 & 200 & 100 \\
\hline
QCD background &$7 \cdot 10^9$ & 4'200 & 1'400 \\
\hline 
\end{tabular}}
\caption{Expected results for $\mathcal{L}=100\,fb^{-1}$ for a 600~GeV Higgs
(NLO cross sections) in the $H \fl ZZ \fl \nu \bar{\nu} jj$
channel. We give the expected number of events after all cuts.
Once the mass of the
tagging jets is asked to be higher than 800~GeV and once it is
asked to be higher than 1200~GeV.}
\label{reszzjjnunu}
\end{center}
\end{table}

What is interesting to notice here is that if the QCD background is not taken into
account, a signal can be isolated
(130 signal events against 200 background events).
Further studies need to be done to see if 
additional cuts against the QCD background could be found.
Then, there would be
a chance to observe a Higgs signal also in that channel.
For example, the cut on the angle allowed already
to get rid of the half of the QCD background.

Note that it would be relevant, not only for that channel, 
to know better the kinematic properties of the
QCD background.
This background could reduce the significance of other channels, 
if, for example, the jets are misidentified as 
missing energy or leptons.

	\chapitre{Discussion}
\markboth{Discussion}{Discussion}

In the following, we will combine the results obtained for the different channels
studied, namely:
$H \fl ZZ \fl \ell \ell \ell \ell$, $H \fl ZZ \fl \ell \ell \nu \nu$, 
$H \fl WW \fl \ell \nu \ell \nu$, $H \fl WW \fl \ell \nu jj$,
$H \fl ZZ \fl \ell \ell jj$ and $H \fl ZZ \fl \nu \nu jj$. 

We concentrate on the following questions:
\begin{itemize}
\item How can a heavy Higgs, with a mass between 300 and 600~GeV be discovered 
at the LHC~? Which channels can give significant results for this purpose~?
We will especially consider what the jet tagging technique
can bring.
\item Then we will discuss how precisely the ratio between the two
decay branching fractions
$H \fl WW$ and $H \fl ZZ$ can be measured.
\item Finally, we will ask ourselves with which statistical precision the
weak boson fusion and gluon fusion cross sections
can be measured.
\end{itemize}
As already said, these two last measurements 
are crucial tests of the theoretical predictions about
how the Higgs couples to top quarks and $W$ and $Z$ bosons.

\section{Discovery luminosity for heavy Higgs}

First of all, we want to find out with which luminosity a heavy Higgs boson can be
discovered at the LHC.
Two things are important to consider for a channel to give a
signal: Firstly, the signal to background ratio has to be high enough
to reduce the effect of the systematic errors. Secondly,
enough statistics have to be left after the selection cuts, to
minimize the statistical errors which are proportional to
$\sqrt{N}$.
In that study,
only the statistical errors are taken into account.
However, only the signatures
where signal to background ratio was high enough ($>0.5$) are
kept. This gives a safety margin on the results, if
the systematic errors are subsequently taken into account.

A summary of the expected
discovery luminosities for every channel studied is made in
Table \ref{alldl}. 
Always two values for the
discovery luminosity were obtained: A first 
one without using the jet tagging technique
and a second one using the jet tagging technique.
We chose to give only
the most significant result of these two.

\begin{table}[p]
\begin{center}
\begin{tabular}{|l|r|r|r|c|c|}
\hline
&&&& Signal & Discovery\\
\ctabi{Channel} & \ctab{$S$}
& \ctab{$B$} &\ctab{$S/B$} & significance & luminosity\\
&&&& $S/\sqrt{B}$ & ($fb^{-1}$)\\
\cline{2-6}
& \multicolumn{5}{c|}{Luminosity: }\\
 & \multicolumn{3}{c|}{$100\,fb^{-1}$} 
 & $5\,fb^{-1}$ &  $\mathcal{L}_{disc}$\\
\hline
\hline
\multicolumn{6}{|c|}{} \\
\multicolumn{6}{|c|}{$m_{Higgs}=300\,GeV$} \\
\multicolumn{6}{|c|}{} \\
\hline
$pp \fl H \fl ZZ \fl \ell\ell \ell \ell$& 140 & 4 & 35.00 &  $\approx$ 6.0 & 4\\ 
\hline
\hline
$qq \fl qqH \fl ZZ \fl \ell\ell \nu \nu$& 
62 & 66 & 0.94 & $\approx$ 1.5 & 43 \\
\hline
$qq \fl qqH \fl WW \fl \ell \nu \ell \nu$& 90 & 90 & 1.00 & $\approx$ 2.0 & 28 \\
\hline
$qq \fl qqH \fl WW \fl \ell \nu jj$& 330 & 280 & 1.18  & $\approx$ 4.5 & 6\\
\hline
$qq \fl qqH \fl ZZ \fl \ell\ell jj$& 53 & 46 & 1.15 
 & $\approx$ 2.0 & 41\\
\hline
\hline
\multicolumn{4}{|l}{Combined} & $\approx$ 8.0 & 2-3\\
\hline
\hline
\multicolumn{6}{|c|}{} \\
\multicolumn{6}{|c|}{$m_{Higgs}=450\,GeV$} \\
\multicolumn{6}{|c|}{} \\
\hline
$pp \fl H \fl ZZ \fl \ell \ell \ell \ell$ & 90 & 6 & 15.00 & $\approx$ 3.5 & 8\\ 
\hline
$pp \fl H \fl ZZ \fl \ell\ell \nu \nu$& 
810 & 1'313 & 0.62 & $\approx$ 5.0 & 5 \\
\hline
\hline
$qq \fl qqH \fl WW \fl \ell \nu \ell \nu$ & 150 & 90 & 1.67 & $\approx$ 3.5 & 12\\
\hline
$qq \fl qqH \fl WW \fl \ell \nu jj$ & 420 & 200 & 2.10 & $\approx$ 7.0 & 3\\
\hline
$qq \fl qqH \fl ZZ \fl \ell\ell jj$ & 75 & 50 & 1.50  
& $\approx$ 2.0 & 22 \\
\hline
\hline
\multicolumn{4}{|l}{Combined} & $\approx$ 10.0 & 2\\
\hline
\hline
\multicolumn{6}{|c|}{} \\
\multicolumn{6}{|c|}{$m_{Higgs}=600\,GeV$}\\
\multicolumn{6}{|c|}{} \\
\hline
$pp \fl H \fl ZZ \fl \ell\ell \ell \ell $& 40 & 10 & 4.00  
& $\approx$ 2.0 & 30\\
\hline
$pp \fl H \fl ZZ \fl \ell\ell \nu \nu$ & 
410 & 400 & 1.03 & $\approx$ 4.5 & 6\\
\hline
\hline
$qq \fl qqH \fl WW \fl \ell \nu \ell \nu$& 130 & 90 & 1.44 & $\approx$ 3.0 & 13\\
\hline
$qq \fl qqH \fl WW \fl \ell \nu jj$& 260 & 150 & 1.73 & $\approx$ 5.0 & 6\\
\hline
$qq \fl qqH \fl ZZ \fl \ell\ell jj$& 75 & 25 & 3.00  
& $\approx$ 2.5& 20\\ 
\hline
\hline
\multicolumn{4}{|l}{Combined} & $\approx$ 8.0 & 2-3\\
\hline
\end{tabular}
\caption{Expected discovery luminosities 
($\mathcal{L}_{disc}$)
for the studied channels,
without systematic errors taken into account. The signal ($S$)
and background ($B$) 
are given for a luminosity of $100 \,fb^{-1}$.
The third column shows the signal to background ratio ($S/B$).
We also give the significance the
signals reach with a luminosity of $5\,fb^{-1}$, ie. corresponding to 
half a year of
LHC low luminosity running. The total significance reached
for this luminosity is also calculated.
For every Higgs masses, we give first the
channels which were more significant without jet tagging ($pp \fl H$)
and then the channels which were
more significant with the jet tagging ($qq \fl qqH$).}
\label{alldl}
\end{center}
\end{table}

\begin{figure}[p]
\begin{center}
\includegraphics*[width=0.9\textwidth]{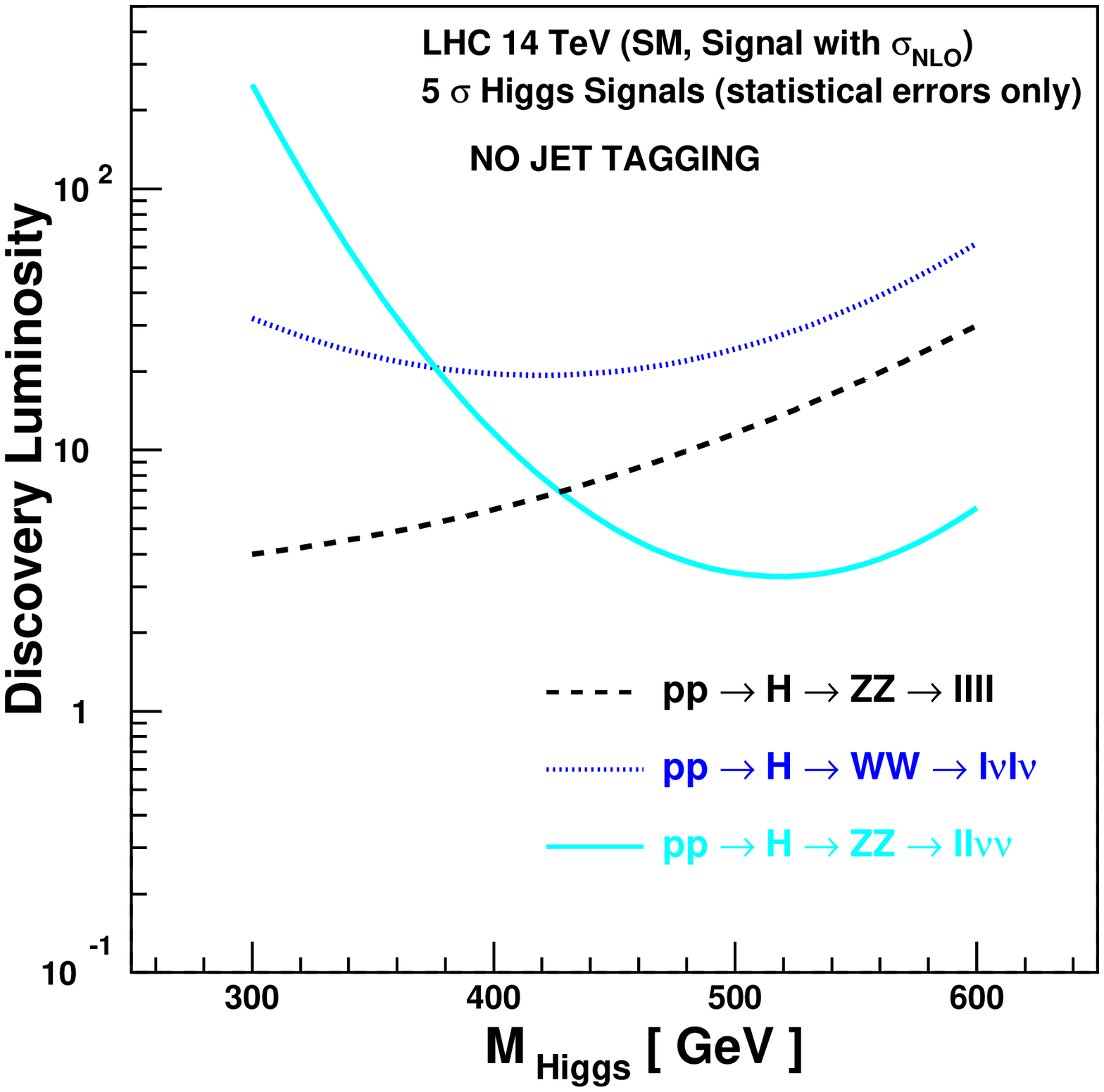}
\caption{Expected luminosity for a 5 $\sigma$ standard deviation,
for the 3 channels that give the most significant contribution
when no jet tagging is made. Note that in the  $H \fl WW \fl \ell \nu \ell \nu$
channel, the signal to background ratio is low (about 0.15).
For the signal, NLO cross sections are used.}
\label{lumidiscnew1}
\end{center}
\end{figure}

\begin{figure}[p]
\begin{center}
\includegraphics*[width=0.9\textwidth]{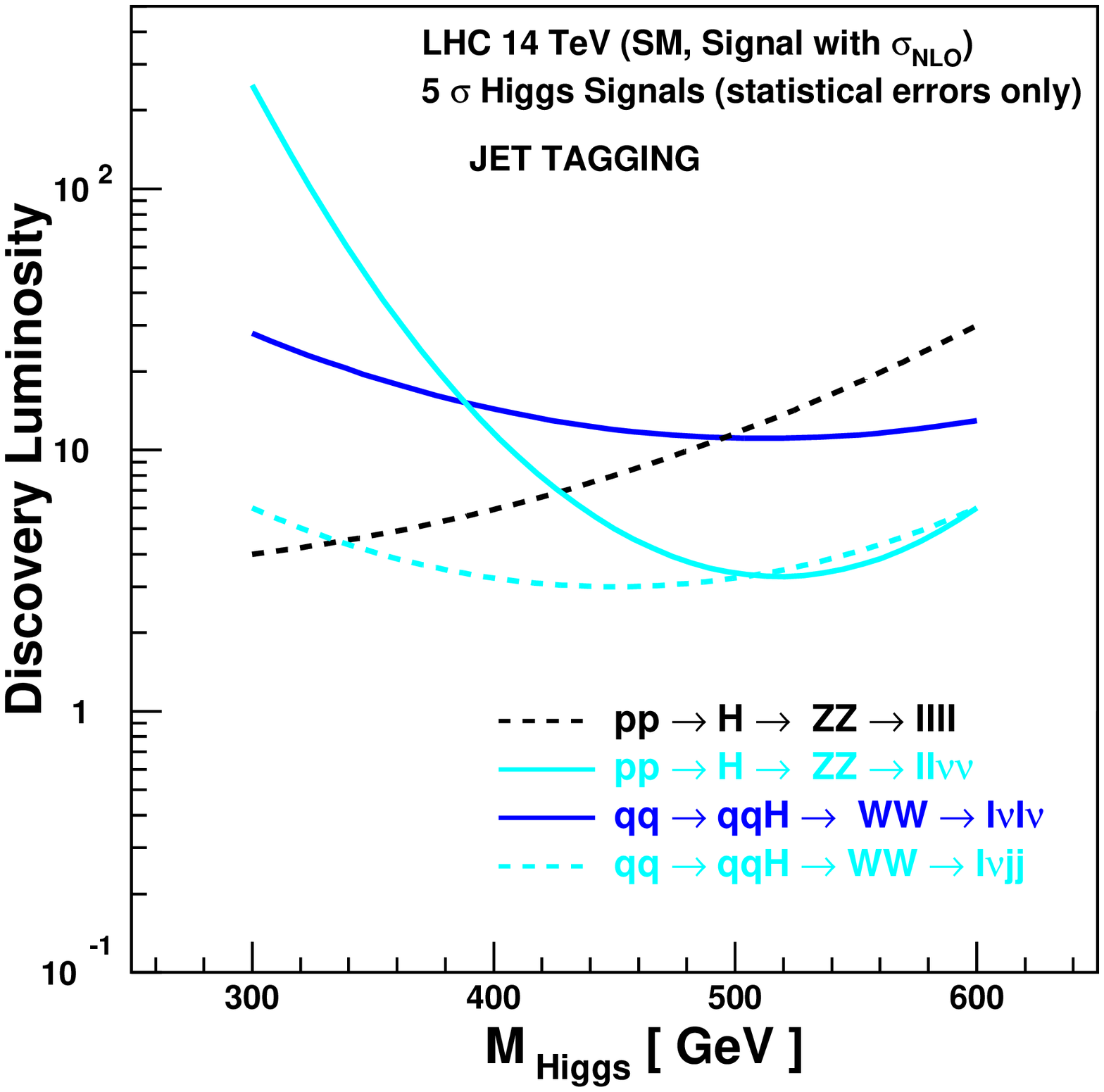}
\caption{Expected luminosity for a 5 $\sigma$ standard deviation, for the
channels that can give an interesting contribution. The channels
where the jet tagging technique is used are taken here into account.
For the signal, NLO cross sections are used.}
\label{lumidiscnew2}
\end{center}
\end{figure}

We see that we have mainly two types of channels. The first ones, 
which are more significant when no jet tagging is used, namely:
$H \fl ZZ \fl \ell \ell \ell \ell$ and $H \fl ZZ \fl \ell \ell \nu \nu$.
The second ones, which only give a significant
signal when the jet tagging technique
is used:
$H \fl ZZ \fl \ell \ell jj$,
$H \fl WW \fl \ell \nu \ell \nu$ and $H \fl WW \fl \ell \nu jj$. 

The forward jet tagging technique
allows to get good
signal to background ratios in several channels.
However, the price to pay as soon as it is used,
is a strong reduction of signal events. This would be a problem for channels
with a small branching ratio, typically the ones where only leptonic
decays are present.

Therefore, even if the jet tagging technique selects essentially a particular way of 
producing the Higgs, it can
lead to a more significant Higgs signal
than a signal obtained when no particular
production mechanism of the Higgs is favored. 
This is especially the case when hadronic decays of
$W$ and $Z$ are present in the signature.

As can be seen from Table \ref{alldl},
the four leptons channel is giving very good results
for Higgs masses of 
300 and 450~GeV, leading to an expected discovery luminosity of 4 and
$8\,fb^{-1}$ respectively. 
The ability of this signature to give significant results relies essentially
on the possibility to reconstruct a narrow signal peak in the 
$ZZ$ mass spectrum.
However as for high Higgs masses, the natural width of the Higgs becomes
wider, this signature is not as powerful as for lower Higgs masses. Furthermore,
its branching fraction begins to be very small for high Higgs masses.

The $H \fl ZZ \fl \ell \ell \nu \nu$ has a good significance for
higher Higgs masses. An expected discovery luminosity for
that channel of 5 and $6\,fb^{-1}$ for a 450 and 600~GeV Higgs 
respectively is expected.
This channel works well for a Higgs with a high mass, because
the decay products of a Higgs which has a high mass receive more $p_t$ 
and therefore more missing $p_t$ is present in the
signal events. This will allow harder cuts on the missing $p_t$ to be done,
leading to a good signal significance, as the background has a 
missing $p_t$ spectrum which steeply falls down.

Finally the third channel which can bring a good significance for
a Higgs discovery is the $H \fl WW \fl \ell \nu q q$
channel, having
expected discovery luminosities of 6, 3 and $6 \,fb^{-1}$ for a
300, 450 and 600~GeV Higgs respectively.
Note that this last signature is obtain with the jet tagging
technique. As it has a rather high branching fraction, this method
works well here.

The two last channels $H \fl WW \fl \ell \nu \ell \nu$ and 
$H \fl ZZ \fl \ell \ell jj$ are giving smaller contributions to the total
significance. 

The last column of Table \ref{alldl} shows the signal significance
obtained for a luminosity of $5\,fb^{-1}$, which corresponds to
half a year of LHC running at low luminosity or, expressed in another
way, one year at 50\% efficiency.
We see that as for the discovery luminosity, almost all the
significance comes from the four leptons channel,
the $H \fl ZZ \fl \ell \ell \nu \nu$ channel
and the $H \fl WW \fl \ell \nu jj$ channel.

If these significances 
are combined, we get
8, 10 and 8 standard deviations for a Higgs with a mass of
300, 450 and 600~GeV, respectively. 

In Table \ref{alldl}, the expected 
discovery luminosities for the combined channels
are also given. We find that
a Higgs with has a mass between 300 and 600~GeV and SM-like couplings,
should be discovered with a luminosity between 2 and $3\,fb^{-1}$
and thus should be discovered already
after one year of LHC low-luminosity running.
Moreover, even if the gluon fusion process would not exist, such a Higgs
could be observed within about a year, using then only the channels
where that signal is isolated with the jet tagging technique.

Figures \ref{lumidiscnew1} and \ref{lumidiscnew2}
give the expected required luminosity to
discover a SM Higgs with a statistical significance of
five standard deviations. 
The first plot is made with the expected discovery luminosities
in the channels when no jet tagging technique is used.
On the second plot, the corresponding results
are shown with the channels where the jet tagging 
technique was used to get a signal. As said before, these alternative signatures,
which are isolated with the jet tagging technique 
($H \fl WW \fl \ell \nu \ell \nu$ and $H \fl WW \fl \ell \nu jj$),
can contribute to
lower the discovery luminosity brought by the standard Higgs discovery
channels ($H \fl ZZ \fl \ell \ell \ell \ell$ 
and $H \fl ZZ \fl \ell \ell \nu \nu$).

\begin{figure}[htb]
\begin{center}
\includegraphics[width=0.8\textwidth]{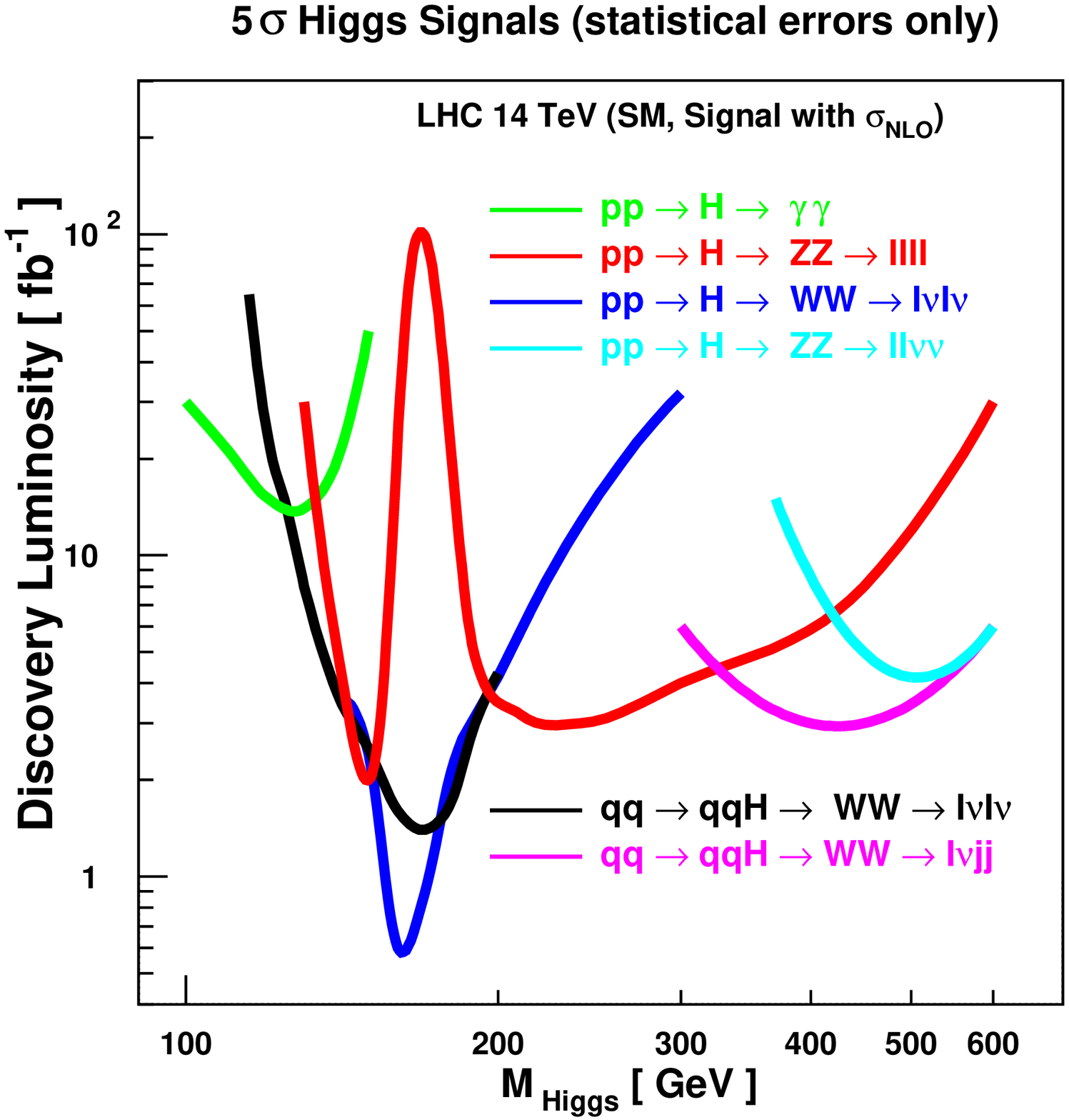}
\caption{Expected luminosity for a 5 $\sigma$ standard deviation,
as presented in the CMS meeting of 9th March 2001.
For the signal, NLO cross sections are used.}
\label{alllumi}
\end{center}
\end{figure}

Figure \ref{alllumi} gives the updated curve for the discovery luminosity
over the all mass range taking into account the results brought by this study and
as presented in the CMS meeting of 9th March 2001.

Finally, one last thing needs to be taken into account when one want
to discuss the ability of a channel to discover the Higgs.
Namely,
how, in a potential discovery channel, the
Higgs mass can be reconstructed.
For the four leptons channels, it is not a problem, as
it is directly obtained from the mass peak.
That means that for
masses up to 450~GeV, where this channel
gives a good signal, the Higgs mass can be measured 
with the discovery luminosity.
As already discussed,
the Higgs mass peak can be also
reconstructed with the $H\fl WW \fl \ell \nu jj$ signature,
but slightly worse. 

In contrast, in the
$H \fl ZZ \fl \ell\ell \nu \nu$ channel, no mass peak can be reconstructed.
One could still obtain the Higgs mass indirectly
from a fit on the missing $p_t$ spectrum 
or from the transverse mass spectrum. 

Taking into account the results of the previous studies, we see that
from 200~GeV up to 300~GeV, the four leptons channels contributes almost alone to
the significance.
It will then not be a problem to measure the
Higgs mass if it lies in that interval.
For higher masses, the
$qq \fl qqH \fl WW \fl \ell \nu jj$
channel can then be used to reconstruct a mass peak.

\section{Measuring the ratio between the two
decay branching fractions of a heavy SM Higgs,
$H \fl WW$ and $H \fl ZZ$}
\markboth{Discussion}{Measuring $\frac{BR(H \fl WW)}{BR(H \fl ZZ)}$}

From now on, the assumption that the Higgs was discovered
and that its mass is known will be made. 
It will be assumed that the available luminosity is $100\,fb^{-1}$, ie.
the luminosity corresponding to one year of LHC
running at full capacity.

The next question we would like to study is how precisely one could measure the 
$\frac{BR(H \fl WW)}{BR(H \fl ZZ)}$ ratio.
As said before, a Standard Model 
Higgs with a mass between 300 and 600~GeV
will decay essentially in $W$'s and $Z$'s. The branching
ratio (BR) for a Higgs decay in $W$ is approximatively 2 times
the  branching
ratio for a Higgs decay in $Z$'s.

It is favorable to take a ratio rather than a branching ratio alone as 
some hypothetical errors on cuts efficiencies 
caused by the simulation can be canceled.
Let's take an example:
We consider the cut on the tagging jet mass system
and make the hypothesis that
the rapidity of the tagging jets was not well simulated. It we assume that
the same error in the $H\fl WW$ channel than in the $H \fl ZZ$ 
channel is introduced, then this error will be canceled when the 
ratio is taken. Actually, as soon as
we divide the branching fractions obtained for these two channels, 
the efficiency of this cut does not really influence the final result.

By the way, we can compare the efficiencies of the tagging jets cuts in the
different channels, which is done in Table \ref{effmjj}. We see that 
the efficiency of the cut
asking the invariant mass of the tagging jets being
higher than 800 GeV is varying only by about 10\% in the different channels.
The efficiency of this cut will
then not influence the branching fractions rate.

\begin{table}[htb]
\begin{center}
\begin{tabular}{|l|c|c|}
\hline
Channel & $m_{Higgs}=300\;\mathrm{GeV}$ & $m_{Higgs}=600\;\mathrm{GeV}$ \\
\hline
$qq \fl qqH \fl ZZ \fl \ell\ell\ell\ell$ & 0.52 & 0.60 \\
\hline
$qq \fl qqH \fl ZZ \fl \ell\ell \nu\nu$ & 0.50 & 0.67 \\
\hline
$qq \fl qqH \fl WW \fl \ell \nu \ell \nu$ & 0.47 & 0.65 \\
\hline
$qq \fl qqH \fl WW \fl \ell \nu jj$ & 0.42 & 0.57 \\
\hline
$qq \fl qqH \fl ZZ \fl \ell\ell jj$ & 0.47 & 0.58 \\
\hline
\end{tabular}
\caption{Efficiency of the cut on the tagging jets mass, for the different
channels studied, for weak boson fusion process.}
\label{effmjj}
\end{center}
\end{table}

Let's see how we can get a cross section:

Assuming that our simulation gives an efficiency of the
cuts close to the reality, the cross section for a the process
is given by the following formula:
\begin{equation}
\label{sigma}
\sigma=\frac{S}{\mathcal{L}\cdot \varepsilon_{cuts}}
\end{equation}
Where $\mathcal{L}$ is the luminosity,
$\varepsilon_{cuts}$ the efficiency of the cuts and
$S\,(=N-B)$ the number of signal events,
$N$ the total number of events
measured which survived the selection cuts minus and $B$ the expected number of 
background events given by the simulation.

Given that formula, the error measured cross section depends 
these parameters: 
\begin{itemize}
\item $\Delta N$, which is given by
$\pm \sqrt{S+B}$ when no systematical errors are taken into account,
\item $\Delta \varepsilon_{cuts}$, which depends on the
simulation used,
\item $\Delta B$, which in some cases depends strongly on 
the Monte Carlo used but which can be tuned through a estimation
of the backgrounds in the regions where no signal is expected,
using data events
(this idea was discussed for the 
$H\fl ZZ \fl \ell\ell \nu\nu$ channel, see page \pageref{norm}),
\item and $\Delta \mathcal{L}$.
\end{itemize}
The only source of error being take into account in that study is the
statistical error on $N$, $\pm \sqrt{S+B}$. It will give us
a lower bound of the precision that could be reached.

To get the $\frac{BR(H \fl WW)}{BR(H \fl ZZ)}$ ratio, 
we have to measure separately 
the two ratios
$\frac{\sigma(gg \fl H \fl WW)}{\sigma(gg \fl H \fl ZZ)}$
and $\frac{\sigma(qq \fl qqH \fl WW)}{\sigma(qq \fl qqH \fl ZZ)}$ 
and subsequently combine them,
as the two Higgs production mechanisms are not the same.\\

We want to determine in a first step the
$\frac{\sigma(qq \fl qqH \fl WW)}{\sigma(qq \fl qqH \fl ZZ)}$
ratio. The signal events are then only events where the Higgs was
produced through weak boson fusion and we will count the
events with a Higgs produced though gluon fusion as background
events.

We showed in the previous sections that the forward jet tagging technique allowed
to isolate signal events containing 
Higgs produced through weak boson fusion as well from the
background events, as the signal events produced through gluon fusion.
Depending on the signatures considered, we find that after the 
jet tagging cuts, there was five to ten times more $qq \fl qqH$ events than
$gg \fl H$ events.

Only the results from channels having a signal to background ratio
higher than 0.1 are kept for the analysis\footnote{
The minimal value required for the signal to
background ratio is in that case smaller (0.1)
than the value that was taken for the discovery (0.5).
Actually, as we are working with higher luminosities,
a better estimation of the background can be then assumed.}.
Table \ref{resqqh} gives the number of weak boson fusion
signal events ($S$) and background events ($B$) expected
for every signature studied, as well as their ratio ($S/B$).
The statistical error on the measure of a particular branching fraction
($\sqrt{S+B}/S$) is given in the last column.

\begin{table}[p]
\begin{center}
\begin{tabular}{|l||r|r|r|r||c|}
\hline
\ctabi{Channel} & \ctab{S}
& \ctab{B} & S/B & $\sqrt{S+B}$ & Stat. error\\
\hline
\multicolumn{6}{|c|}{} \\
\multicolumn{6}{|c|}{300 GeV Higgs, $qq \fl qqH$} \\
\multicolumn{6}{|c|}{} \\
\hline
$qq \fl qqH \fl ZZ \fl \ell\ell\ell\ell$ & 15 & 2 & 7.50 & 4 & 27\%\\
\hline
$qq \fl qqH \fl ZZ \fl \ell\ell \nu\nu$ & 55 & 73 & 0.75 & 11 & 21\%\\
\hline
$qq \fl qqH \fl WW \fl \ell \nu \ell \nu$ & 80 & 100 & 0.80 & 13 & 17\% \\
\hline
$qq \fl qqH \fl WW \fl \ell \nu jj$ & 290 & 320 & 0.91 & 25 & 9\%\\
\hline
$qq \fl qqH \fl ZZ \fl \ell\ell jj$ & 50 & 49 & 1.02 & 10 & 20\%\\
\hline
\hline
\multicolumn{6}{|c|}{} \\
\multicolumn{6}{|c|}{450 GeV Higgs, $qq \fl qqH$} \\
\multicolumn{6}{|c|}{} \\
\hline
$qq \fl qqH \fl ZZ \fl \ell\ell\ell\ell$ & 6 & 1 & 6.00 & 2 & 33\% \\
\hline
$qq \fl qqH \fl ZZ \fl \ell\ell \nu \nu$ & 55 & 43 & 1.28 & 10 & 18\% \\
\hline
$qq \fl qqH \fl WW \fl \ell \nu \ell \nu$ & 140 & 100 & 1.40 & 15 & 11\% \\
\hline
$qq \fl qqH \fl WW \fl \ell \nu jj$ & 330 & 290 & 1.14 & 25 & 8\% \\
\hline
$qq \fl qqH \fl ZZ \fl \ell\ell jj$ & 60 & 65 & 0.92 & 11 & 18\% \\
\hline
\hline
\multicolumn{6}{|c|}{} \\
\multicolumn{6}{|c|}{600 GeV Higgs, $qq \fl qqH$} \\
\multicolumn{6}{|c|}{} \\
\hline
$qq \fl qqH \fl ZZ \fl \ell\ell\ell\ell$ & 5 & 0 & - & 2 & 40\% \\
\hline
$qq \fl qqH \fl ZZ \fl \ell\ell \nu\nu$ & 40 & 19 & 2.11 & 8 & 20\% \\
\hline
$qq \fl qqH \fl WW \fl \ell \nu \ell \nu$ & 110 & 110 & 1.00 & 15 & 14\% \\
\hline
$qq \fl qqH \fl WW \fl \ell \nu jj$ & 220  & 190 & 1.16 & 20 & 9\% \\
\hline
$qq \fl qqH \fl ZZ \fl \ell\ell jj$ & 70 & 30 & 2.33 & 10 & 14\% \\
\hline
\end{tabular}
\caption{Expected statistical errors on the measurement 
of different branching fractions
of the weak boson fusion process, for a luminosity of $100\,fb^{-1}$.
The Higgs events produced with gluon fusion are counted here as 
background events.}
\label{resqqh}
\end{center}
\end{table}

We can combine the different channels to get the final error.
The formula used 
for the total error $\sigma_{tot}$, with
the particular channels errors, $\sigma_{i}$ is:
\begin{equation}
\label{combform}
\sigma_{tot}=\frac{1}{\sqrt{w}}, \: \mathrm{where} \:
w=\sum_i \frac{1}{\sigma_i^2}
\end{equation}
We first combine the channels to get the error on the
$qq \fl qqH \fl ZZ$ and $qq \fl qqH \fl WW$
branching fractions.
Expected results are given in Table \ref{resqqhcomb}.
Depending on the mass, we see that we have errors between 6.5\% and 12.8\%.

\begin{table}[htb]
\begin{center}
\begin{tabular}{|l|r|r|r|}
\hline
\multicolumn{4}{|c|}{Combined results for weak boson fusion}\\
\hline
& \multicolumn{3}{c|}{Statistical error}\\
\cline{2-4}
\ctabi{Channel} & \multicolumn{3}{c|}{Higgs mass} \\
 & 300 GeV & 450 GeV & 600 GeV \\
\hline
$qq \fl qqH \fl ZZ$ & 12.8\% & 11.9\% & 11.0\% \\
\hline
$qq \fl qqH \fl WW$ & 8.0\% & 6.5\% & 7.6\% \\
\hline
\hline
$qq \fl qqH$ & 6.8\% & 5.7\% & 6.2\% \\
\hline
\end{tabular}
\caption{Expected statistical errors when
all channels contributions are combined and
for luminosity of $100\,fb^{-1}$ on the measurement 
of two Higgs branching fractions when it is produced through
weak boson fusion. We also give the total cross
section for weak boson fusion.}
\label{resqqhcomb}
\end{center}
\end{table}

The second step is to determine the $\frac{\sigma(gg \fl H \fl WW)}
{\sigma(gg \fl H \fl ZZ)}$ ratio.
The results of the Higgs search when no jet tagging 
is applied can be used, as
the contribution of this process is always larger than the contribution
from weak boson fusion.
The events coming from the weak boson fusion are considered 
as background events.
Also in that case, only the channels where the signal to background
ratio is higher than 10\% are kept.

The results for the different channels are given in Table \ref{resggh}.
Like for the weak boson fusion, we combine the results of the 
different channels to get the total error. Results are given 
in Table \ref{resgghcomb}. We get errors between 4.3\% and 17\%.

\begin{table}[p]
\begin{center}
\begin{tabular}{|l||r|r|r|r||c|}
\hline
\ctabi{Channel} & \ctab{S} 
& \ctab{B} & \ctab{S/B} & $\sqrt{S+B}$ & Stat. error\\
\hline
\multicolumn{6}{|c|}{} \\
\multicolumn{6}{|c|}{300 GeV Higgs, $gg \fl H$} \\
\multicolumn{6}{|c|}{} \\
\hline
$gg \fl H \fl ZZ \fl \ell\ell\ell\ell$ & 90 & 54 & 1.67 & 12 & 13\% \\
\hline
$gg \fl H \fl WW \fl \ell \nu \ell \nu$ & 460 & 2'807 & 0.16  & 57 & 12\% \\
\hline
\hline
\multicolumn{6}{|c|}{} \\
\multicolumn{6}{|c|}{450 GeV Higgs, $gg \fl H$} \\
\multicolumn{6}{|c|}{} \\
\hline
$gg \fl H \fl ZZ \fl \ell\ell\ell\ell$ & 70 & 26 & 2.69 & 10 & 14\% \\
\hline
$gg \fl H \fl ZZ \fl \ell\ell \nu\nu$ & 700 & 1'423 & 0.49 & 46 & 7\% \\ 
\hline
$gg \fl H \fl ZZ \fl \ell\ell jj$ & 2'290 &  18'760 & 0.12 & 145 & 6\% \\
\hline
$gg \fl H \fl WW \fl \ell \nu \ell \nu$ & 530 & 2'555 & 0.21 & 56 & 10\% \\
\hline
\hline
\multicolumn{6}{|c|}{} \\
\multicolumn{6}{|c|}{600 GeV Higgs, $gg \fl H$} \\
\multicolumn{6}{|c|}{} \\
\hline
$gg \fl H \fl ZZ \fl \ell\ell\ell\ell$ & 25 & 25 & 1.00 & 7 & 28\% \\
\hline
$gg \fl H \fl ZZ \fl \ell\ell \nu\nu$ & 330 & 480 & 0.69 & 28 & 9\% \\ 
\hline
$gg \fl H \fl WW \fl \ell \nu \ell \nu$ & 310 & 2'560 & 0.12 & 54 & 17\% \\
\hline
\end{tabular}
\caption{Expected statistical errors on the measurement 
of different branching fractions when the Higgs is produced through
gluon fusion, for a luminosity of $100\,fb^{-1}$.
The Higgs events produced with weak boson fusion are counted here as 
background events.}
\label{resggh}
\end{center}
\end{table}

\begin{table}[p]
\begin{center}
\begin{tabular}{|l|r|r|r|}
\hline
\multicolumn{4}{|c|}{Combined results for gluon fusion}\\
\hline
& \multicolumn{3}{c|}{Statistical error}\\
\cline{2-4}
\ctabi{Channel} & \multicolumn{3}{c|}{Higgs mass} \\
 & 300 GeV & 450 GeV & 600 GeV \\
\hline
$gg \fl H \fl ZZ$ & 13.0\% & 4.3\% & 8.6\% \\
\hline
$gg \fl H \fl WW$ & 12.0\% & 10.0\% & 17.0\% \\
\hline
\hline
$gg \fl H$ & 8.9\% & 4.0\% & 7.7\% \\
\hline
\end{tabular}
\caption{Expected statistical errors on the measurement 
of the gluon fusion cross section,
when all channels contributions are combined, for a luminosity of $100\,fb^{-1}$.
We first combine the channels to get the error on the
$gg \fl H \fl ZZ$ and $gg \fl H \fl WW$
branching fractions. Then we combine all channels
to get the $gg \fl H$ cross section.}
\label{resgghcomb}
\end{center}
\end{table}

We have now everything we need to make a prediction on the statistical error
we can get on the measurement of the
$\frac{BR(H \fl WW)}{BR(H \fl ZZ)}$ ratio. Let's see how we can get this
information for a 450~GeV Higgs:

If we refer to Tables \ref{resqqhcomb} and \ref{resgghcomb}, we find that
the branching ratio for $gg \fl H \fl ZZ$ 
for a 450~GeV Higgs
is known up to
a statistical error of 4.3\%
and the cross section for $gg \fl H \fl WW$ up to a statistical error
of 10\%\footnote{It is sufficient
to measure a given decay mode of the $W$'s, as the branching ratios
of the different $W$'s decay modes
are known from LEP.}.
That means that in the best of the cases, we
could measure the $\frac{\sigma(gg \fl H \fl WW)}{\sigma(gg \fl H \fl ZZ)}$
ratio for a 300~GeV Higgs up to an error of 
11\% in the first LHC year of full running.
The formula used to get the error, $\Delta BR$,  on the ratio is
\begin{equation}
\Delta BR=\sqrt{\sum \Delta BR_i^2}
\end{equation}
We can now repeat the same procedure for the weak boson fusion, where we
have errors of 6.5\% for $H\fl WW$ and 11.9\% for $H\fl ZZ$, which gives a 
statistical error on the ratio of 14\%.
We combine the two results with the same method than the one used in the previous
section, using the Formula \ref{combform} given on page
\pageref{combform}, and we find a final statistical error
of 9\%. 

A summary of all the results is given in Table \ref{zzww}.

\begin{table}[htb]
\begin{center}
\begin{tabular}{|c|c|c|c|}
\hline
Higgs mass & \multicolumn{2}{c|}{Error determined in the process:} & Combined  \\
& $\frac{\sigma(gg \fl H \fl WW)}{\sigma(gg \fl H \fl ZZ)}$ &
$\frac{\sigma(qq \fl qqH \fl WW)}{\sigma(qq \fl qqH \fl ZZ)}$&
$\frac{BR(H \fl WW)}{BR(H \fl ZZ)}$  \\
\hline
300~GeV & 18\%  & 15\% & 12\% \\
\hline
450~GeV & 11\% & 14\% & 9\% \\
\hline
600~GeV & 19\% & 13\% & 11\% \\
\hline
\end{tabular}
\caption{Expected statistical precision that could be reached with a 
luminosity of $100\,fb^{-1}$ on the measurement of the  
$\frac{BR(H \fl WW)}{BR(H \fl ZZ)}$
ratio.}
\label{zzww}
\end{center}
\end{table}

From this, we can conclude that
the $\frac{BR(H \fl WW)}{BR(H \fl ZZ)}$ ratio could be measured
with a statistical precision of about 10\%,
when only statistical errors are taken into account and after one year of high 
luminosity LHC running.

\section{Measuring the weak boson production 
and the gluon fusion cross sections}
\markboth{Discussion}{Measuring $\sigma (qq\fl qqH)$ and $\sigma (gg\fl H)$}

We now estimate how precisely
the cross section for weak boson fusion can be measured, with a 
luminosity of $100\,fb^{-1}$.
We can actually directly refer to the Table \ref{resqqh}. We only need to
combine the results of the different branching ratios we have to get the
total error, using Formula \ref{combform}.

We find a statistical error on the measurement of the weak
boson fusion cross section of 6.8\%, 5.7\% and 6.2\%
for a 300, 450 and 600~GeV Higgs respectively, see Table \ref{resqqhcomb}.

The channel which gives the smallest error for the 
weak boson fusion cross section is the $H \fl WW \fl \ell \nu jj$
channel, with statistical errors of
9\%, 8\% and 9\%
for a 300, 450 and 600~GeV Higgs respectively
and largely contributes to improve the total resolution. 

We can also estimate the statistical precision on a measurement of the gluon
fusion cross section. This time, we refer to Tables
\ref{resggh} and \ref{resgghcomb}.
The main channel which contribute to this measurement
cross section is the
$H \fl ZZ \fl \ell \ell \nu\nu$ channel,
which gives statistical errors of
7\% and 9\%
for a 450 and 600~GeV Higgs respectively and largely contributes to
lower the total error.
The $H \fl ZZ \fl \ell \ell jj$ gives a contribution of 6\%
for a 450~GeV Higgs, but its signal to background ratio is small (0.12).
Note that the four leptons channel does not bring a main
contribution for that measurement, due
to its too small branching ratio.
We find a statistical error on the measurement of the gluon
fusion cross section of 8.9\%, 4.0\% and 7.7\%
for a 300, 450 and 600~GeV Higgs respectively.
\newpage
As a summary of the results obtained, 
we would like to present the errors we got for the
different cross sections and branching ratios applied
to the Standard Model predictions.
If we
assume that the values measured 
for the Higgs production through gluon fusion and weak boson fusion
cross sections
as well as the ratio between the two Higgs decay modes
are the ones predicted by the Standard Model,
which statistical error would they have~?
Here is what was obtained for the different Higgs
masses studied\footnote{The values for the different cross sections
and branching ratios are taken from \cite{crossec}.}:
\begin{center}
\fbox{\begin{Beqnarray*}
\hspace*{0.32cm}m_{Higgs}=300\,\mathrm{GeV}:&& \\
 \frac{BR(H \fl WW)}{BR(H \fl ZZ)}& =\quad  2.3 & \pm \quad 0.3 \\
 \sigma (qq \fl qqH)& = \quad 1.5 & \pm \quad 0.1 \quad pb\hspace*{0.32cm} \\
\sigma (gg \fl H)& = \quad  7.8 & \pm \quad  0.7 \quad pb 
\end{Beqnarray*}}
\\[0.5cm]
\fbox{\begin{Beqnarray*}
\hspace*{0.3cm}m_{Higgs}=450\,\mathrm{GeV}:&& \\
\frac{BR(H \fl WW)}{BR(H \fl ZZ)}& = \quad  2.1 & \pm  \quad 0.2 \\ 
\sigma (qq \fl qqH)& = \quad  0.65 & \pm  \quad  0.04 \quad  pb\hspace*{0.3cm} \\
\sigma (gg \fl H)& = \quad   2.7 & \pm   \quad 0.1 \quad  pb 
\end{Beqnarray*}}
\\[0.5cm]
\fbox{\begin{Beqnarray*}
\hspace*{0.3cm}m_{Higgs}=600\,\mathrm{GeV}:&& \\
\frac{BR(H \fl WW)}{BR(H \fl ZZ)}& = \quad 2.1 & \pm \quad 0.2 \\
\sigma (qq \fl qqH)& = \quad 0.40 & \pm \quad 0.02 \quad pb\hspace*{0.3cm}\\
\sigma (gg \fl H)& = \quad  1.8 & \pm \quad 0.1 \quad pb 
\end{Beqnarray*}}
\end{center}

We see that the statistical precision reached
with a luminosity of $100\,fb^{-1}$ is already quite good~!\\

A last point we would like to treat is 
how can this study, developed
here within the Standard Model, 
be used in other theoretical Models.
Actually the possibility to use the forward emitted jets produced in the
weak boson fusion process depends only on the coupling of the
Higgs to vector bosons, as the requirement for a Higgs to
be produced through weak boson fusion is that it can
couple to vectors bosons. Moreover the Higgs signatures 
studied were always characterized by a Higgs
decaying into vector bosons. So if the Higgs can be produced through
weak boson fusion, it can also decay in the studied modes.

The jet tagging method will then be usable in the frame of
every theory assuming
a Higgs coupling at vector bosons without a too small branching fraction.

Zeppenfeld and \textit{al.} developed for example
a theoretical method based on
the forward emitted jets coming from weak boson fusion processes
to isolate light MSSM Higgs \cite{zepp}.

Note finally that
the Higgs produced through gluon fusion relies on the Higgs
coupling to the top quarks. If we assume a theory where a
Higgs can couple in the same time to vector bosons and top quarks,
a similar study can be repeated, taking into account that
the ratio between the gluon fusion and weak boson cross section can 
be different from the one of the Standard Model (about 2 for high Higgs
masses).

	\chapitre*{Summary}
\addcontentsline{toc}{chapter}{Summary}
\markboth{Summary}{Summary}

A possibility to detect a Higgs, with Standard Model - like 
couplings, having a mass lying between
300 and 600~GeV was discussed for the following signatures:
\begin{equation}
\nonumber
\begin{split}
H &\fl ZZ \fl \ell \ell \ell \ell\\
H &\fl ZZ \fl \ell \ell \nu \nu\\ 
H &\fl WW \fl \ell \nu \ell \nu\\
H &\fl WW \fl \ell \nu jj\\
H &\fl ZZ \fl \ell \ell jj\\
H &\fl ZZ \fl \nu \nu jj
\end{split}
\end{equation}
We could isolate a Higgs signal in all of these channels except
for the $H \fl ZZ \fl \nu \nu jj$ where
too high QCD background was present.
We found that a very powerful selection cut for the channels
where hadronic decays were present was to use the two forward
emitted jets present in the weak boson fusion Higgs production
process. \\

It was shown that a 300~GeV Higgs signal, with the NLO cross section,
could be detected already with a luminosity between 2 and $3\,fb^{-1}$. The
two best channels were
$H \fl ZZ \fl \ell \ell \ell \ell$ and $qq\fl qqH \fl WW \fl \ell \nu jj$.
A 450~GeV Higgs could be discovered with a luminosity
of $2\,fb^{-1}$, using mainly the following processes:
$H \fl ZZ \fl \ell \ell \nu \nu$
and $qq\fl qqH \fl WW \fl \ell \nu jj$. 
Finally, a 600~GeV Higgs could be discovered with a luminosity
between 2 and $3\,fb^{-1}$, primarily from the
$H \fl ZZ \fl \ell \ell \nu \nu$ and
$qq\fl qqH \fl WW \fl \ell \nu jj$ channels.\\

With the use of the jet tagging technique,
the weak boson fusion process was isolated.
It was shown that the cross section for this process
could be measured with a statistical
precision of 6.8\%, 5.7\% and 6.2\% for a 
300, 450 and 600~GeV Higgs respectively, with a luminosity of $100\,fb^{-1}$.
The gluon fusion cross section could be measured with a statistical
precision of 8.9\%, 4.0\% and 7.7\% for a
300, 450 and 600~GeV Higgs respectively.
The ratio between the two decay branching fractions for a heavy SM Higgs,
$H \fl WW$ and $H \fl ZZ$
could be measured with a statistical precision of about 10\%,
with a luminosity of $100\,fb^{-1}$.\\

These results need be refined in the future.
The simulations did not
include a full detector simulation.
Hence no systematic errors were taken into account. 
For example, it would be interesting 
to consider the error on the energy or momentum measurement 
in the simulation and
to study how well the detector is actually able to reconstruct jets. 
Especially for the high luminosity phase, 
many proton proton interactions take place in one bunch crossing.
Furthermore proton proton interactions in the previous bunch crossing influence
the measurement (pile-up effects). They affect in particular the detection of the
Higgs from the decay modes involving jets.
Another point is that the
Monte Carlo programs which simulate the backgrounds are based only on LO
calculations. The next step would be to include NLO corrections.\\[0.5cm]

Although the reconstruction of the Higgs from the hadronic decay mode 
is more difficult than that from the leptonic decay mode, 
it increases significantly the signal statistics. This not only enhances
the chance of the discovery but also allows to study the
Higgs production and decay modes.\\

What I am asking myself after these four months of work
on that topic is if there
is in Nature a Higgs with a mass between 300
and 600~GeV and Standard Model - like couplings...
To get an answer to this question, see you in five years~!



     	\chapitre*{Acknowledgments}
\addcontentsline{toc}{chapter}{Acknowledgments}
\markboth{Acknowledgments}{Acknowledgments}

For these acknowledgments, I prefer to switch to French. \\

Ces quatre mois de travail de dipl\^ome ont \'et\'e pour moi 
tr\`es int\'eressants et riches en d\'ecouvertes. 
Je tiens \`a remercier de tout mon 
coeur toutes les personnes qui ont contribu\'e \`a
rendre cette exp\'erience possible.\\

Tout d'abord, j'aimerais
remercier Felicitas Pauss pour la possibilit\'e
qu'elle m'a donn\'ee de faire mon travail de dipl\^ome dans le groupe de
l'ETHZ au CERN.\\

Ensuite, mes remerciements vont sp\'ecialement \`a 
Michael Dittmar, qui m'a ``supervis\'ee'' tout
au long de ce travail et qui a su me guider et me conseiller, en faisant
preuve d'une patience sans bornes.
Merci d'avoir contribu\'e \`a m'ouvrir de nouveaux 
horizons aussi vari\'es que la physique du Higgs, la grammaire allemande 
(jetzt wei\ss~ich, da\ss~man nicht sagt: ``das fastfertige Kriminalroman''...)
ou encore le th\'e\^atre de marionnettes~!\\

J'aimerais aussi remercier tous
les gens du groupe de l'ETHZ et sp\'ecialement
Steve pour la salsa et les pauses caf\'e: ``salade de fruits, 
jolie, jolie, jolie...'', Dario pour les inoubliables soupers du mercredi
soir, Radek et Andr\'e pour l'aide inconditionnelle \`a  faire fonctionner
les ordinateurs (maintenant je crois que je sais rebooter le PC
de Michael les yeux ferm\'es...).\\

Merci enfin \`a ma famille qui a su me soutenir 
tout au long de mes \'etudes
et m'encourager dans la voie, parfois p\'enible, que j'ai choisie.\\

Encore deux mentions sp\'eciales \`a St\'ephane qui m'a donn\'e
l'id\'ee du titre 
lors d'une m\'emorable sortie \`a peaux de phoque
et \`a Olivier pour sa relecture attentive de l'introduction.

\end{document}